\documentclass[12pt,titlepage]{utarticle}
\usepackage{amsmath}
\usepackage{mathrsfs}
\usepackage{amsfonts}
\usepackage{amssymb}
\usepackage{amsthm}
\usepackage{mathtools}
\usepackage{color}
\usepackage{cite}
\usepackage{enumitem}
\usepackage{ucs}
\usepackage{xparse}
\usepackage{tikz}
\usetikzlibrary{cd}
\usepackage{hyperref}
\usetikzlibrary{decorations.pathreplacing,calligraphy}
\usepackage{comment}
\usepackage{float}
\usepackage{soul}
\usepackage{longtable}

\usepackage{booktabs}  
\usepackage{tabularx}  
\usepackage{array}     
\usepackage{xcolor}    
\usepackage{makecell}  
\newcolumntype{P}{>{\raggedright\arraybackslash}m{0.22\textwidth}}
\newcolumntype{E}{>{\centering\arraybackslash}m{0.19\textwidth}}

\usepackage{kpfonts}
\usepackage[scaled=0.90]{helvet} 
\usepackage{courier} 
\normalfont
\usepackage[T1]{fontenc}

\definecolor{aqua}{rgb}{0, 1.0, 1.0}
\definecolor{fuschia}{rgb}{1.0, 0, 1.0}
\definecolor{gray}{rgb}{0.502, 0.502, 0.502}
\definecolor{lime}{rgb}{0, 1.0, 0}
\definecolor{maroon}{rgb}{0.502, 0, 0}
\definecolor{navy}{rgb}{0, 0, 0.502}
\definecolor{olive}{rgb}{0.502, 0.502, 0}
\definecolor{purple}{rgb}{0.502, 0, 0.502}
\definecolor{silver}{rgb}{0.753, 0.753, 0.753}
\definecolor{teal}{rgb}{0, 0.502, 0.502}

\definecolor{midgreen}{RGB}{15,170,15}

\makeatletter
\newdimen\itex@wd%
\newdimen\itex@dp%
\newdimen\itex@thd%
\def\itexspace#1#2#3{\itex@wd=#3em%
\itex@wd=0.1\itex@wd%
\itex@dp=#2ex%
\itex@dp=0.1\itex@dp%
\itex@thd=#1ex%
\itex@thd=0.1\itex@thd%
\advance\itex@thd\the\itex@dp%
\makebox[\the\itex@wd]{\rule[-\the\itex@dp]{0cm}{\the\itex@thd}}}
\makeatother

\makeatletter
\newif\if@sup
\newtoks\@sups
\def\append@sup#1{\edef\act{\noexpand\@sups={\the\@sups #1}}\act}%
\def\reset@sup{\@supfalse\@sups={}}%
\def\mk@scripts#1#2{\if #2/ \if@sup ^{\the\@sups}\fi \else%
  \ifx #1_ \if@sup ^{\the\@sups}\reset@sup \fi {}_{#2}%
  \else \append@sup#2 \@suptrue \fi%
  \expandafter\mk@scripts\fi}
\def\tensor#1#2{\reset@sup#1\mk@scripts#2_/}
\def\multiscripts#1#2#3{\reset@sup{}\mk@scripts#1_/#2%
  \reset@sup\mk@scripts#3_/}
\makeatother

\makeatletter
\newbox\slashbox \setbox\slashbox=\hbox{$/$}
\def\itex@pslash#1{\setbox\@tempboxa=\hbox{$#1$}
  \@tempdima=0.5\wd\slashbox \advance\@tempdima 0.5\wd\@tempboxa
  \copy\slashbox \kern-\@tempdima \box\@tempboxa}
\def\slash{\protect\itex@pslash}
\makeatother

\def\clap#1{\hbox to 0pt{\hss#1\hss}}

\def\mathrlap{\mathpalette\mathrlapinternal}
\def\mathclap{\mathpalette\mathclapinternal}

\def\mathrlapinternal#1#2{\rlap{$\mathsurround=0pt#1{#2}$}}
\def\mathclapinternal#1#2{\clap{$\mathsurround=0pt#1{#2}$}}

\let\oldroot\root
\def\root#1#2{\oldroot #1 \of{#2}}
\renewcommand{\sqrt}[2][]{\oldroot #1 \of{#2}}

\DeclareSymbolFont{symbolsC}{U}{txsyc}{m}{n}
\SetSymbolFont{symbolsC}{bold}{U}{txsyc}{bx}{n}
\DeclareFontSubstitution{U}{txsyc}{m}{n}

\DeclareSymbolFont{stmry}{U}{stmry}{m}{n}
\SetSymbolFont{stmry}{bold}{U}{stmry}{b}{n}

\DeclareFontFamily{OMX}{MnSymbolE}{}
\DeclareSymbolFont{mnomx}{OMX}{MnSymbolE}{m}{n}
\SetSymbolFont{mnomx}{bold}{OMX}{MnSymbolE}{b}{n}
\DeclareFontShape{OMX}{MnSymbolE}{m}{n}{
    <-6>  MnSymbolE5
   <6-7>  MnSymbolE6
   <7-8>  MnSymbolE7
   <8-9>  MnSymbolE8
   <9-10> MnSymbolE9
  <10-12> MnSymbolE10
  <12->   MnSymbolE12}{}

\makeatletter
\def\re@DeclareMathSymbol#1#2#3#4{%
    \let#1=\undefined
    \DeclareMathSymbol{#1}{#2}{#3}{#4}}
\re@DeclareMathSymbol{\neArrow}{\mathrel}{symbolsC}{116}
\re@DeclareMathSymbol{\neArr}{\mathrel}{symbolsC}{116}
\re@DeclareMathSymbol{\seArrow}{\mathrel}{symbolsC}{117}
\re@DeclareMathSymbol{\seArr}{\mathrel}{symbolsC}{117}
\re@DeclareMathSymbol{\nwArrow}{\mathrel}{symbolsC}{118}
\re@DeclareMathSymbol{\nwArr}{\mathrel}{symbolsC}{118}
\re@DeclareMathSymbol{\swArrow}{\mathrel}{symbolsC}{119}
\re@DeclareMathSymbol{\swArr}{\mathrel}{symbolsC}{119}
\re@DeclareMathSymbol{\nequiv}{\mathrel}{symbolsC}{46}
\re@DeclareMathSymbol{\Perp}{\mathrel}{symbolsC}{121}
\re@DeclareMathSymbol{\Vbar}{\mathrel}{symbolsC}{121}
\re@DeclareMathSymbol{\sslash}{\mathrel}{stmry}{12}
\re@DeclareMathSymbol{\boxslash}{\mathrel}{stmry}{27}
\re@DeclareMathSymbol{\boxbslash}{\mathrel}{stmry}{28}
\re@DeclareMathSymbol{\boxast}{\mathrel}{stmry}{24}
\re@DeclareMathSymbol{\boxcircle}{\mathrel}{stmry}{29}
\re@DeclareMathSymbol{\boxbox}{\mathrel}{stmry}{30}
\re@DeclareMathSymbol{\obslash}{\mathrel}{stmry}{20}
\re@DeclareMathSymbol{\obar}{\mathrel}{stmry}{58}
\re@DeclareMathSymbol{\olessthan}{\mathrel}{stmry}{60}
\re@DeclareMathSymbol{\ogreaterthan}{\mathrel}{stmry}{61}
\re@DeclareMathSymbol{\bigsqcap}{\mathop}{stmry}{"64}
\re@DeclareMathSymbol{\biginterleave}{\mathop}{stmry}{"6}
\re@DeclareMathSymbol{\invamp}{\mathrel}{symbolsC}{77}
\re@DeclareMathSymbol{\parr}{\mathrel}{symbolsC}{77}
\makeatother

\makeatletter
\def\Decl@Mn@Delim#1#2#3#4{%
  \if\relax\noexpand#1%
    \let#1\undefined
  \fi
  \DeclareMathDelimiter{#1}{#2}{#3}{#4}{#3}{#4}}
\def\Decl@Mn@Open#1#2#3{\Decl@Mn@Delim{#1}{\mathopen}{#2}{#3}}
\def\Decl@Mn@Close#1#2#3{\Decl@Mn@Delim{#1}{\mathclose}{#2}{#3}}
\Decl@Mn@Open{\llangle}{mnomx}{'164}
\Decl@Mn@Close{\rrangle}{mnomx}{'171}
\Decl@Mn@Open{\lmoustache}{mnomx}{'245}
\Decl@Mn@Close{\rmoustache}{mnomx}{'244}
\Decl@Mn@Open{\llbracket}{stmry}{'112}
\Decl@Mn@Close{\rrbracket}{stmry}{'113}
\makeatother

\makeatletter
\DeclareRobustCommand\widecheck[1]{{\mathpalette\@widecheck{#1}}}
\def\@widecheck#1#2{%
    \setbox\z@\hbox{\m@th$#1#2$}%
    \setbox\tw@\hbox{\m@th$#1%
       \widehat{%
          \vrule\@width\z@\@height\ht\z@
          \vrule\@height\z@\@width\wd\z@}$}%
    \dp\tw@-\ht\z@
    \@tempdima\ht\z@ \advance\@tempdima2\ht\tw@ \divide\@tempdima\thr@@
    \setbox\tw@\hbox{%
       \raise\@tempdima\hbox{\scalebox{1}[-1]{\lower\@tempdima\box
\tw@}}}%
    {\ooalign{\box\tw@ \cr \box\z@}}}
\makeatother

\makeatletter
\NewDocumentCommand\mathraisebox{moom}{%
\IfNoValueTF{#2}{\def\@temp##1##2{\raisebox{#1}{$\m@th##1##2$}}}{%
\IfNoValueTF{#3}{\def\@temp##1##2{\raisebox{#1}[#2]{$\m@th##1##2$}}%
}{\def\@temp##1##2{\raisebox{#1}[#2][#3]{$\m@th##1##2$}}}}%
\mathpalette\@temp{#4}}
\makeatletter

\makeatletter
\def\udots{\mathinner{\mkern2mu\raise\p@\hbox{.}
\mkern2mu\raise4\p@\hbox{.}\mkern1mu
\raise7\p@\vbox{\kern7\p@\hbox{.}}\mkern1mu}}
\makeatother

\newcommand{\lt}{<}

\theoremstyle{plain}

\newtheorem{prop}{Proposition}

\theoremstyle{definition}
\newtheorem{defn}{Definition}

\theoremstyle{remark}

\newcounter{example}

\newcommand{\Kc}{K_{\mathcal{C}/\overline{\mathcal{M}}}}
\newcommand{\KcD}{\Kc(D)}

\begin{document}

\preprint{UTWI-34-2025\\}

\title{Families of Hitchin Systems in Type-D}

\author{
 Aswin Balasubramanian%
 \address{ School of Mathematics and Natural Sciences, \\ Chanakya University, Devanahalli,\\ Bengaluru, KA 562165, India \\
     \email{aswinb.phys@gmail.com}
},
 Jacques Distler%
 \address{
     Theory Group\\
     Department of Physics,\\
     University of Texas at Austin,\\
     Austin, TX 78712, USA \\
      \email{distler@golem.ph.utexas.edu}
 },\\
Ron Donagi%
\address{
    Department of Mathematics and\\ Department of Physics\\
    University of Pennsylvania\\
    Philadelphia, PA 19104-6395, USA\\
     \email{donagi@upenn.edu}\\
},
Carlos Perez-Pardavila${}^{\mathrm{b}\mathrlap{,}}$
\address{
    Beyond: Center for Fundamental Concepts in Science, \\
    Arizona State University, \\
    Tempe, AZ 85287, USA\\
     \email{cjp3247@utexas.edu}
}
}

\Abstract{
The Coulomb branch geometry of a 4d $\mathcal{N}=2$ SCFT is encoded in the data of a complex integrable system. In class-S, this is the Hitchin System (of ADE type) on the punctured curves $C$ on which we compactified from 6d to 4d. As we vary the complex structure of $C$, these fit together to form a (nontrivial!) bundle of Hitchin systems over the moduli space of complex structures of $C$ (the ``conformal manifold'' of the family of SCFTs). We carry out that construction for type-D. Compared to the type-A case, the construction is much more complicated because of local constraints at the punctures. Those local constraints were studied in \cite{Balasubramanian:2023iyx}. Here, we work out their implications for the global bundle of spectral (Seiberg-Witten) curves.
 } 

\date{October 30, 2025}
\maketitle 
\renewcommand{\baselinestretch}{0.90}\normalsize
\pagestyle{empty}
\tableofcontents
\renewcommand{\baselinestretch}{1.15}\normalsize
 \vfill\eject

\pagestyle{plain}
\setcounter{page}{1}
\section*{Summary of Notation}

\begin{longtable}{|l|p{3.5cm}|p{10cm}|}
\hline
\textbf{Symbol} & \textbf{First Occurence} & \textbf{Definition} \\
\hline
\endhead 
$B$ & Def in Eq. \ref{firstbase} & The base of the Hitchin integrable system for a fixed Riemann surface $C$. \\
\hline
$\mathcal{B}$ & \S\ref{families_section_type_A} ($A_N$) and \S\ref{families_section_type_D} ($D_N$) & The global version of $B$ which fits together as a family to form a graded holomorphic vector bundle over the compactified moduli space $\overline{\mathcal{M}}_{g,n}$ . \\
\hline 
$B_{naive}$ & Def in Eq. \ref{naivebase} & The ``naïve Hitchin base'' (on a fixed curve) before local constraints at the punctures are imposed. Its global version (over $\overline{\mathcal{M}}_{g,n}$) is the coherent sheaf introduced in Eq \ref{naivebasefamily}. \\ 
\hline 
$L_{2k}$ & \S\ref{Type_D_Global_Hitchin_Def} & These are the base line bundles in type D. They are defined as $L_{2k}=K_{C}(D)^{2k}\otimes\mathcal{O}(-\sum_{a}\chi_{2k}^{(a)}p_{a})$ analogous to Eq \ref{firstbase} for type A. \\
\hline
$\tilde{L}$ & \S\ref{Type_D_Global_Hitchin_Def} & This is the square root of the base line bundle $L_{2N}$. \\
\hline
$\mathcal{S}_{p,k}$ & Eq. \ref{defn_S} & Torsion sheaves on the universal curve $C$ introduced when solving even type constraints. \\
\hline
$\mathcal{J}_{p,k}$ &  Eq. \ref{Jprops} & Line bundles on the universal curve $C$ meant to replace the torsion sheaves $\mathcal{S}_{p,k}$ so that a bundle of spectral curves can be constructed. .  \\
\hline
$\mathcal{F}$ & Eq. \ref{veryevenSES} & A rank-2 vector bundle on $\mathcal{C}_{g,n}$ that serves as the kernel in the exact sequence in Eq \ref{veryevenSES}. It arises when dealing with very-even partitions with a single distinct even part. \\
\hline
$\mathcal{V}$ & Eq. \ref{naivebase} and Eq. \ref{V_explicit}  & It is define to be the coherent sheaf on $C$ that replaces the direct sum of line bundles in the ``naïve Hitchin base'' after constraints are imposed. The Hitchin base is then recovered as $B=H^0(C,\mathcal{V})$. The explicit form of $\mathcal{V}$ is given in Eq. \ref{V_explicit}. \\
\hline
$\mathcal{V}^{\prime}$ & Eq. \ref{V_prime_explicit} & The twisted version of $\mathcal{V}$ that ensures the direct image is a locally-free vector bundle. \\
\hline
\end{longtable}
\section{Introduction}

The Coulomb branch geometry of a 4d $\mathcal{N}=2$ supersymmetric quantum field theory is governed by a complex integrable system \cite{Seiberg:1994rs, Seiberg:1994aj,Martinec:1995by,Donagi:1995cf,Freed:1997dp}. For a 4d \emph{superconformal} field theory (SCFT), the Coulomb branch has a $\mathbb{C}^\times$ scaling symmetry (the complexification of the $U(1)_r$ R-symmetry of the SCFT), and so takes the form of a complex cone, with the SCFT at the tip. If the Coulomb branch chiral ring is freely-generated (which we will assume), this is a graded vector space $B=\oplus_k B_k$, where the subscript denotes the $\mathbb{C}^\times$ weight of each graded component.

The superconformal case has another feature, namely that such theories frequently occur in families (there is a finite-dimensional ``conformal manifold'' of exactly-marginal deformations of the SCFT). In all known cases, there is a limit where the family can be interpreted as the gauging (with semi-simple gauge group $G$) of an isolated $\mathcal{N}=2$ SCFT. The marginal deformations, at least near that limit, are interpreted as the exactly-marginal complex gauge couplings ($\tau_i=\frac{\theta_i}{2\pi} +\frac{4\pi i}{g_i^2}$) of the gauge theory. There can be multiple such ``weak coupling limits'' leading to \emph{dual} descriptions of the same abstract family of SCFTs.

There is a particularly interesting class of such $\mathcal{N}=2$ SCFTs, dubbed ``class-S'' \cite{Gaiotto:2009we,Gaiotto:2009hg}, in which (for present purposes, we restrict ourselves to the case of ``regular punctures'' in untwisted class-S) the ``conformal manifold'' is the moduli space 
$\overline{\mathcal{M}}_{g,n}$ of Deligne-Mumford stable punctured curves, and, for each point $C_{g,n}\in \mathcal{M}_{g,n}$, the complex integrable system is a Hitchin system on the punctured Riemann surface $C_{g,n}$. This naturally suggests that we consider \emph{families} of Hitchin systems as we vary the complex structure of $C$. These families extend to the boundary. As we approach the boundary, some simple gauge group $H$ becomes weakly-coupled. Weak-coupling is where physicists have good control over the behaviour of the theory, and that insight tells us how to extend our family of Hitchin systems over the locus where $C$ becomes a nodal curve.

In the zero-coupling limit, $\dim(H)$ $\mathcal{N}=2$ vector multiplets decouple and what remains\footnote{To be really precise, what remains is the $H$-invariant subsector of $\mathcal{T}$.} is an SCFT $\mathcal{T}$ with global symmetry group $F$ of which we had previously weakly-gauged $H\subset F$. If we rescale the Coulomb branch (using the $\mathbb{C}^\times$ scaling symmetry) to zoom in on the Coulomb branch of $\mathcal{T}$, the (rescaled) VEVs of the complex scalars in Cartan of $H$ remain as non-dynamical mass parameters, even as the fluctuating parts of the vector multiplets decouple. In the Hitchin System language, this means that the symplectic integrable system determines a Poisson integrable system, where the Casimirs are the aforementioned $\operatorname{rank}(H)$ VEVs. More precisely, as we approach the boundary, $\operatorname{rank}(H)$ directions in the fiber of the Hitchin system decompactify. Let $v_i= \{b_i,\cdot\},\; i=1,\dots,\operatorname{rank}(H)$ be the Hamiltonian vector fields which generate translations in those noncompact directions. The quotient of our original symplectic integrable system by the flow generated by the $v_i$ is a Poisson integrable system, where the $b_i$ are Casimirs. That Poisson integrable system governs the Coulomb branch geometry of the physical theory $\mathcal{T}$ from which the $\mathcal{N}=2$ vector multiplets have decoupled.

The SCFTs of class-S are labeled by a choice of ADE Lie algebra $\mathfrak{j}$ and a collection of $n$ nilpotent orbits $O_a \subset \mathfrak{j}$. We refer to this collection of nilpotents $\mathcal{O}_a$ as the \textit{Hitchin nilpotents}. The complex integrable system which determines the Coulomb branch geometry is a $\mathfrak{j}$-Hitchin System on $C$, the moduli space of meromorphic Higgs bundles where the Higgs field has a simple pole at each $p_a$ with residue $\in O_a$ \cite{markman1994spectral,bottacin1995symplectic,MR3545326}.  The Coulomb branch, the aforementioned graded vector space $B$, is the base of the Hitchin system. As we vary the complex structure of $C$, these fit together to form a graded holomorphic vector bundle $\mathcal{B}$ over $\mathcal{M}_{g,n}$ which (if we incorporate, as dictated by the physics, a certain twisting at the boundary in our definition) extends to a holomorphic vector bundle $\mathcal{B}$ over all of $\overline{\mathcal{M}}_{g,n}$.

That vector bundle is nontrivial. Its transition functions tell us how to relate the local operators which parametrize the Coulomb branch in one region (say, near a point where we have one weakly-coupled description of the SCFT) to those which parametrize the Coulomb branch in another region (say, near some other weakly-coupled description). The dictionary between the two descriptions of the SCFT involves some horrible non-local change-of-variables; but in this subsector (of operators parametrizing the Coulomb branch) the change-of-variables is both \emph{local} and \emph{calculable}. In fact, not only do we get a vector bundle of Hitchin bases, we are able to construct a bundle of \emph{spectral curves} (AKA Seiberg-Witten curves), from which the Coulomb branch geometry can be recovered. That bundle of spectral curves is also nontrivial.

We carried out this program for class-S theories of type-A in \cite{Balasubramanian:2020fwc}. We explained the twisting at the boundary, the construction of a ``universal family'' of spectral curves over $\overline{\mathcal{M}}_{g,n}$, and 
a very concrete formula for the vector bundle $\mathcal{B}$ in the case of $\overline{\mathcal{M}}_{0,4}$. We review some of that construction in \S\ref{AFamilies}.

In \cite{Balasubramanian:2023iyx}, we studied the local behaviour of the Hitchin base in type-D. There are many subtle differences from the type-A case, as we review in \S\ref{sec:local_hitchin_base}. The chief difference is the presence of \emph{constraints} \cite{Chacaltana:2011ze,MR3815160}. Let $t$ be a local coordinate at the puncture. If we compute $\det(\Phi(t)-w\Bid)$, where $\Phi(t)$ is the Higgs field, it has an expansion in symmetric polynomials $\phi_k(t)$. In type-D, the leading Laurent coefficients of the $\phi_k(t)$ obey polynomial relations (``constraints''), whereas in type-A they are independent. This has rather intricate implications for the global case and it is the purpose of this paper to unravel those implications.

The most na\"\i ve version of the global construction would be to ignore the constraints, pretend we were working in $\mathfrak{j}=\mathfrak{su}(2N)$ instead of $\mathfrak{so}(2N)$, and then set to zero all of the Hitchin base parameters in the $\phi_k$ for $k$ odd and tune the parameters in $\phi_{2N}$ so that it is a perfect square: $\phi_{2N}=\tilde{\phi}^2$. That is, indeed, the correct procedure if all of the nilpotent orbits $O_a$ are the regular nilpotent orbit $[2N-1,1]$.  

\subsection{What is done in this paper}

When we work in families, we construct a collection of line bundles $\mathcal{L}_k$ over the universal curve $\mathcal{C}_{g,n}\xrightarrow{\;\pi\;} \overline{\mathcal{M}}_{g,n}$ and $\mathcal{B}_k=\pi_*(\mathcal{L}_k)$. If we replace some of the $[2N-1,1]$ orbits by smaller nilpotent orbits, even in type-A, we need to ``twist,'' i.e. replace the $\mathcal{L}_k$ by $\mathcal{L}'_k=\mathcal{L}_k(-\sum_S n_S C_S)$ where the $C_S$ are certain divisors supported over the boundary of $\overline{\mathcal{M}}_{g,n}$ in order for $\pi_*(\mathcal{L}'_k)$ to be a vector bundle. The same is true in type-D, except that --- whenever the local Hitchin base associated to $O_a$ has constraints --- we need to solve those constraints \emph{first}, before twisting. How to impose the constraints involves a delicate interplay between the local behaviour near a puncture and global considerations. That (rather involved) discussion, in \S\ref{imposing}, is the heart of this paper.

The main distinction is between \emph{even-type} constraints, which decrease the dimension of the Hitchin base, and \emph{odd-type} constraints which don't. Both generically\footnote{The exceptions are the constraints associated to an even part (with even multiplicity) at the beginning of the partition or a very even partition with just one or two distinct even parts.} involve introducing new line bundles $\mathcal{J}$ over the universal curve, whose direct image(s) 
$\pi_*(\mathcal{J})$ contribute to the bundle of Hitchin bases. For genus-0, we conjecture that such a $\mathcal{J}$ always exists. As we shall see, however, it may not be unique. Different choices of $\mathcal{J}$ correspond to different choices of fiber coordinates on $\mathcal{B}\to\overline{\mathcal{M}}_{0,n}$.

Our work in this paper is confined to genus-0. This has the advantage that 
\begin{itemize}
\item The moduli spaces $\overline{\mathcal{M}}_{0,n}$ are smooth complex varieties. We don't have to deal  with the complications associated to the stacky nature of $\overline{\mathcal{M}}_{g,n}$ for $g>0$.
\item $\operatorname{Pic}(\overline{\mathcal{M}}_{0,n})$ is discrete, so we have excellent control over the line bundles that appear in our construction.
\end{itemize}
We hope to return to the higher genus case in another publication.

We mostly work in the conformal limit (where the Hitchin system on a smooth curve $C$ is symplectic). This is what obtains when the residues of the Higgs field at the punctures are nilpotent. One can also deform the residues onto a sheet\footnote{A sheet is an irreducible component of the union of equidimensional adjoint orbits. See \cite{Balasubramanian:2018pbp} for a more precise statement about sheets and mass deformations of the SCFT.} in $\mathfrak{j}$. This corresponds to turning on a relevant perturbation (``mass deformation'') of the SCFT. In type-A, there is a unique sheet attached to each nilpotent orbit in $A_{N-1}$. In type-D, this is no longer true. But as we saw in \cite{Balasubramanian:2023iyx}, there is a 1-1 correspondence between subgroups of a certain finite group $\overline{A}_b(\mathcal{O}_H)$ attached to the Hitchin nilpotent $\mathcal{O}_H$ and the ``Nahm'' nilpotent orbits $\{O_N\}$ in the dual special piece\footnote{These are the nilpotent orbits that map to $O_H$ under the Spaltenstein-Barbasch-Vogan map $d_S$. See \cite{Chacaltana:2012zy} for more details and a more complete explanation of the relation between Nahm and Hitchin nilpotent orbits.}. The different choices of Nahm orbit correspond to \emph{different} SCFTs whose associated Hitchin systems are closely-related, but not identical. The distinction is sometimes hard to see by examining the Hitchin systems directly in the conformal (symplectic) limit, but becomes clear once we turn on the mass deformations. We explore this a bit in Appendix \ref{massappendix}, where we show how mass deformations distinguish different choices of Nahm orbits, and hence of the associated special sheets, that may be obscured in the conformal limit. We do not construct the general mass-deformed global Hitchin system here, deferring a more complete discussion to another publication.

In \S\ref{AFamilies} we recap the basic formalism for constructing families of Hitchin systems in type-A ($\mathfrak{j}=\mathfrak{su}(N)$) at genus-0. Much of this formalism carries over directly to type-D ($\mathfrak{j}=\mathfrak{so}(2N)$). \S\ref{sec:local_hitchin_base} reviews \cite{Balasubramanian:2023iyx} on the local behaviour in type-D. \S\ref{GlobalD} aims to put these two strands together. After some preliminaries about the families of Hitchin systems on smooth curves, \S\ref{nilpnode} classifies the possible degeneration limits of the SCFT when the curve $C$ develops a node. \S\ref{imposing} then gets down to the nitty-gritty of how to impose the constraints of \S\ref{sec:local_hitchin_base} in our global setting. \S\ref{even_easy_cases} - \S\ref{veryevensingle} treat the case of even type constraints while \S\ref{oddtype} treats the case of odd-type constraints.  \S\ref{twisting} explains the twisting at the boundary of the moduli space of the line bundles on the universal curve required such that their direct image --- the bundle of Hitchin bases $\mathcal{B}$ --- extends as a bundle over the boundary. In \S\ref{directimage}, we explain how to compute that direct image, giving very concrete formul\ae\ for $\overline{\mathcal{M}}_{0,4}$. \S\ref{sec:examples} illustrates this whole (admittedly rather involved) construction with examples for $n=3,4,5$ punctures. If those feel insufficient, Appendix \ref{so8complete} gives a complete prescription for constructing the spectral curves for all possible combinations of nilpotent orbits  in $\mathfrak{j}=\mathfrak{so}(8)$ for the 4-punctured sphere.

%

\section{Review}
\subsection{Families of Hitchin systems in Type-A}\label{AFamilies}
\label{families_section_type_A}
Let us start by reviewing the construction of \cite{Balasubramanian:2020fwc}. Let $C$  be a closed stable smooth curve with $n$ marked points\footnote{$n\geq 3$ for $g=0$ and $n\geq 1$ for $g=1$.}. We study meromorphic Higgs bundles on $C$ where, at each marked point $p_a$, we fix the residue of the Higgs field to lie in some given nilpotent conjugacy class $O_{H,a}$. That is, we study the moduli space $\text{\sl Higgs}$ of pairs $(V,\Phi)$, where $V$ is an $SL_N$ principal bundle and $\Phi\in H^0(C,ad(V)\otimes K_C(D))$, where $D=\sum_a p_a$ and we restrict $\Phi|_{p_a}\in O_{H,a}$. There is a natural map, the  Hitchin map
\[
\mu: \text{\sl Higgs}\to B
\]
where\footnote{Denoting the nilpotent orbit $O_H$ by a partition $[P]$ of $N$,
\[
\chi_k =l,\qquad \text{s.t.} \sum_{i=1}^{l-1}P_i< k \leq \sum_{i=1}^{l}P_i.
\] The function $\chi_k ([P])$ is sometimes called the \textit{level function} associated to the partition $[P]$ \cite{su2022parabolic,lee2025relative}. }
\begin{equation}\label{firstbase}
B=\bigoplus_{k=2}^{N} H^0(C,L_k),\qquad L_k\coloneqq K_C(D)^{\otimes k}\otimes \mathcal{O}\bigl(-\sum_{a=1}^n \chi_k^{(a)}p_a\bigr)
\end{equation}
whose generic fiber is a complex Lagrangian torus. In other words, the pair $(\text{\sl Higgs},\mu)$ is a complex symplectic integrable system.

We restrict ourselves to ``OK'' systems, where $H^1(C,L_k)=0$. This is trivially satisfied for $g>0$ or, if $g=0$, for $n>2(N-1)$. For $g=0$ and $3\leq n\leq2(N-1)$, it is a \emph{condition} on the nilpotent orbits at the punctures.

Now we wish to work in families. As we vary a smooth curve $C\in \mathcal{M}_{g,n}$, the base line bundles $L_k$ fit together to form holomorphic line bundles $\mathcal{L}_k$ over the universal curve $\mathcal{C}_{g,n}\xrightarrow{\;\pi\;}\overline{\mathcal{M}}_{g,n}$. These line bundles extend over the boundary, where $C$ develops a node. However it may happen that when $C$ develops a node, $H^1(C,\mathcal{L}_k)$ can jump. 

To fix this, we have to ``twist'' the line bundles $\mathcal{L}_k$, as in \cite{Balasubramanian:2020fwc}.  The Deligne-Mumford compactification, $\overline{\mathcal{M}}_{g,n}$, contains boundary components corresponding to both separating and non-separating nodes. At the former, $C$ is a reducible curve $C=C_S \cup C_{S^\vee}$, where $C_S$ has genus $g_{C_S}$ and contains a subset $S\subset\{p_1,p_2,\dots,p_n\}$ of the marked points\footnote{For stability, $S$ must contain at least two points  when $g_{C_S}=0$. When $g_{C_S}>0$, there's no condition on the number of points in $S$. Similarly, for $g-g_{C_S}>0$, there's no condition on the number of points in $S^\vee$.}
and $C_{S^\vee}$ has genus $g-g_{C_S}$ and contains the complementary set of marked points.
We define 
\begin{equation}\label{nksdef}
n_k^S\coloneqq\max\Bigl(0,k-1- g_{C_S}(2k-1)-\sum_{p_a\in S}(k-\chi_k^{(a)})\Bigr)
\end{equation}
This vanishes for $g_{C_S}>0$, so we might as well restrict to the case where $C_S$ has genus-0. Let $\mathcal{C}_S$ be the Cartier divisor on $\mathcal{C}_{g,n}$ corresponding to $C_S$.
 The line bundles over the universal curve
\begin{equation}\label{Lprimedef}
\mathcal{L}'_k=\mathcal{L}_k\otimes\mathcal{O}(-\sum_Sn_k^S\mathcal{C}_S)
\end{equation}
have the property that they are isomorphic to $\mathcal{L}_k$ when restricted to a smooth fiber of the universal curve, and, when restricted to any fiber (nodal or smooth), $H^1(C,\mathcal{L}'_k)=0$. The Hitchin bases $B$, for each fixed curve $C$, now fit together to form a graded holomorphic vector bundle $\mathcal{B}\to \overline{\mathcal{M}}_{g,n}$.
\[
\mathcal{B}= \bigoplus_{k=2}^N\mathcal{B}_k,\qquad \mathcal{B}_k\coloneqq\pi_*(\mathcal{L}'_k)
\]
As we shall see, $\mathcal{B}$ is a \emph{nontrivial} vector bundle over $\overline{\mathcal{M}}_{g,n}$.

Pulling back $\mathcal{B}$ to the universal curve, we form a bundle\footnote{For each $C\in\mathcal{M}_{g,n}$, we have a space of Higgs bundles whose base is $\mathcal{B}|_C$ and whose fiber is $\text{Jac}(\Sigma)$.} of spectral curves\goodbreak\noindent $\Sigma\to\pi^*(\mathcal{B})\to\mathcal{C}_{g,n}$
\begin{equation}\label{slNspectral}
\begin{split}
\Sigma&=\bigl\{0=\det(\Phi-w\Bid)\bigr\}\\
&= \bigl\{0=w^N-\textstyle{\sum_{k=2}^N}  w^{N-k}\phi_k\bigr\}\subset \KcD.
\end{split}
\end{equation}
Here $w$ is a fiber coordinate on the $D$-twisted relative cotangent bundle\footnote{$\Kc$ is the line bundle on $\mathcal{C}_{g,n}$ whose restriction to any fiber $C$ is $K_C$. $D=\sum E_a$ is the divisor on $\mathcal{C}_{g,n}$ whose intersection with any fiber $C$ is the divisor $D=\sum p_a$ on $C$.} (dualizing sheaf) $\KcD\to \mathcal{C}_{g,n}$, and $\phi_k\in H^0(C,\mathcal{L}'_k)$. Locally, near $p_a$,
\[
\phi_k(t) =t^{\chi^{(a)}_k}(c_k +O(t))
\]
where $t$ is a local coordinate centered at $p_a$, and the $\chi^{(a)}_k$ are the coefficients appearing in \eqref{firstbase}.

Since $\mathcal{B}\to \overline{\mathcal{M}}_{g,n}$ is nontrivial, so too is the bundle of spectral curves $\Sigma\to \overline{\mathcal{M}}_{g,n}$. When we write explicit formul\ae, they will always be understood as being with respect to a local trivialization of $\mathcal{B}$. In particular for $\overline{\mathcal{M}}_{0,4}=\mathbb{CP}^1$, we can choose trivializations good in the northern hemisphere ($\lambda\neq\infty$), and in the southern hemisphere ($\lambda\neq 0$).

Mostly, we will restrict ourselves to genus-0 and low numbers of punctures. For these, we have very explicit models.

\subsubsection{\texorpdfstring{$\mathcal{C}_{0,n}$}{C₀,ₙ}}\label{sec:C0n}
For any $n$, $\mathcal{C}_{0,n}=\overline{\mathcal{M}}_{0,n+1}$ is a smooth projective variety of dimension $n-2$. It is birational to $(\mathbb{CP}^1)^{n-2}$. Our methods readily extend to the $n$-punctured sphere, though concrete formul\ae\ are harder to write down.

Sean Keel, in his PhD thesis \cite{KeelThesis1992},  showed how $\overline{\mathcal{M}}_{0,n}$ can be constructed by a succession of smooth blowups. As a consequence

\begin{itemize}
\item $H^{\text{odd}}(\overline{\mathcal{M}}_{0,n})=0$
\item $H^{2i}(\overline{\mathcal{M}}_{0,n})=A^i(\overline{\mathcal{M}}_{0,n})$
\end{itemize}

\noindent
and Keel constructed the Chow ring $A^\bullet(\overline{\mathcal{M}}_{0,n})$. In particular, $Pic(\overline{\mathcal{M}}_{0,n})=H^2(\overline{\mathcal{M}}_{0,n})=A^1(\overline{\mathcal{M}}_{0,n})$ is freely-generated of dimension
\[
\dim(A^1(\overline{\mathcal{M}}_{0,n}))= 2^{n-1} -n(n-1)/2 -1
\]
It is generated by the Chow classes $D_S$ of the boundary divisors, subject to the linear relations
\begin{equation}\label{relations}
\begin{gathered}
D_{S} = D_{S^\vee}\\
\sum_{\mathclap{\begin{smallmatrix}a,b\in S\\ c,d\in S^\vee\end{smallmatrix}}} D_S =
\sum_{\mathclap{\begin{smallmatrix}a,c\in S\\b,d\in S^\vee\end{smallmatrix}}} D_S =
\sum_{\mathclap{\begin{smallmatrix}a,d\in S\\b,c\in S^\vee\end{smallmatrix}}} D_S,\qquad \forall\; a,b,c,d\; \text{distinct}
\end{gathered}
\end{equation}
where $S\subset\{1,2,\dots,n\}$ and $|S|,|S^\vee| \geq 2$. The total class of the boundary, $\delta$, is most easily written by singling out one of the points (say, the $n^{\text{th}}$ one) and writing
\begin{equation}
\delta = \sum_{n\in S} D_S
\end{equation}

Let $\sigma_a: \overline{\mathcal{M}}_{0,n}\to \mathcal{C}_{0,n}$ be the $n$ sections. The point line bundles, $I_a\coloneqq\sigma_a^*(\Kc)$, have first Chern classes (Lemma 7.4 of \cite{ACGHvolII})
\begin{equation}
\psi_a= c_1(I_a)=\sum_{\mathclap{\begin{smallmatrix}a\in S\\b,c\in S^\vee\end{smallmatrix}}} D_S
\end{equation}
for $a\neq b\neq c$.

The bundle of Hitchin bases,  $\pi_*(\mathcal{L}_k)$ is a holomorphic vector bundle on $\overline{\mathcal{M}}_{0,n}$.  For $k=2$ it is the log-cotangent bundle $\mathcal{T}^\vee = T^*(\overline{\mathcal{M}}_{0,n})(\log\delta)$ which fits into the exact sequence
\begin{equation}\label{logTangent}
0\to T^*(\overline{\mathcal{M}}_{0,n}) \to T^*(\overline{\mathcal{M}}_{0,n})(\log\delta) \to \bigoplus_{n\in S} \mathcal{O}_{D_S}\to 0
\end{equation}
For  $k>2$ and  $n>4$, the classification of holomorphic bundles on $\overline{\mathcal{M}}_{0,n}$ is \emph{complicated}. So, in those cases, we will content ourselves with computing their topological type (i.e., their Chern classes).

\subsubsection{\texorpdfstring{$\mathcal{C}_{0,3}$}{C₀,₃}}\label{C03}

$\overline{\mathcal{M}}_{0,3}$ is a point; $\mathcal{C}_{0,3}=\mathbb{CP}^1$, on which we choose homogeneous coordinates $x,y$. The three punctures are located at $p_1=(1,0)$, $p_2=(0,1)$ and $p_3=(1,1)$. The polynomial equation for $\Sigma$ is bihomogeneous, where $w$ has $(\mathbb{C}^\times)^2$ weights $(1,1)$, $x$ and $y$ have weights $(1,0)$, the fiber coordinates on $\mathcal{B}_k$ have weights $(0,k)$, $\Sigma$ has weight $(N,N)$ and the Seiberg-Witten differential $\lambda_{\text{SW}}$ has weights $(0,1)$.

\subsubsection{\texorpdfstring{$\mathcal{C}_{0,4}$}{C₀,₄}}\label{C04}

$\overline{\mathcal{M}}_{0,4}=\mathcal{C}_{0,3}=\mathbb{CP}^1$ which we take to have homogeneous coordinates $\lambda_1,\lambda_2$. Set $\lambda=\lambda_1/\lambda_2$.  $\mathcal{C}_{0,4}$ is a complex surface, dP${}_4$. Let $x,y,z$ be the homogeneous coordinates on $\mathbb{CP}^2$. We blow up 4 points
\[
E_{1}\to (1,0,0),\quad 
E_{2}\to (0,1,0),\quad 
E_{3}\to (0,0,1),\quad 
E_{4}\to (1,1,1)
\]
to obtain $\mathcal{C}_{0,4}$. The projection
\[
\pi: \mathcal{C}_{0,4}\to \overline{\mathcal{M}}_{0,4}
\]
is given by the solution\footnote{This is written as a (rational) map from $\mathbb{CP}^2$ to $\mathbb{CP}^1$ which is undefined at the 4 $E_i$, but composing with the blow-up it becomes an everywhere defined morphism dP${}_4\to\mathbb{CP}^1$. } to
\begin{equation}\label{C04fiber}
\lambda_1 x(y-z)+\lambda_2 y(z-x)=0.
\end{equation}
Over the interior  of $\mathcal{M}_{0,4}$ the fiber  $C_\lambda= \pi^{-1}(\lambda)$ is a smooth genus-0 curve which intersects each of the point divisors $E_{a}$ once. At the boundary, $\lambda=0,1,\infty$, \eqref{C04fiber} degenerates to a pair of lines which intersect at the node and each of which intersects two of the four point divisors $E_a$.

\vbox{
\begin{displaymath}
C_0=\{y(z-x) = 0\},\quad\text{(with the node at}\,n_0=(1,0,1)\,)
\end{displaymath}
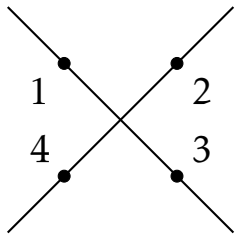
\begin{center}
\begin{tikzpicture}
\begin{scope}[scale=.75]
\draw[thick,-] (0,0) -- (4,-4);
\draw[thick,-] (4,0) -- (0,-4);
\filldraw (1,-1) circle (3pt) node[anchor=north east, scale=1.25] {$1$};
\filldraw (3,-3) circle (3pt) node[anchor=south west, scale=1.25] {$3$};
\filldraw (3,-1) circle (3pt) node[anchor=north west, scale=1.25] {$2$};
\filldraw (1,-3) circle (3pt) node[anchor=south east, scale=1.25] {$4$};
\end{scope}
\end{tikzpicture}
\end{center}
}

\vbox{
\begin{displaymath}
C_1=\{z(x-y) = 0\},\quad\text{(with the node at}\,n_1=(1,1,0)\,)
\end{displaymath}
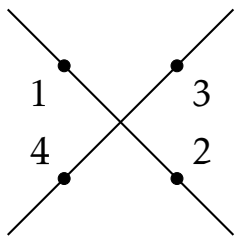
\begin{center}
\begin{tikzpicture}
\begin{scope}[scale=.75]
\draw[thick,-] (0,0) -- (4,-4);
\draw[thick,-] (4,0) -- (0,-4);
\filldraw (1,-1) circle (3pt) node[anchor=north east, scale=1.25] {$1$};
\filldraw (3,-3) circle (3pt) node[anchor=south west, scale=1.25] {$2$};
\filldraw (3,-1) circle (3pt) node[anchor=north west, scale=1.25] {$3$};
\filldraw (1,-3) circle (3pt) node[anchor=south east, scale=1.25] {$4$};
\end{scope}
\end{tikzpicture}
\end{center}
}

\vbox{
\begin{displaymath}
C_{\infty}=\{x(y-z) =0\},\quad\text{(with the node at}\,n_{\infty} = (0,1,1)\,)
\end{displaymath}
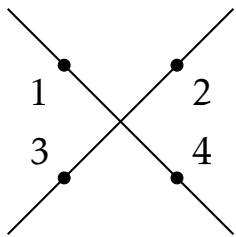
\begin{center}
\begin{tikzpicture}
\begin{scope}[scale=.75]
\draw[thick,-] (0,0) -- (4,-4);
\draw[thick,-] (4,0) -- (0,-4);
\filldraw (1,-1) circle (3pt) node[anchor=north east, scale=1.25] {$1$};
\filldraw (3,-3) circle (3pt) node[anchor=south west, scale=1.25] {$4$};
\filldraw (3,-1) circle (3pt) node[anchor=north west, scale=1.25] {$2$};
\filldraw (1,-3) circle (3pt) node[anchor=south east, scale=1.25] {$3$};
\end{scope}
\end{tikzpicture}
\end{center}
}

Again, $\Sigma$ is given by a bihomogeneous polynomial of degree $(N,N)$ where $w$ has $(\mathbb{C}^\times)^2$ weights $(1,1)$, $x,y,z$ have weights $(1,0)$, the fiber coordinates on $\mathcal{B}_k$ have weights $(0,k)$ and the Seiberg-Witten differential $\lambda_{\text{SW}}$ has weights $(0,1)$.

Every holomorphic vector bundle on $\mathbb{CP}^1$ splits as a direct sum of line bundles labeled by their degrees. So, in particular,
\begin{equation}\label{degrees}
\mathcal{B}_k= \bigoplus_i \mathcal{O}(d_{i,k})
\end{equation}
We can trivialize in the northern hemisphere ($\lambda\neq\infty$) and in the southern hemisphere ($\lambda\neq 0$). The degrees in \eqref{degrees} encode the transition functions for $\mathcal{B}$ and hence for the family $\Sigma\to \overline{\mathcal{M}}_{0,4}$ of spectral curves. We found an explicit formula for those degrees in \cite{Balasubramanian:2020fwc}, which we review in \S\ref{directimage}.

\subsubsection{\texorpdfstring{$\mathcal{C}_{0,5}$}{C₀,₅}}
We already discussed the geometry of $\overline{\mathcal{M}}_{0,5}=\text{dP}_4$ in \S\ref{C04}.  The boundary $\delta$ consists of 10 lines $D_{ab}$  with intersection numbers:
\[
\begin{split}
D_{ab}\cap D_{ab}&=-P\\
D_{ab}\cap D_{cd}&=P\qquad a,b,c,d\;\text{distinct}\\
D_{ab}\cap D_{cd}&=0\qquad \text{otherwise}
\end{split}
\]
where $P$ is the class of a point and we use $\cap$ for multiplication in the Chow ring. We easily see that $\delta=\sum  D_{ab}$ is twice an integer class and $(\tfrac{1}{2}\delta)\cap (\tfrac{1}{2}\delta) = 5 P$. 

The homology ring of $\mathcal{C}_{0,5}=\overline{\mathcal{M}}_{0,6}$ is freely-generated by the classes $E_a\coloneqq D_{a6}$, $\mathcal{C}_{ab}\coloneqq D_{ab6}$ and
\[
K\coloneqq c_1(\Kc) = D_{ab}+D_{cd6}+D_{ce6}+D_{de6}-D_{a6}-D_{b6},\qquad a,b,c,d,e\; \text{distinct}
\]
The projection $\pi: \mathcal{C}_{0,5}\to\overline{\mathcal{M}}_{0,5}$ is the forgetful map that forgets the $6^{\text{th}}$ point. The $E_a$, $a=1,\dots,5$,  are the divisors on $\mathcal{C}_{0,5}$ which are the images of the 5 sections.

Concretely\footnote{An alternative presentation of $\overline{\mathcal{M}}_{0,6}(\simeq \mathcal{C}_{0,5})$ is to take the Segr\'e cubic, given by the homogeneous equations
\[
\begin{split}
z_1+z_2+z_3+z_4+z_5+z_6&=0\\
z_1^3+z_2^3+z_3^3+z_4^3+z_5^3+z_6^3&=0
\end{split}
\]
in $\mathbb{CP}^5$, and blow up the 10 nodes. The 10 exceptional divisors from that blowup are the $\mathcal{C}_{ab}$.
}, $\mathcal{C}_{0,5}$ is a blowup of $\mathbb{CP}^3$ (see \cite{MR1237834} Theorem 4.33):
\begin{itemize}
\item Pick 5 points $p_a$ in general position\footnote{I.e., no three points are colinear and no four points are coplanar.}. Without loss of generality, we can choose the points $p_1=(1,0,0,0)$, $p_2=(0,1,0,0)$, $p_3=(0,0,1,0)$, $p_4=(0,0,0,1)$ and $p_5=(1,1,1,1)$. Let $\ell_{ab}$ be the line containing $p_a,p_b$.
\item Blow up the 5 points $p_a$. These yield the divisors $E_a$.
\item Blow up the proper transforms\footnote{Before blowing up, 4 lines meet at each point $p_a$. Their proper transforms meet the exceptional divisor $E_a$ at 4 distinct points. Blowing up the proper transform of $\ell_{ab}$ turns its intersection with $E_a$ into a $\mathbb{CP}^1$. That is, we turn $E_a$ from a $\mathbb{CP}^2$ into a dP${}_4$: $\mathbb{CP}^2$ blown up at the 4 points of intersection.} of the 10 lines $\ell_{ab}$. These yield the divisors $\mathcal{C}_{ab}$.
\item The (pullback of) the hyperplane class $H=K+\sum_a E_a$. Together $H, E_a$ and $\mathcal{C}_{a b}$ freely generate $H^2(\mathcal{C}_{0,5})$.
\end{itemize}

\noindent
Denoting the homogeneous coordinates on $\mathbb{CP}^3$ as $(y_1,y_2,y_3,y_4)$, the projection\goodbreak\noindent $\pi:\mathcal{C}_{0,5}\to\overline{\mathcal{M}}_{0,5}$ is given by the simultaneous solution to the quadrics
\begin{equation}
\begin{split}
0 &= x_1 y_1(y_2-y_4) + x_2 y_2(y_4-y_1)\\
0&= x_2 y_2(y_3-y_4)+x_3 y_3(y_4-y_2)\\
0 &=x_3 y_3(y_1-y_4)+ x_1 y_1(y_4-y_3)
\end{split}
\end{equation}
This map can be extended over the $E_a$ provided that we also blow up the $\mathbb{CP}^2$,  whose homogeneous coordinates are $(x_1,x_2,x_3)$, at the 4 points described in \S\ref{C04}, thus turning it into $\overline{\mathcal{M}}_{0,5}$. Once we do that, the map extends over the $\mathcal{C}_{ab}$ which  project to the 10 boundary divisors $D_{ab}\subset \overline{\mathcal{M}}_{0,5}$. The fiber over a generic point of $\overline{\mathcal{M}}_{0,5}$ is a twisted cubic\footnote{Explicitly,
\[
\begin{split}
y_1(s,t) &= (s-t)(sx_1+t(x_2-x_1))(s(x_1-x_3)+t(x_2-x_1))\\
y_2(s,t) &= -t(sx_1+t(x_2-x_1))(s(x_1-x_3)+t(x_2-x_1))\\
y_3(s,t) &= -t(s-t)(x_1-x_2)(sx_1+t(x_2-x_1))\\
y_4(s,t) &= -t(s-t)(x_1-x_2)(s(x_1-x_3)+t(x_2-x_1))
\end{split}
\]
}.

For the line bundle
$$
\mathcal{L}= \Kc\bigl(\sum_a E_a\bigr)^{\otimes k}\otimes\mathcal{O}\Bigl(-\sum_a\chi^{(a)}_k E_a-\sum_{a<b} \chi^{(ab)}_k \mathcal{C}_{ab}\Bigr)
$$
we have
\begin{equation}
\begin{split}
ch(\pi_*\mathcal{L})
&= \bigl(3k+1-\sum_a\chi^{(a)}_k\bigr)\\
&\qquad +
\tfrac{1}{6}\Bigl[
3\bigl(k(k-1)-\sum_a \chi^{(a)}_k(\chi^{(a)}_k-1)+\sum_{a<b} (\chi^{(a b)}_k)^2\bigr)(\tfrac{1}{2}\delta)\\
&\qquad\qquad+
\sum_{a\lt b}\Bigl(\sum_{c\neq a,b}\bigl(\chi^{(c)}_k(\chi^{(c)}_k-1)-(\chi^{(a c)}_k)^2-(\chi^{(b c)}_k)^2\bigr)
-2\sum_{\mathclap{\begin{smallmatrix}c,d\neq a,b\\c<d\end{smallmatrix}}}(\chi^{(c d)}_k)^2\\
&\qquad\qquad 
-\chi^{(a b)}_k(2\chi^{(a b)}_k-2\chi^{(a)}_k-2\chi^{(b)}_k+2k-1)\Bigr)D_{a b}
\Bigr]\\
&\qquad+\tfrac{1}{12}\Bigl[k(k-1)(2k-1) -\sum_a \chi^{(a)}_k(\chi^{(a)}_k-1)(2\chi^{(a)}_k-1) \\
&\qquad\qquad
-\sum_{a\lt b} \bigl(\chi^{(a b)}_k(\chi^{(a b)}_k+1)(4\chi^{(a b)}_k-1) -6\chi^{(a b)}_k(\chi^{(a)}_k+\chi^{(b)}_k-k)\bigr)\Bigr]P
\end{split}
\end{equation}
The details of the computation are left to  Appendix \ref{notes_on_}.

\subsection{The local Hitchin base in type-D}
\label{sec:local_hitchin_base}
Before proceeding to global considerations, we should first review the results of \cite{Balasubramanian:2023iyx,MR3815160} on the local behaviour in Type-D. Let $D=\sum_{a=1}^n p_a$ be the divisor on $C$ corresponding to the $n$ punctures. Let $t$ be a local coordinate centered at one of the punctures. If we take the Higgs field to be a holomorphic section of $ad(V)\otimes K_C(D)$ (instead of the slightly more conventional $ad(V)\otimes K_C$) then  in local coordinates it takes the form
\begin{equation}
    \Phi(t)= \mathfrak{n} + t( g+ O(t))
    \label{higgslocalform}
\end{equation}
where $\mathfrak{n}\in\mathfrak{j}$ is a nilpotent element lying in a \emph{special} nilpotent orbit $O_H$, and $g\in\mathfrak{j}$ is an arbitrary element of the Lie algebra.  

As in \cite{Balasubramanian:2023iyx,MR3815160}, we \textit{define} the local Hitchin base/image to be the image of the Hitchin map $\mu$ evaluated on Higgs pairs $(V,\Phi)$ on a punctured disc with the Higgs field $\Phi$ having the local behaviour as in \eqref{higgslocalform}. More precisely, we take the coefficients of both the Higgs field and its characteristic polynomial to be formal power series in $t$.

We are interested in such Higgs fields modulo (complexified) gauge transformations, so it is convenient to express the gauge-invariant data in $\Phi(t)$ in terms of the invariant traces (equivalently, symmetric polynomials). For $\mathfrak{j}=\mathfrak{so}(2N)$, these take the form
\[
\begin{split}
\det(\Phi(t)-w\Bid)&=w^{2N}+ \sum_{k=1}^Nw^{2(N-k)}\phi_{2k}(t)\\
&=w^{2N}+ \sum_{k=1}^{N-1}w^{2(N-k)}\phi_{2k}(t) +\tilde{\phi}(t)^2
\end{split}
\]
where
\begin{equation}\label{philocalexp}
\begin{split}
\phi_{2k}(t)&={t^{\chi_{2k}}}( c_{2k} +O(t))\\
\tilde{\phi}(t)&={t^{\chi_{2N}/2}}( \tilde{c} +O(t)).
\end{split}
\end{equation}
The formula for the leading power of $t$ is very simply determined\footnote{The same algorithm holds in the $\mathfrak{sl}(N)$ case, except that $[P]$ is a partition of $N$ and $\det(\Phi(t)-\lambda\Bid)=\lambda^N- \sum_{k=2}^N\lambda^{N-k}\phi_{k}(t)$. } from the partition $[P]$ corresponding to $O_H$:
\begin{itemize}
\item Define the partial sums $k_j = \sum_{i=1}^{j} P_i$.
\item Let $j$ be the integer such that $k_{j-1}< 2k \leq k_j$.
\end{itemize}

Then $\chi_{2k}=j$.

A-priori, the $c_{2k}$, thus-defined, are polynomial functions on $\mathfrak{j}=\mathfrak{so}(2N)$. The salient \emph{difference} from the $\mathfrak{sl}(N)$ case is that they are not all independent; rather they obey some polynomial relations (``constraints''). The most obvious (and well-known) constraint is that $c_{2N}=\tilde{c}^2$, where $\tilde{c}$ is another polynomial function on $\mathfrak{g}$. But there are additional constraints on the $c_{2k}$, which can be characterized as follows\footnote{Here it will be important to us that $O_H$ is \emph{special} which, in the present context, means that $[P]$ has an even number of odd parts between any even parts (including at the beginning and end of the partition).}

\begin{prop}\label{localconstraints}
{~} 
\begin{itemize}
\item[Even:]\label{localconstraintseven} For an even part with multiplicity $2l$, i.e.~$P_{2j+1}=\dots=P_{2(j+l)}=2s$, let $2m=\sum_{i=1}^{2j }P_i$. Introduce a formal variable $u$. Then we have a relation of the form
\begin{equation}\label{BK}
c_{2m} u^{2l}+ c_{2(m+s)}u^{2l-1}+\dots+ c_{2(m+2ls)}= \bigl(\alpha_{m} u^l+\alpha_{m+2s} u^{l-1}+\dots+\alpha_{m+2ls}\bigr)^2
\end{equation}
where the $\alpha_k$ are also polynomials on the Lie algebra. If $j=0$, \emph{i.e.}~if the even part occurs at the start of the partition, then we define $c_0=\alpha_0=1$. Note, in particular that $c_{2m}=\alpha_m^2$ and $c_{2(m+2ls)}=\alpha_{m+2ls}^2$, \emph{i.e.}~these polynomials on the Lie algebra are perfect squares.
\item[Odd:]\label{localconstraintsodd} For every ``marked pair'' of odd parts of the form $P_{2j}=2r+1$, $P_{2j+1}=2s+1$, with $r>s$, let $2k=\sum_{i=1}^{2j}P_i$. We have a relation of the form
\begin{equation}\label{oddconstraint}
 c_{2k}=\alpha_k^2
\end{equation}
where, again, the $\alpha_k$ are polynomials on $\mathfrak{g}$.

Unlike the even constraints, here we have \emph{a choice}. Consider a partition with $l$ marked pairs. For each marked pair, we can choose to parametrize the Hitchin base with either the corresponding $c_{2k}$ or $\alpha_k$. There are thus $2^l$ choices for the local Hitchin base. There are also $2^l$ nilpotent orbits in the dual (``Nahm'') special piece. In \cite{Balasubramanian:2023iyx}, we established a 1--1 correspondence between a choice of Nahm nilpotent orbit in the special piece and the choice of local Hitchin base.
\end{itemize}
\end{prop}

The above result is a consequence of Theorems 2 and 3 from \cite{Balasubramanian:2023iyx} and the work of \cite{Distler:2024ckw} which proved Conjecture 1 from \cite{Balasubramanian:2023iyx}. 

Note that the constraints involve only the \emph{leading} coefficients in the $t$-expansion \eqref{philocalexp}. This is in contradistinction to the $E_{6,7,8}$ cases, where the constraints involve also the subleading coefficients\footnote{In fact, the constraints in the $E_8$ case are only partially known.} \cite{Chacaltana:2014jba,Chacaltana:2017boe,Chacaltana:2018vhp}.

\section{The Global Story for Type-D}\label{GlobalD}
\subsection{Global Hitchin base on a smooth curve \texorpdfstring{$C$}{C}}
\label{Type_D_Global_Hitchin_Def}
As before, we first consider a fixed curve $C$. Let $L_{2k}= K_C(D)^{2k}\otimes \mathcal{O}(-\sum_a \chi^{(a)}_{2k} p_a)$ and\goodbreak\noindent $\tilde{L}=K_C(D)^N\otimes  \mathcal{O}(-\tfrac{1}{2}\sum_a \chi^{(a)}_{2N} p_a)$. The Hitchin base for the Hitchin system on $C$ is obtained by taking the ``naïve Hitchin base'',
\begin{equation}\label{naivebase}
B_{\text{na\"\i ve}} = H^0\Bigl(C, \bigl(\bigoplus_{k=1}^{N-1}L_{2k}\bigr)\; \oplus\tilde{L}\Bigr)
\end{equation}
and imposing the constraints (Prop.\ref{localconstraints}) obtained from each of the punctures.  This replaces the direct sum of line bundles in \eqref{naivebase} with a certain coherent sheaf $\mathcal{V}$ on $C$.

\begin{defn}
We say that the theory is \emph{bad} if $H^1(C,\mathcal{V})\neq 0$. Conversely, the theory is \emph{OK} if $H^1(C,\mathcal{V})= 0$. \footnote{This is identical to how we defined the OK/Bad dichotomy for type A Hitchin systems in \cite{Balasubramanian:2020fwc} which, in turn, was a simplified version of a Good/Ugly/Bad trichotomy that appeared first in the study of certain 3d $\mathcal{N}=4$ theories \cite{Gaiotto:2008ak}. } In what follows, we will only consider OK theories.
\end{defn}

How the construction of $\mathcal{V}$ works in practice depends somewhat on where the string of even parts occurs.

\subsection{Families of Hitchin bases}
\label{families_section_type_D}

The line bundles $L_{2k}$ and $\tilde{L}$ on each smooth fiber $C$ fit together to form\footnote{There is an additional twisting by a certain component of the boundary divisor of $\mathcal{C}_{g,n}$, which we will, for the moment, ignore.} line bundles $\mathcal{L}_{2k}$ and $\tilde{\mathcal{L}}$ on the universal curve $\mathcal{C}_{g,n}$:
\begin{equation}\label{theLs}
\mathcal{L}_{2k}= \KcD^{\otimes 2k}\otimes \mathcal{O}(-\sum_a \chi^{(a)}_{2k} E_{a}),\qquad
\tilde{\mathcal{L}}=\KcD^N\otimes  \mathcal{O}(-\tfrac{1}{2}\sum_a \chi^{(a)}_{2N} E_{a})
\end{equation}
Here, the marked points $p_a$ are replaced by the ``point divisors'' $E_{a}$ which intersect each fiber once, and $D=\sum_a E_{a}$.

We will use repeatedly in what follows the fact that the restrictions to the point divisors are
\begin{equation}\label{pointrestrictions}
\mathcal{L}_{2k}\otimes \mathcal{O}_{E_{a}}= \mathcal{O}_{E_{a}}(-\chi^{(a)}_{2k} E_{a}),\qquad
\tilde{\mathcal{L}}\otimes \mathcal{O}_{E_{a}}= \mathcal{O}_{E_{a}}(-\tfrac{1}{2}\chi^{(a)}_{2N} E_{a})
\end{equation}
This follows immediately from noting that $\left.\KcD\right|_{E_a}= \left.\Kc(E_a)\right|_{E_a}=N^*_{E_a/\mathcal{C}}(E_a)$, where $N^*_{E_a/\mathcal{C}}$ is the conormal bundle to $E_a$ in $\mathcal{C}$. By the Adjunction Formula, $N^*_{E_a/\mathcal{C}} =\mathcal{O}_{E_a}(-E_a)$ and hence $\left.\KcD\right|_{E_a}=\mathcal{O}_{E_a}$.

The ``na\"\i ve Hitchin base'' is a coherent sheaf  the moduli space
\begin{equation}\label{naivebasefamily}
\mathcal{B}_{\text{na\"\i ve}} = \pi_*\Bigl( \bigl(\bigoplus_{k=1}^{N-1}\mathcal{L}_{2k}\bigr)\; \oplus\tilde{\mathcal{L}}\Bigr)
\end{equation}
This is not quite what we want, for two reasons
\begin{enumerate}
\item While $\mathcal{B}_{\text{na\"\i ve}} $ is locally-free over the interior of the moduli space, $H^0$  and $H^1$ of the line bundles, $L=\mathcal{L}\vert_C$, can jump when $C$ degenerates to a nodal curve.
\item We have not incorporated the constraints of Proposition \ref{localconstraints}.
\end{enumerate}

To fix this, we need to first impose the constraints and then twist the resulting line bundles by a certain linear combination of components of the boundary divisor of $\mathcal{C}$. When the dust has settled, the family of Hitchin bases will form a graded vector bundle (graded by the $\mathbb{C}^\times$ weights) over the compactified moduli space: $\mathcal{B}\to \overline{\mathcal{M}}_{g,n}$.

\subsection{The Hitchin system on a fixed curve \texorpdfstring{$C$}{C}}

Before launching into the details of imposing the constraints, let us outline the rest of the construction. Over each fixed curve $C$, we construct the spectral curve
\[
\Sigma = \bigl\{w^{2N}+\sum_{k=1}^{N-1}\phi_{2k} w^{2(N-k)}\; +\tilde{\phi}^2\bigr\}\subset K_C(D)
\]
The coefficients $\phi_{2k}$ are holomorphic sections of $K_C(D)^{\otimes 2k}\otimes \mathcal{O}(-\sum \chi_a p_a)$ and hence homogeneous polynomials on $B=\mathcal{B}\vert_C$.  $\Sigma$ has an involution $\sigma: \Sigma \circlearrowleft$, under which $\sigma^*w=-w$. The fiber of the Hitchin system over the point $b\in B$ is the Prym variety with respect to the involution $\sigma$ (of the normalization of $\Sigma$) .

These spectral curves fit together into  a family over the total space of the vector bundle $\mathcal{B}\to \overline{\mathcal{M}}_{0,n}$.
\begin{equation}\label{allinthefamily}
\begin{tikzcd}
\Sigma \arrow[rd, dashrightarrow]\arrow[d] &\\ 
\pi^*(\mathcal{B}) \arrow[r] & \mathcal{C}_{g,n}\arrow[d, "\pi"] \\
\mathcal{B} \arrow[r]& \overline{\mathcal{M}}_{\mathrlap{0,n}}
\end{tikzcd}
\end{equation}

The vector bundle $\mathcal{B}$ is nontrivial. In the particular case of $\overline{\mathcal{M}}_{0,4}$, we can calculate it very explicitly; it is a direct sum of line bundles. We can trivialize it in patches and, in that trivialization, write explicit polynomial formul\ae\ for the family $\Sigma$ of spectral curves. On patch overlaps, $\mathcal{B}$, and hence $\Sigma$, have transition functions (see \S\ref{directimage} below). 

\subsection{The nilpotent at the node}\label{nilpnode}

As already alluded to, the behaviour as we approach the boundary of $\overline{\mathcal{M}}_{{g,n}}$ is an important part of our story. This is the regime in which some gauge group factor in the physics becomes weakly-coupled. The physics analysis gives us important insight into the behaviour of the family \eqref{allinthefamily} as $C$ develops a node. In the limit, some number of $\mathcal{N}=2$ vector multiplets (transforming in the adjoint of some simple gauge group $H$) decouple from the rest of the (family of) SCFT(s). The symptom of this, in the Hitchin system, is that $\operatorname{rank}(H)$ of the fiber directions decompactify. To reproduce the decoupling that takes place in the physics, we quotient by the action of the Hamiltonian vector fields $\{b_i,\cdot\}$ which generate translations in these noncompact directions of the fiber. The quotient space is a Poisson integrable system where the $b_i$ play the role of Casimirs.  To make a more precise statement, we need to track the behaviour of the Higgs field in the nodal limit.

Sections of the dualizing sheaf are differentials with simple poles at the nodes so generically the Higgs field has a simple pole at the node with residue $\in\mathfrak{so}(2N)$.

The generic  behaviour is that the residue is regular semi-simple. We call this ``the Standard Node.''  Tuning $N$ parameters (i.e., going to a codimension-$N$ subspace of $B$) turns the residue into one lying in the regular nilpotent orbit. More precisely, since $B$ is a graded vector space, we tune one parameter in each $\mathbb{C}^\times$ degree $k$, for $k=2,4,6,\dots, 2N-2;N$ (the degrees of the independent Casimirs of $\mathfrak{so}(2N)$). Over that subspace, we have the spectral curve/symplectic Hitchin system associated to the normalization $\tilde{C}$ of $C$, with regular nilpotents at the pre-images of the node.

That is the generic behaviour. However, when the node is a separating node (so that $\tilde{C}=C_L\amalg C_R$) and $C_L$  has genus-0 and the number of punctures on $C_L$ is sufficiently small and the nilpotent orbits at those punctures are of sufficiently low dimension, the behaviour at the node can be more exotic.

Specifically, two things can happen:
\begin{itemize}
\item The residue at the node can fail to be a generic element of $\mathfrak{so}(2N)$. It is still possible to tune parameters to reach the regular nilpotent orbit, but the number of parameters that need to be tuned is smaller. Instead of the Casimirs of $\mathfrak{so}(2N)$, they form the Casimirs of some simple $\mathfrak{h}\subset\mathfrak{so}(2N)$. The symplectic complex integrable system that we thus obtain is the Cartesian product of a Hitchin system on $C_R$ (the one obtained by putting the regular nilpotent at the pre-image of the node on $C_R$) with some other complex integrable system associated to the puncture data on $C_L$, but which \emph{is not} an OK Hitchin system associated to any choice of pole with residue in $\mathfrak{so}(2N)$ at the pre-image of the node on $C_L$. We denote this ``nilpotent at the node'' by the pair $([2N-1,1],\mathfrak{h})$. Here $\mathfrak{h}$ is the (complexified) Lie algebra of the gauge group $H$ which goes to zero coupling in the nodal limit, as described in the first paragraph of this section.
\item There may be no tuning of parameters that yields the regular nilpotent. Instead, the Hitchin system on $C_R$ is the one associated to putting some other special nilpotent orbit $O$ at the pre-image of the node. The parameters that we have to tune are the independent Casimirs of some simple $\mathfrak{h}\subset \mathfrak{f}$, where $\mathfrak{f}$ is the ``flavour symmetry'' algebra associated to the Nahm-dual of $O$. Since the Nahm special piece corresponding to a Hitchin orbit $O_H$ can contain more than one orbit, we need to specify the Nahm-dual orbit $O_N$ of which  $\mathfrak{f}$ is the flavour symmetry algebra. Equivalently, we need to specify the Sommers-Achar subgroup $\Gamma\subset \overline{A}_b(O_H)$ (see \S\ref{oddtype} and Def. 8 from \cite{Balasubramanian:2023iyx} ). So the nilpotent at the node is a triple, $(O_H,\Gamma, \mathfrak{h})$. We denote the level (normalized as in \cite{Argyres:2007cn}), $k$, of $\mathfrak{h}$ by a subscript. Of course we must have\footnote{The $\beta$-function for simple gauge group factor (whose Lie algebra is $\mathfrak{h}$) vanishes just in case the level $k$ of the $\mathfrak{h}$ current algebra for the SCFT on $C_R$ and the level $k'$ of the $\mathfrak{h}$ current algebra for the SCFT on $C_L$ add up to $4\check{h}(\mathfrak{h})$.} $k\leq 4\check{h}(\mathfrak{h})$, which restricts the possible $\mathfrak{h}\subset \mathfrak{f}$. 

\end{itemize}

The possible choices of nilpotent at the node for $\mathfrak{so}(8)$ were catalogued\footnote{Note, however, that the entries in the table in section 3.1 of \cite{Chacaltana:2011ze} are labeled by the Nahm nilpotent orbits and the levels reported in the table are the conjugate levels, \mbox{$k'=4\check{h}(\mathfrak{h})-k$}.} in \cite{Chacaltana:2011ze}.  The main peculiarity that distinguishes $\mathfrak{so}(8)$ from $\mathfrak{so}(2N)$ with $N\geq 5$ is that there are three non-conjugate embeddings $\mathfrak{so}(7)\hookrightarrow \mathfrak{so}(8)$, instead of just one\footnote{They are distinguished by which 8-dimensional irrep of $\mathfrak{so}(8)$ decomposes as $7\oplus 1$ under $\mathfrak{so}(7)$. Our notation for the ``additional'' embeddings denotes which of the two spinor representations of $\mathfrak{so}(8)$ is the one that decomposes as $7\oplus 1$.}.

In the table below, we catalogue the possibilities for the nilpotent at the node for arbitrary  $\mathfrak{so}(2N)$. Here $O_N$ is the Nahm nilpotent orbit, $O_H$ is the Hitchin nilpotent orbit (where the residue of the Higgs field at the node lies) and $\mathfrak{h}_k$ is the flavour symmetry and level. For $\mathfrak{so}(8)$, we have denoted the two ``additional'' embeddings by $\color{red}{\mathfrak{so}(7)}$, $\color{blue}{\mathfrak{so}(7)}$. In the $3^{\text{rd}}$ from last entry, $O_N$ is the non-special orbit which is the Kraft-Procesi small degeneration \cite{kraft1989special} of the special Nahm orbit $[2N-4n-1,3,2^{2(n-1)},1^2]_N$. The corresponding Hitchin orbit is the one with Sommers-Achar subgroup $\mathbb{Z}_2$. In the last four entries of the table, $\mathfrak{sp}(l)_k$ arises from a $2^{2n}$ subpartition of the partition $O_N$. In the remaining entries, $\mathfrak{h}_k$ arises from a $1^{l}$ subpartition.


{
\scriptsize
\renewcommand{\arraystretch}{1.9}
\renewcommand{\tabcolsep}{1pt}

\begin{longtable}{|c|c|c|c|}
\hline
$O_N$&$O_H$&$\mathfrak{h}_k$&Restrictions\\
\hline 
\endhead
$[1^{2N}]$&$[2N-1,1]$&$\mathfrak{so}(n)_{4(N-1)}$&$N+1\leq n\leq 2N$\\
&&$\mathfrak{su}(N)_{4(N-1)}$&\\
&&$\mathfrak{su}(N-1)_{4(N-1)}$&\\
&&${(\mathfrak{g}_2)}_{16}$&$N=5$\\
&&$\color{red}{\mathfrak{so}(7)}_{12}$&$N=4$\\
&&$\color{blue}{\mathfrak{so}(7)}_{12}$&$N=4$\\
&&${(\mathfrak{g}_2)}_{12}$&$N=4$\\
\hline
$\bigl[p_1,\dots,p_{2l},1^{2N-\sum p_i}\bigr]$&$\bigl[2(N+l-r)-1,2l+1,q_1,\dots,q_m\bigr]$&${\mathfrak{so}(n)}_{4(N+l-r-1)}$&\mbox{\shortstack{$\sum_{i=1}^{2l} p_i=2r$,\\$\sum_{i=1}^m q_i=2r-4l$,\\$2l\leq r\leq N-1 $,\\$N+l-r+1\leq n\leq 2(N-r)$}}\\
\hline
$\bigl[p_1,\dots,p_{2l+1},1^{2N-\sum p_i}\bigr]$&$\bigl[2(N+l-r)-1,2l+1,q_1,\dots,q_m\bigr]$&${\mathfrak{so}(n)}_{4(N+l-r-1)}$&\mbox{\shortstack{$\sum_{i=1}^{2l+1} p_i=2r+1$,\\$\sum_{i=1}^m q_i=2r-4l$,\\$2l+1\leq r\leq N-1$,\\$N+l-r+1\leq n\leq 2(N-r)-1$}}\\
\hline
$[{(2n)}^2,1^{2N-4n}]$&$[2N-4n+1,3,2^{2n-2}]$&${\mathfrak{su}(N-2n)}_{4(N-2n)}$&$2\leq 2n\leq N-2$\\
&&${(\mathfrak{g}_2)}_{16}$&$N-2n=4$\\
\hline
$[(2r+1),(2s+1),1^{2N-2r-2s-2}]$&$[2N-2r-2s-1,3,2^{2s-2},1^{2r-s+2}]$&$\mathfrak{su}(N-r-s-1)_{4(N-r-s-1)}$&$2\leq 2s\leq 2r\leq N-3$\\
&&${(\mathfrak{g}_2)}_{16}$&$N-r-s=5$\\
\hline
$[2N-2n-1,1^{2n+1}]$&$[2n+1,1^{2N-2n-1}]$&$\mathfrak{so}(2n+1)_{4n}$&$2\lt n\lt N-1$\\
&&$\mathfrak{su}(n)_{4n}$&$2\lt n\lt N-1$\\
\hline
$[2N-3,1^3]$&$[3,1^{2N-3}]$&${\mathfrak{so}(3)}_8$&\\
\hline
${\color{red}[2^{2n}]},{\color{blue}[2^{2n}]}$&${\color{red}[{(2n)}^2]},{\color{blue}[{(2n)}^2]}$&$\mathfrak{sp}(n)_{4n}$&$2n=N$\\
&&$\mathfrak{sp}(n-1)_{4n}$&$2n=N$\\
\hline
$[2N-4n-1,2^{2n},1]$&$\bigl([(2n+1)^2,1^{2N-4n-2}],\mathbb{Z}_2\bigr)$&$\mathfrak{sp}(n)_{4n+3}$&$2n\leq N-2$\\
\hline
${\color{red}[(N-2n)^2,2^{2n}]},{\color{blue}[(N-2n)^2,2^{2n}]}$&${\color{red}[(2n+2)^2,2^{N-2n-2}]},{\color{blue}[(2n+2)^2,2^{N-2n-2}]}$&$\mathfrak{sp}(n)_{4(n+1)}$&$2n\leq N-4$, $N$ even\\
\hline
$[2r+1,2s+1,2^{2n}]$&$[(2n+2)^2,2^{2(s-1)},1^{2(r-s+1)}]$&$\mathfrak{sp}(n)_{4(n+1)}$&$2n=N-r-s-1$\\
\hline
\end{longtable}

}
%

\subsection{Imposing the constraints}\label{imposing}

Except for the slightly more delicate case where we have multiple very-even partitions each with a single distinct even part (which we discuss in \S\ref{veryevensingle}) we can solve the constraints associated to the nilpotent orbits at each of the marked points one-at-a-time, and the order in which we solve them does not matter. Solving the constraints at each marked point will modify the line bundles \eqref{theLs} whose direct image is the bundle of Hitchin bases, and we will use those modified line bundles as input when solving the constraints at the next marked point. Finally, when we are done, we will generically need to twist by some component of the boundary (see \S\ref{twisting}) before taking the direct image.

Solving the even-type constraints requires the most elaborate explanation, so we start with that.

\subsubsection{Even-type constraints --- the easy cases}
\label{even_easy_cases}

Each even part (with even multiplicity), in the partition $[P]$ labeling the nilpotent residue at a puncture $p$, imposes constraints which locally have the form \eqref{BK}. The purpose of the present section is to explain how to solve those constraints globally.  How that works depends rather sensitively on where in the partition $[P]$ the even part occurs. Broadly, we can distinguish three cases: an even part at the beginning, middle or end of the partition. When the even part occurs at the beginning, the $\alpha$'s in \eqref{BK} can be solved-for.  Doing so will modify the $\mathcal{L}_{2k}$s without introducing any torsion sheaves. For even parts in the middle or end of the partition, solving the constraints will involve introducing torsion sheaves $\mathcal{S}_{p,k}$ on the universal curve $\mathcal{C}$ and the story becomes much more complicated. Moreover, \emph{very even} partitions (consisting of only even parts) require yet-more refined treatment. 

In what follows, we will extensively use the  parameterization of nilpotent orbits in $\mathfrak{so}(2N)$ by D-type partitions of N (see \cite{Chacaltana:2010ks} and \cite{CollingwoodMcGovern}). Every D-partition labels a unique nilpotent orbit of $\mathfrak{so}(2N)$ except in the case of a very even D-partition (this is a partition that contains only even parts). In this case, there are two distinct nilpotent orbits in $\mathfrak{so}(2N)$ corresponding to every very even partition. Following the convention used in \cite{Chacaltana:2010ks}, we will denote these as the {\color{red}red} and {\color{blue} blue} nilpotent orbits associated to corresponding very even D-partition (see, for ex, in \S\ref{veryevensingle} below). 

Let us begin, then, with the easy case(s), where no torsion sheaves are needed.

\paragraph*{Even part at the beginning of the partition:} Let the residue of the Higgs field at the marked point $p$ lie in the nilpotent orbit $[P]=[(2s)^{2l}, \dots]$, where the ``$\dots$'' begins with an odd part. In this case, we can solve the local constraints for the $\alpha$s without introducing any torsion sheaves $\mathcal{S}_{p,k}$ (or the associated line bundles $\mathcal{J}_{p,k}$). The effect is to replace
\[
\mathcal{L}_{2s}\oplus\mathcal{L}_{4s}\oplus\mathcal{L}_{6s}\oplus \dots\oplus\mathcal{L}_{4ls}
\]
by
\[
\mathcal{L}_{2s}\oplus\mathcal{L}_{4s}\oplus \dots\oplus \mathcal{L}_{2ls}\oplus\mathcal{L}_{2(l+1)s}(-E_p)\oplus\dots\oplus\mathcal{L}_{4ls}(-E_p)
\]
as the input to the next stage in the process.

\paragraph*{Very even partition $[(2s)^{2l},(2r)^{2}]$:} Here, again, the constraints can be solved without introducing any torsion sheaves. Locally, we have
\[
\begin{gathered}
u^{2l}+c_{2s} u^{2l-1}+c_{4s}u^{2l-2}+\dots+c_{4ls} = (u^l +\alpha_{2s}u^{l-1}+\dots+\alpha_{2ls})^2\\
c_{4ls}v^2+c_{4ls+2r}v+\tilde{c}^2=(\alpha_{2ls}v\pm \tilde{c})^2
\end{gathered}
\]
where the $\pm$ give the two nilpotent orbits corresponding to this very-even partition. For example, consider the partition $[4^2,2^2]$ in $\mathfrak{so}(12)$. \eqref{BK} gives
\[
\begin{split}
u^2 +c_4 u +c_8 &=(u+\alpha_4)^2\\
c_8 v^2+c_{10} v +\tilde{c}^2 &= (\alpha_4 v + \alpha_6)^2
\end{split}
\]
which yields
\[
\begin{split}
c_8 &=\tfrac{1}{4}c_4^2\\
c_{10} &= \pm c_4\tilde{c}
\end{split}
\]
depending on whether we choose $\alpha_6 =\pm \tilde{c}$.

As in the previous case, we can solve for all the $\alpha$s and the effect is to replace
\[
\mathcal{L}_{2s}\oplus\mathcal{L}_{4s}\oplus\mathcal{L}_{6s}\oplus \dots\oplus\mathcal{L}_{4ls}\oplus\mathcal{L}_{4ls+2r}
\]
by
\[
\mathcal{L}_{2s}\oplus\mathcal{L}_{4s}\oplus \dots\oplus \mathcal{L}_{2ls}\oplus\mathcal{L}_{2(l+1)s}(-E_p)\oplus\dots\oplus\mathcal{L}_{4ls}(-E_p)\oplus\mathcal{L}_{4ls+2r}(-E_p)
\]

\paragraph*{Very even partition $[(2s)^{2l}]$:} This case also does not introduce any torsion sheaves, but is more subtle. The constraint equations constrain a linear combination $\tfrac{1}{2}c_N\mp  \tilde{c}=\dots$. We will discuss this case separately in \S\ref{veryevensingle}.

In all other cases, solving for the $\alpha$s requires the introduction of torsion sheaves $\mathcal{S}_{p_a,k}$ which are supported on the point divisors $E_a$. At least in genus-0, we will be able to replace these by line bundles $\mathcal{J}_{p_a,k}\to \mathcal{C}$, which will allow us to construct families of spectral curves over $\overline{\mathcal{M}}_{0,n}$, just as we were able to do in type-A. 

\subsubsection{The torsion sheaves \texorpdfstring{$\mathcal{S}_{p,k}$}{Sₚ,ₖ}}
In order for the even-type constraints to make sense, we need the restrictions of the $\mathcal{L}$s to the point divisors $E_a$ to admit a square-root. Let $[P]$ be the partition describing the residue of the pole at the point $p\in C$.
As before, let
\[
2k_j=\sum_{i=1}^{2j}P_i
\]
and note that $\chi_{2k_j}=2j$. Then, for every even part of $[P]$ (with even multiplicity):
\[
P_{2j+1}=P_{2j+2}=\dots=P_{2j+2l}=2s
\]
we have, by virtue of \eqref{pointrestrictions}, a canonical square-root of the restriction
\begin{equation}\label{defn_S}
\left.\mathcal{L}_{2k_j+4ms}\right\vert_{E_p}= \mathcal{S}_{p,k_j+2ms}^{\otimes 2}\;,\qquad m=0,1,\dots,l
\end{equation}
where, from\eqref{theLs}, the $\mathcal{S}_{p,k_j+2ms}=\mathcal{O}_{E_p}(-(j+m)E_p)$ are line bundles on $E_p$, which we can extend by zero to torsion sheaves on $\mathcal{C}$. Similarly, for every marked pair
\[
P_{2j}=2r+1,\qquad P_{2j+1}=2s+1,\qquad r>s
\]
we have
\begin{equation}
\left.\mathcal{L}_{2k_j}\right\vert_{E_p}= \mathcal{S}_{p,k_j}^{\otimes 2}
\end{equation}
with $\mathcal{S}_{p,k_j}=\mathcal{O}_{E_p}(-j E_p)$, which will allow us to solve the odd-type constraints in \S\ref{oddtype}.

The constraints \eqref{BK} then become quadratic relations on the holomorphic sections of the $\mathcal{L}_{2k}$ and of the $\mathcal{S}_k$. Specifically, let
\[
P_{2j+1}=P_{2j+2}=\dots=P_{2j+2l}=2s
\]
as before. Let
\[
\begin{split}
W&= \left(\bigoplus_{m=0}^{2l} H^0(C,\mathcal{L}_{2k_j+2ms})\right)\oplus
\left(\bigoplus_{m=0}^{l}H^0(C,\mathcal{S}_{p,k_j+2ms})\right)\\
Z&= \bigoplus_{m=0}^{2l} H^0(C,\mathcal{L}_{2k_j+2ms}\otimes\mathcal{O}_{E_p})
\end{split}
\]
The constraints define a subvariety $V=\kappa^{-1}(0)$, where $\kappa: W\to Z$ is the nonlinear map given by
\begin{equation}\label{kappasubspace}
\begin{split}
\kappa(f_{2k_j}, f_{2(k_j+s)},&\dots, f_{2(k_j+2ls)};\;g_{k_j},g_{k_j+2s},\dots,g_{k_j+2ls})\\=
&\bigl(f_{2k_j}\vert_p-g_{k_j}^2,\; f_{2(k_j+s)}\vert_p-2g_{k_j}g_{k_j+2s},\;\dots\;,\; f_{2(k_j+2ls)}\vert_p-g_{k_j+2ls}^2\bigr)
\end{split}
\end{equation}
In families, this naturally extends to a (nonlinear) bundle map $\kappa: \mathcal{W}\to \mathcal{Z}$, with $\pi_*(\cdot)$ replacing $H^0(C,\cdot)$. The OK condition, \emph{after } imposing the constraint, implies that $R^1\pi_*(\mathcal{L}(-E_p))=0$. When this holds, the short exact sequence of coherent sheaves on $\mathcal{C}$
\[
0\to \mathcal{L}(-E_p) \to \mathcal{L} \to \left.\mathcal{L}\right\vert_{E_p}\to 0
\]
induces a short exact sequence of coherent sheaves on $\overline{\mathcal{M}}_{g,n}$
\begin{equation}\label{hopeitsplits}
0\to\pi_*(\mathcal{L}(-E_p))\to  \pi_*(\mathcal{L}) \to \pi_*(\mathcal{L}\otimes \mathcal{O}_{E_p})\to 0.
\end{equation}
If this sequence splits, i.e. if
\[
\pi_*(\mathcal{L})=\pi_*(\mathcal{L}(-E_p))\oplus \pi_*(\mathcal{L}\otimes \mathcal{O}_{E_p})
\]
then we can identify
\begin{equation}\label{kappasplit}
\kappa^{-1}(0) \simeq \left(\bigoplus_{m=0}^{2l} \pi_*\bigl(\mathcal{L}_{2k_j+2ms}(-E_p)\bigr)\right)\oplus
\left(\bigoplus_{m=0}^{l}\pi_*(\mathcal{S}_{p,k_j+2ms})\right).\\
\end{equation}
For $\pi:\mathcal{C}_{0,4}\to \overline{\mathcal{M}}_{0,4}$, we can compute explicitly as in Appendix \ref{foursplitting}, and show that the sequence splits\footnote{This is a property of the bundles $\pi_*(\mathcal{L})$ that appear, not the fact that we are on $\overline{\mathcal{M}}_{0,4}=\mathbb{CP}^1$. Indeed, the canonical example of a short exact sequence of vector bundles that doesn't split is drawn from this case:
\[
0\to \mathcal{O}(n)\to\mathcal{O}(n+1)\oplus \mathcal{O}(n+1)\to \mathcal{O}(n+2)\to 0
\]
doesn't split because $H^1\Bigl(\mathbb{CP}^1, Hom\bigl(\mathcal{O}(n+2),\mathcal{O}(n)\bigr)\Bigr)$ is 1-dimensional and the obstruction is a nontrivial element of that group. }. We have also checked in explicit examples that the splitting of the sequence \eqref{hopeitsplits} continues to hold for $g=0,n=5$(see \S\ref{fivepuncturedexamples}). However, beyond those examples, we don't have a general proof and the situation is even more unclear for general $g,n$.

The obstruction to such a splitting lies in the cohomology group 
$$H^1\Bigl(\overline{\mathcal{M}}_{g,n}, Hom\bigl(\pi_*(\mathcal{L}\otimes\mathcal{O}_{E_p}), \pi_*(\mathcal{L}(-E_p))\bigr)\Bigr)
$$
There are certainly cases where that cohomology group can be shown to vanish. But it doesn't always vanish (even in examples on the 4-punctured sphere, where we know the sequence splits). Since the constructions that follow below are only applicable when we are in one of the cases where the sequence \eqref{hopeitsplits} does split, we will mostly restrict ourselves in this paper to $g=0,n=4$.

When \eqref{kappasplit} holds, since we are working at $g=0$, we can actually do better and replace the torsion sheaves $\mathcal{S}_{p,k}$ with \emph{line bundles} $\mathcal{J}_{p,k}$ on the universal curve. In order to write down a (family of) spectral curve(s), we need to know how to take a section of $\mathcal{S}_{p,k}$ and manufacture a section of $\mathcal{L}_k$. The line bundle $\mathcal{J}_{p,k}$ provides the information needed to do that.

\subsubsection{The line bundles \texorpdfstring{$\mathcal{J}_{p,k}$}{Jₚ,ₖ}}\label{Jpk}

For each $\mathcal{S}_{p,k}$, we seek to replace it by a line bundle $\mathcal{J}_{p,k}\to \mathcal{C}$ which will go into constructing our bundle of spectral curves. To that end, we impose\footnote{The twist \eqref{Jdeftwist} is required so that $\pi_*(\mathcal{J}_{p,k})$ is locally-free.}
\begin{subequations}\label{Jprops}
\begin{align}
\mathcal{J}_{p,k} &= \KcD^{\otimes k} \mathcal{O}\bigl(-\sum_a \tau_a E_{a} -\sum_S n_S \mathcal{C}_S\bigr)\label{Jdef}\\
\intertext{where}
\tau_{p}&= \chi^{(p)}_{2k}/2\label{Jdeftaup}\\
\tau_{p'}&\geq \chi^{(p')}_{2k}/2,\quad \forall p'\neq p\label{Jdeftaunotp}\\
\sum_{a=1}^n \tau_a &=k(n-2)\label{Jdefsum}\\
\intertext{and}
n_S&=\max\Bigl(0,-(|S|-1)k-1+\sum_{p_a\in S}\tau_a\Bigr)\label{Jdeftwist}
\end{align}
The resulting line bundles $\mathcal{J}_{p,k}$ have the following properties
\begin{itemize}
\item $\mathcal{J}_{p,k}^{\otimes 2}$ is a subsheaf of $\mathcal{L}_{2k}$. This is assured by \eqref{Jdeftaup},\eqref{Jdeftaunotp}.
\item $\left.\mathcal{J}_{p,k}\right\vert_{E_p}=\mathcal{S}_{p,k}$. This is assured by \eqref{Jdeftaup}, \emph{provided}
\begin{equation}
n_S=0\; \text{for}\; S\ni p.
\end{equation}
That's another condition on the $\tau_a$.
\end{itemize}
\end{subequations}
\begin{itemize}
\item Under the conditions of \eqref{Jprops}, the direct image $\pi_*(\mathcal{J}_{p,k})$ is a subsheaf of $\pi_*(\mathcal{S}_{p,k})$:
\begin{equation}\label{JsubsheafS}
\pi_*(\mathcal{J}_{p,k})=\pi_*(\mathcal{S}_{p,k}) \otimes\mathcal{O}(-\sum_{S\ni p}\ell_{S} D_{S})
\end{equation}
 where
 \begin{equation}\label{ellDef}
 \ell_{S} \coloneqq \max\bigl(0,-(|S|-1)k+\sum_{p_a\in S}\tau_a \bigr)
 \end{equation}
 and $D_S$ is the component of the boundary divisor over which $C$ degenerates to a (genus-0) component containing the points in $S$, and rest of the surface (containing the points in $S^\vee$). \eqref{Jdeftwist} ensures that $\pi_*(\mathcal{J}_{p,k})$ is locally-free; \eqref{Jdefsum} ensures that it's a line bundle. That it's \emph{this particular} line bundle \eqref{JsubsheafS} is a nontrivial statement.
\end{itemize}

The existence of a $\mathcal{J}_{p,k}$ is quite dramatic. We \emph{know} what the $\pi_*(\mathcal{S}_{p,k})$ are. Recall the \emph{point line bundle}, $I_p$. It is the line bundle\footnote{The first Chern class of the point line bundle $I_p$ is customarily denoted by $\psi_p=c_1(I_p)$.} over $\overline{\mathcal{M}}_{g,n}$ whose fiber over $C\in\overline{\mathcal{M}}_{g,n}$ is $\left.K_C\right\vert_{p}$. We have
\begin{equation}\label{pointlineimage}
\pi_* (\mathcal{O}_{E_p}(-j E_p))= I_p^{\otimes j}
\end{equation}
$\pi_*(\mathcal{J}_{p,k})$ is supposed to be  a subsheaf of $I_p^{\otimes\tau_p}$ twisted by a certain component of the boundary as in \eqref{JsubsheafS}. We conjecture that such a $\mathcal{J}$ always exists, for any $\mathcal{C}_{0,n}$, though it will be clear in examples that it need-not be unique. As will be clear, different choices of $\mathcal{J}$, when they are available, correspond to different choices of fiber coordinates on the bundle of Hitchin bases.


\subsubsection{Even parts in the middle or end of the partition}

Having seen the general framework for solving the constraints associated to even parts in the middle or at the end of the partition, we  return to fill in the rest of the details.

\paragraph*{Successive even parts:} Obviously, if we have successive distinct even parts,
\[
\begin{split}
P_{2j+1}=P_{2j+2}=\dots=P_{2j+2l}&=2r\\
P_{2j+2l+1}=P_{2j+2l+2}=\dots=P_{2j+2l+2m}&=2s\\
\end{split}
\]
for $r>s$, then $\mathcal{S}_{p,k_j+2lr}$ (and the associated $\mathcal{J}_{p,k_j+2lr}$) is common to both sets of constraints.

For even parts \emph{not} at the beginning or at the end of the partition, the net effect of imposing the even-type constraint corresponding to $[\dots,(2s)^{2l},\dots]$ at $p$ is to replace
\[
\mathcal{L}_{2k_j}\oplus\mathcal{L}_{2k_j+2s}\oplus\dots\oplus\mathcal{L}_{2k_j+4ls}
\]
by
\begin{equation} \label{eq:evencons}
\mathcal{L}_{2k_j}(-E_p)\oplus\mathcal{L}_{2k_j+2s}(-E_p)\oplus\dots\oplus\mathcal{L}_{2k_j+4ls}(-E_p)\oplus
\mathcal{J}_{p,k_j}\oplus \mathcal{J}_{p,k_j+2s}\oplus \dots\oplus \mathcal{J}_{p,k_j+2ls}
\end{equation}

\paragraph*{Even part at the end of the partition:} For an even part at the end of the partition, $[P]=[\dots,(2s)^{2l}]$, the constraint looks like
\[
c_{2N-4ls}u^{2l}+c_{2N-2(2l-1)s} u^{2l-1}+\dots+c_{2N-2s}u+\tilde{c}_{N}^2=(\alpha_{N-2ls}u^l+\alpha_{N-2(l-1)s}u^{l-1}+\dots+\alpha_N)^2
\]
If $[P]$ is not very even, we can without loss of generality take $\alpha_N=\tilde{c}_N$ so that solving the constraints yields the replacement of
\[
\mathcal{L}_{2N-4ls}\oplus \mathcal{L}_{2N-2(2l-1)s}\oplus\dots\oplus\mathcal{L}_{2N-2s}\oplus\tilde{\mathcal{L}}_N
\]
by
\[
\mathcal{L}_{2N-4ls}(-E_p)\oplus \mathcal{L}_{2N-2(2l-1)s}(-E_p)\oplus\dots\oplus\mathcal{L}_{2N-2s}(-E_p)\oplus\tilde{\mathcal{L}}_N\oplus \mathcal{J}_{p,N-2ls}\oplus\dots\oplus \mathcal{J}_{p,N-2s}
\]
If $[P]$ is very-even, then there are two independent solutions $\alpha_N=\pm\tilde{c}_N$, corresponding to the two distinct nilpotent orbits.

\subsubsection{Very-even partition with a single distinct even part}\label{veryevensingle}
When the very-even partition has just one distinct even part, $[(2s)^{2l}]$, the constraint reads
\[
c_{2sl}\mp  2\tilde{c}=2\alpha_{2s}\alpha_{2s(l-1)}+\dots
\]
 So, rather than decreasing the degree of one or the other of the corresponding line bundles by 1, we have an exact sequence of the form (with $N=4ls$)
\begin{equation}\label{veryevenSESone}
0\to F\to L_N\oplus \tilde{L}\to R_p\to 0
\end{equation}
where $R_p$ is a skyscraper sheaf supported at $p$.

More generally, we might have $n$ such punctures and an exact sequence of the form
\begin{equation}\label{veryevenSES}
0\to F\to L_N\oplus \tilde{L}\xrightarrow{\beta} R_{p_1}\oplus \dots\oplus R_{p_n}\to 0
\end{equation}
In families, this becomes
\begin{equation}\label{veryevenveryglobalSES}
0\to\mathcal{F}\to \mathcal{L}_N\oplus \tilde{\mathcal{L}}\xrightarrow{\beta} \mathcal{O}_{E_{1}}(-l_1E_{1})\oplus \dots\oplus  \mathcal{O}_{E_{n}}(-l_nE_{n})\to 0
\end{equation}
where the partition at the $a^{\text{th}}$ puncture is $[(2s_a)^{2l_a}]$  and $\mathcal{L}_N\vert_{E_{a}}\simeq \tilde{\mathcal{L}}\vert_{E_{a}}\simeq \mathcal{O}_{E_{a}}(-l_aE_{a})$. At the $a^{\text{th}}$ puncture, the map $\beta$ is
\[
\beta_a(f,\tilde{f})= f\vert_{E_{a}}\mp 2 \tilde{f}\vert_{E_{a}}
\]
The kernel $\mathcal{F}$ is a rank-2 vector bundle on $\mathcal{C}_{g,n}$. In constructing $\mathcal{F}$, we should first impose all of the constraints on $\mathcal{L}_N$ and $\tilde{\mathcal{L}}$ arising from the \emph{other} punctures (not of the form $[(2s)^{2l}]$) before imposing \eqref{veryevenveryglobalSES}. The OK condition then requires that restricted to any smooth fiber, $H^1(C,\mathcal{F}|_C)=0$.

When $\mathcal{L}_{N}\simeq \tilde{\mathcal{L}}$, conjugation by $\tfrac{1}{2}\left(\begin{smallmatrix}1&-1\\1&1\end{smallmatrix}\right)$ diagonalizes the $\beta$ action with the result that $\mathcal{F}$ splits as a direct sum of line bundles
\begin{equation}\label{FsplitIsom}
\mathcal{F}= \mathcal{L}_N \otimes \Bigl(
\mathcal{O}\bigl(\quad-\sum_{\mathclap{O_{p_a}={\color{red}[(2s_a)^{2l_a}]}}}E_{a}\quad\bigr)\oplus
\mathcal{O}\bigl(\quad-\sum_{\mathclap{O_{p_a}={\color{blue}[(2s_a)^{2l_a}]}}}E_{a}\quad\bigr)
\Bigr)
\end{equation}

From \eqref{FsplitIsom}, it is clear that some choices of red-versus-blue can lead to bad theories. As an example, consider the 4-punctured sphere with nilpotents ${\color{red}[2^4]}{\color{red}[2^4]}[2^2,1^4][7,1]$ versus ${\color{red}[2^4]}{\color{blue}[2^4]}[2^2,1^4][7,1]$. After imposing the constraint from the $[2^2,1^4]$ at $E_{3}$, we have
\[
\tilde{\mathcal{L}}\simeq\mathcal{L}_4 = \KcD^{\otimes{4}}\otimes\mathcal{O}(-2E_{1}-2E_{2}-3E_{3}-E_{4})
\]
So for ${\color{red}[2^4]}{\color{red}[2^4]}[2^2,1^4][7,1]$ \eqref{FsplitIsom} says
\[
\mathcal{F}= \KcD^{\otimes{4}}\otimes\mathcal{O}(-3E_{1}-3E_{2}-3E_{3}-E_{4}) \oplus \KcD^{\otimes{4}}\otimes\mathcal{O}(-2E_{1}-2E_{2}-3E_{3}-E_{4})
\]
which is \emph{bad}, since $h^1(C,\mathcal{F})=1$. On the other hand, for ${\color{red}[2^4]}{\color{blue}[2^4]}[2^2,1^4][7,1]$, we have
\[
\mathcal{F}= \KcD^{\otimes{4}}\otimes\mathcal{O}(-3E_{1}-2E_{2}-3E_{3}-E_{4}-\mathcal{C}_{13})\oplus \KcD^{\otimes{4}}\otimes\mathcal{O}(-2E_{1}-3E_{2}-3E_{3}-E_{4}-\mathcal{C}_{23})
\]
(here, we've included the twisting at the boundary) which is OK. Implementing the constraints thus yields
\[
\begin{aligned}
\mathcal{L}_2&= \KcD^2\otimes \mathcal{O}(-E_{1}-E_{2}-E_{3}-E_{4})\\
\mathcal{F}&=\KcD^4\otimes\left[
\mathcal{O}(-3E_{1}-2E_{2}-3E_{3}-E_{4}-\mathcal{C}_{13})
\oplus 
\mathcal{O}(-2E_{1}-3E_{2}-3E_{3}-E_{4}- \mathcal{C}_{23})
\right]\\
\mathcal{L}'_6&= \KcD^6\otimes \mathcal{O}(-4E_{1}-4E_{2}-4E_{3}-E_{4}-\mathcal{C}_{1 2}-\mathcal{C}_{1 3}-\mathcal{C}_{2 3})
\end{aligned}
\]
Only $\mathcal{L}_2$ has a non-vanishing push-forward and  the spectral curve is
\[
  0=w^4\left(w^2+\tfrac{1}{2}a x(y-z)\right)^2
\]
The resulting physical theory is $SU(2)+4(2)+16(1)$. 

More generally, $\tilde{\mathcal{L}}$ is a subsheaf of $\mathcal{L}_N$, while its restriction to the punctures where $O_{p_a}=[(2s)^{2l}]$ is isomorphic to that of $\mathcal{L}_N$.
\[
\tilde{\mathcal{L}}\subset\mathcal{L}_N,\qquad \tilde{\mathcal{L}}\vert_{E_{a}}\simeq \mathcal{L}_N\vert_{E_{a}}\; \forall O_{p_a}={\color{red}[(2s_a)^{2l_a}]}\; \text{or}\; {\color{blue}[(2s_a)^{2l_a}]}
\] 
Let the number of such  punctures that are red be $n_r$ and the number that are blue be $n_b$. Then
\[
\begin{aligned}
\mathcal{F}&= \mathcal{L}_N \bigl(\quad-\sum_{\mathclap{O_{p_a}={\color{red}[(2s_a)^{2l_a}]}}}E_{a}\quad\bigr)\oplus
\tilde{\mathcal{L}}\bigl(\quad-\sum_{\mathclap{O_{p_a}={\color{blue}[(2s_a)^{2l_a}]}}}E_{a}
\quad\Bigr),&&n_r>n_b\\
\mathcal{F}&= \mathcal{L}_N \bigl(\quad-\sum_{\mathclap{O_{p_a}={\color{blue}[(2s_a)^{2l_a}]}}}E_{a}\quad\bigr)\oplus
\tilde{\mathcal{L}}\bigl(\quad-\sum_{\mathclap{O_{p_a}={\color{red}[(2s_a)^{2l_a}]}}}E_{a}
\quad\Bigr),&&n_b>n_r
\end{aligned} \label{eq:Fs}
\]
When $n_r=n_b$, it's unclear what the formula should be; the two obvious choices are not isomorphic as bundles over the universal curve and, as we shall see in Example \ref{SO12fivepointsplittingexample}, their direct images can be distinct vector bundles over the moduli space. When the direct images differ, it seems that one or the other is the preferred choice but, aside from explicit calculation imposing the constraints, we don't have an a-priori way to decide between them.


\subsubsection{Odd-type constraints} \label{oddtype}
The story with the \emph{odd} constraints is more subtle. For each marked pair,
\[
P_{2j}=2r+1,\quad P_{2j+1}=2s+1,\quad r>s
\]
we have a \emph{choice} of whether to impose \eqref{oddconstraint} (and parametrize the Hitchin base by $\alpha_{k_j}$), or use the ``original'' $c_{2k_j}$ as the Hitchin base parameter.  If we \emph{do} impose the constraint then, analogous to \eqref{kappasubspace}, the constraints define a nonlinear subspace $\kappa^{-1}(0)$ where\goodbreak\noindent $\kappa: H^0(C,\mathcal{L}_{2k_j})\oplus H^0(C,\mathcal{S}_{p,k_j})\to H^0(C,\mathcal{L}_{2k_j}\otimes\mathcal{O}_{E_p})$ is given by
\[
\kappa(f_{2k_j},g_{k_j}) = f_{2k_j}\vert_{p} -g_{k_j}^2
\]
If the sequence \eqref{hopeitsplits} splits then, analogous to \eqref{kappasplit}, we can identify
\begin{equation}\label{oddtypekappasplit}
\kappa^{-1}(0)\simeq \pi_*(\mathcal{L}_{2k_j}(-E_p))\oplus \pi_*(\mathcal{S}_{p,k_j})
\end{equation}
Finally, under the conditions of \S\ref{Jpk}, we can replace the torsion sheaf $\mathcal{S}_{p,k_j}$ in \eqref{oddtypekappasplit} by a line bundle $\mathcal{J}_{p,k_j}$. When all the dust has settled, imposing the odd-type constraint amounts to replacing $\mathcal{L}_{2k_j}$ by $\mathcal{L}_{2k_j}(-E_p)\oplus \mathcal{J}_{p,k_j}$.

For a partition $[P]$ with $n$ marked pairs, these $2^n$ choices are in 1--1 correspondence with the $2^n$ Nahm nilpotent orbits in the special piece dual to the given Hitchin orbit $O_H$ or, alternatively, with the $2^n$ \emph{special sheets}\footnote{The notion of a special sheet was introduced in \cite{Balasubramanian:2018pbp}. See Definition A.3.12 of that paper. To be precise, apart from the restriction to special sheets, a \textit{further} refinement of the notion of sheets is needed to obtain a 1:1 map. This is also discussed in \cite{Balasubramanian:2018pbp}. } in $\mathfrak{j}$ attached to $O_H$. Deforming away from $O_H$ onto a sheet  corresponds to turning on (relevant) mass deformations which yields a Poisson-Hitchin system. The set of allowed mass-deformations is thus neatly-encoded in a choice of Nahm nilpotent orbit $O_N$ in the special piece dual to $O_H$.

As first noted in \cite{Chacaltana:2012zy}, there is a choice of what global gauge group to mod out by, in forming the moduli space of meromorphic Higgs bundles on $C$. Rather than merely fixing the conjugacy class $O_H$ of the residue of $\Phi$ at the marked point and modding out by all gauge transformations $\mathcal{G}=\operatorname{Map}\bigl(C,SO(2N)\bigr)$, we can fix a particular element $e\in O_H$ and mod out by those gauge transformations $\mathcal{G}_e$ which preserve the residue $e$ of $\Phi$ at the marked point. This group is disconnected; its component group is $(\mathbb{Z}_2)^n$, with a preferred set of generators labeled by the $n$ marked pairs in $[P]$. That component group is, in turn, isomorphic to the group $\overline{A}_b(O)\subset \overline{A}(O)$, which we introduced in \cite{Balasubramanian:2023iyx} (see Def. 7). The global form of the gauge group, determined by choosing a subset of those generators (equivalently, a ``Sommers-Achar subgroup'' $\Gamma\subset \overline{A}_b(O)$), is thus in 1--1 correspondence with the choice of Nahm orbit in the dual special piece. Equivalently, it is in 1--1 correspondence with a choice of special sheet containing $O_H$ or, more precisely, to a special  \emph{birational sheet} \cite{MR4359565,losev2021unipotent,mason2023unipotent,westaway2023birational} containing a finite cover of $O_H$. 

Quotienting by these different gauge groups yields Hitchin bases which are (ramified) $(\mathbb{Z}_2)^l$ covers of each other. The special Nahm orbit ($\Gamma=$ trivial) corresponds to modding out by only the identity component of $\mathcal{G}_e$. The maximally-non-special Nahm orbit ($\Gamma=\overline{A}_b(O)$) corresponds to modding out by all of $\mathcal{G}_e$. As discussed in Appendix \ref{MassDeformSpectral}, this stackiness is made evident when we mass-deform\footnote{It is well-known that some information about the space of possible mass-deformations is \emph{necessary} to distinguish $\mathcal{N}=2$ Coulomb branches even when they are indistinguishable in the conformal limit. See, e.g.~\cite{Argyres:2016xua}.}.

On the level of the Hitchin system, the quotienting operation is a bit more delicate. The quotient of the massless system is typically singular, so it needs to be resolved. The new Hitchin system (corresponding to some non-special Nahm orbit in the special piece) is \emph{birational} to the quotient of the original Hitchin system (corresponding to the special Nahm orbit).

The relation between the two Hitchin systems can be seen very concretely in a rank-1 example. Consider the Hitchin systems associated to the rank-1 $E_6$ and $E_8$ Minahan-Nemeschansky SCFTs. These are non-compact elliptic surfaces with constant $j$-invariant and a single singular fiber over the origin (respectively, of Kodaira type $IV^*$ and $II^*$). As in our odd-type constraint situation, the Hitchin base of the latter is evidently a $\mathbb{Z}_2$ quotient of the Hitchin base of the former. With a little work, one can show that, after resolving the singularities of the quotient, the same is true of the corresponding elliptic surfaces. Away from the central fiber, $X_{IV^*}\to X_{II^*}$ is a double cover.  Globally, $X_{II^*}$ is birational to the $\mathbb{Z}_2$ quotient of $X_{IV^*}$. We leave the details to Appendix \ref{Karayayla}.

Finally, note that, unlike the \emph{even} constraints, the \emph{odd} constraints do not change the dimension of the Hitchin base
\[
h^0(C, \mathcal{L}_{2k}(-E_p)\oplus \mathcal{J}_{p,k}) = h^0(C,\mathcal{L}_{2k})
\]
At least, that's the case if $H^1(C, \mathcal{L}_{2k}(-E_p))=0$. Otherwise, imposing the odd constraint would lead to a bad theory.

\subsection{The twisting at the boundary}\label{twisting}

Our OK condition is that $H^1(C,\mathcal{V})=0$ for every smooth $C$. Here $\mathcal{V}$ is the  vector bundle on the universal curve that results after imposing the constraints
\begin{equation}\label{V_explicit}
\mathcal{V}=\bigoplus_k\mathcal{L}_k\oplus\bigoplus_{p,k}\mathcal{J}_{p,k}\oplus\mathcal{F}
\end{equation}

(where $\mathcal{F}$ appears in the presence of very-even punctures with a single distinct even part).

For each fixed smooth $C$, we obtain the corresponding Hitchin base as $B=H^0(C,\mathcal{V})$. These fit together to form a coherent sheaf $\mathcal{B}=\pi_*(\mathcal{V})$. We would like this direct image sheaf to be a vector bundle, which is to say that we don't want $H^0(C,\mathcal{V})$ to jump when $C$ degenerates to a nodal curve. Said differently, we would like the OK condition, $H^1(C,\mathcal{V})=0$, to hold even for a nodal curve $C$.

Assuming $\mathcal{F}$ splits, as in \S\ref{veryevensingle}, $\mathcal{V}$ is a direct sum of line bundles of the form
\[
\mathcal{L}_k=\KcD^{\otimes k}\otimes\mathcal{O}\bigl(-\sum_a \tau^{(a)}_k E_a\bigr)
\]
In general, these will violate the OK condition; we need to twist by a component of the boundary, as in \eqref{nksdef},\eqref{Lprimedef}. As before, let
\[
n_k^S\coloneqq\max\Bigl(0,k-1-\sum_{p_a\in S}(k-\tau_k^{(a)})\Bigr)
\]
and replace each $\mathcal{L}_k$ by $\mathcal{L}'_k=\mathcal{L}_k\otimes\mathcal{O}(-\sum_S n_k^S\mathcal{C}_S)$ (and similarly for the $\mathcal{J}_{p,k}$s). The resulting bundle of Hitchin bases, $\mathcal{B}=\pi_*(\mathcal{V}')$ , is locally-free.

Hence we  can replace $\mathcal{V}$ by:
\begin{equation}\label{V_prime_explicit}
\mathcal{V}'=\bigoplus_k\mathcal{L}'_k\oplus\bigoplus_{p,k}\mathcal{J}'_{p,k}
\end{equation}
which now pushes forward to a vector bundle over $\overline{\mathcal{M}}_{0,n}$.
\[
\mathcal{B}=  \pi_*(\mathcal{V}')
\]

Compared to the ``na\"\i ve'' Hitchin base, $\mathcal{B}_{\text{na\"\i ve}}$, we have $\dim{B}=\dim{B}_{\text{na\"\i ve}}-L$, where $2L$ is the total number of even parts in all of the Hitchin nilpotent orbits at the punctures.

\subsection{The direct image}\label{directimage}

Our task is now to compute the direct image $ \pi_*(\mathcal{V}')$. $\mathcal{V}'$ is a direct sum of line bundles $\mathcal{L}$ of the form
\begin{equation}\label{ourLtoPush}
\mathcal{L}= \KcD^{\otimes k}\otimes \mathcal{O}\bigl(-\sum_a\tau_k^{(a)}E_{p_a}\bigr)\otimes  \mathcal{O}\bigl(-\sum_S n_k^S \mathcal{C}_S\bigr)
\end{equation}
where the $\tau_k^{(a)}$ are integers $1\leq \tau_k^{(a)}< k$ satisfying\footnote{This condition is trivially satisfied for $k\geq 2$, $n\geq 0$ and $g>0$. For $g=0$, (and $n\geq3$), it reduces to $\sum_{a=1}^n \tau^{(a)}\leq (n-2)k+1$.}
\begin{equation}
\sum_{a=1}^{n} \tau^{(a)}_k\leq 2(k-1)(g-1)+nk-1
\end{equation}
and the $n_k^S$ are given by \eqref{nksdef}. Among them are the line bundles $\mathcal{J}_{p,k}$, whose direct images were computed/\emph{defined} in \S\ref{Jpk}. They push forward to \emph{line bundles} on $\overline{\mathcal{M}}_{0,n}$ which are very specific subsheaves of (tensor powers of) the point line bundles. Because $\text{Pic}(\overline{\mathcal{M}}_{0,n})$ is discrete, these are entirely characterized by their first Chern classes 
\[
c_1(\pi_*(\mathcal{J}_{p,k}))=\pi_*\bigl(\tfrac{1}{2}c_1(\mathcal{J})^2-\tfrac{1}{2}c_1(\mathcal{J})\cap K + \text{Td}_2(\pi)\bigr)
\]

For more general  $\mathcal{L}$,  we don't have a general formula for $\pi_*(\mathcal{L})$ on arbitrary $\overline{\mathcal{M}}_{g,n}$. But we can give very concrete expressions for $\overline{\mathcal{M}}_{0,4}\simeq \mathbb{CP}^1$. The result is verbatim the same as in Section 3.3 of \cite{Balasubramanian:2020fwc}. Our line bundles \eqref{ourLtoPush} are specified by a quintuple of integers, $(\tau^{(1)},\tau^{(2)},\tau^{(3)},\tau^{(4)};k)$. The direct image is 
\begin{equation}\label{pfdecomp}
\pi_*(\mathcal{L})=\bigoplus_{j=1}^{\lfloor k/2\rfloor}m_j \mathcal{O}(j)
\end{equation}
where the multiplicities $m_j$ are slightly complicated functions of this quintuple.

\begin{enumerate}
\item If it should happen that $\tau^{(a)}+\tau^{(b)} > k+1$ for some $a,b$, replace $\tau^{(a)}\to \tau^{(a)}-1$, $\tau^{(b)}\to \tau^{(b)}-1$ and $k\to k-1$.
\item Repeat step (1) as often as needed\footnote{This step corresponds \emph{precisely} with the twisting by components of the boundary described in \S\ref{twisting}. The point is that in $\text{Pic}(\mathcal{C}_{0,4})$ we have the isomorphism
$$
\KcD^{-1}\otimes\mathcal{O}(E_{a}+E_{b}) \simeq \mathcal{O}(-\, \mathcal{C}_{ab})
$$
So the operation of step (1) can be seen as tensoring with the line bundle given by \emph{either} the LHS or the RHS.}.
\item Now, with the $\tau^{(a)}$ in the desired range, let
\begin{equation}\label{fidef}
f_i =\max\Bigl(0,\tfrac{1}{2}\bigl[(k-2i+2)(k-2i+1)-\sum_{a=1}^4(\tau^{(a)}-i+1)\max(0,\tau^{(a)}-i)\bigr]\Bigr),\qquad i=1,\dots,\lfloor k/2\rfloor
\end{equation}
Then the multiplicities in \eqref{pfdecomp} are given by
\begin{equation}\label{midef}
\begin{split}
m_{\lfloor k/2\rfloor}&= f_{\lfloor k/2\rfloor}\\
m_{\lfloor k/2\rfloor-i}&= f_{\lfloor k/2\rfloor-i}-\sum_{j=0}^{i-1}(i-j+1) m_{\lfloor k/2\rfloor -j},\qquad i=1,\dots,\lfloor k/2\rfloor-1
\end{split}
\end{equation}

\end{enumerate}

\section{Examples}\label{sec:examples}

In this section, we illustrate the construction outlined above, by way of a series of examples. Example \ref{so12oddtypeexample} involves a 3-punctured sphere, examples \ref{SO12fourpointexample}--\ref{fouralphas} involve 4-punctured spheres, and we examine in detail the degeneration limits as $\lambda$ varies over $\overline{\mathcal{M}}_{0,4}$. Examples \ref{SO12fivepointexample},\ref{SO12fivepointsplittingexample} involve 5-punctured spheres and we compute the bundle of Hitchin bases over $\overline{\mathcal{M}}_{0,5}$.

\begin{itemize}
\item In example \ref{so12oddtypeexample}, we have both even-type and odd-type constraints. We construct two spectral curves, one birational to the double-cover of the other, depending on whether we impose the odd-type constraint.
\item The even-type constraints at $p_2$ in example \ref{SO12fourpointexample} require the introduction of line bundles $\mathcal{J}_{p_2,3}$,$\mathcal{J}_{p_2,5}$, as described in \S\ref{Jpk}. We see that different choices of $\mathcal{J}_{p_2,5}$ merely amount to different choice of coordinates on the Hitchin base.
\item In example \ref{SO8veryevenexample}, we have very-even punctures with a single even part.  There are distinct choices of the rank-2 vector bundle $\mathcal{F}$, but they push-forward to the same thing, and yield identical bundles of Hitchin bases.
\item In example \ref{SO8veryevennilpotentatnodeexample}, the degeneration as $\lambda\to1$ yields a very-even nilpotent at the node.
\item Example \ref{fouralphas} is a more complicated example where different choices of $\mathcal{J}_{a,k}$  yield different coordinatizations of the Hitchin base.
\item In example \ref{SO12fivepointexample}, we compute the bundle of Hitchin bases over $\overline{\mathcal{M}}_{0,5}$, for a particular example of 5-punctured sphere.
\item In example \ref{SO12fivepointsplittingexample}, we explore the splitting of the rank-2 vector bundle $\mathcal{F}\to \mathcal{C}_{0,5}$ which arises, as in \S\ref{veryevensingle}, when we have very-even partition(s) consisting of a single distinct even part. Here, we find that different choices of splitting yield topologically-distinct bundles of Hitchin bases.
 \end{itemize}
 
\subsection{\texorpdfstring{$n=3$}{n=3}}\label{3pointexamples}

Recall from \S\ref{C03}, that $\mathcal{C}_{0,3}\simeq \mathbb{CP}^1$, where we take the homogeneous coordinates to be $x,y$.  For $\mathfrak{so}(2N)$, the spectral curve $\Sigma$ is given as the vanishing locus of a polynomial  in $w,x,y$ 
\[
\Sigma=\Bigl\{ 0=w^{2N} +\bigl(\sum_{k=2}^{N-1}\phi_{2k}(x,y)w^{2(N-k)}\bigr) + \tilde{\phi}_{N}(x,y)^2\Bigr\}
\]
which is bihomogeneous of degree $(2N,2N)$ if we take $w$ to have degree $(1,1)$, $x,y$ to have degrees $(1,0)$ and the coefficients in $\phi_{2k}$ and $\tilde{\phi}$ to have degrees $(0,2k)$ and $(0,N)$ respectively. $\phi_{2k}$ vanishes to order
\begin{itemize}
\item $\chi_{2k}^{(1)}$ at $p_1=\{y=0\}$
\item $\chi_{2k}^{(2)}$ at $p_2=\{x=0\}$
\item $\chi_{2k}^{(3)}$ at $p_3=\{x=y\}$.
\end{itemize}
and similarly for $\tilde{\phi}$.
Subject to those vanishing orders, the coefficients  are further constrained by the constraints of Prop.~\ref{localconstraints}.

\refstepcounter{example}\paragraph{Example \theexample\label{so12oddtypeexample}:} Consider the 3-punctured sphere for $\mathfrak{so}(12)$, with Hitchin nilpotent orbits given by the partitions $[4^2,3,1]$, $[3^2,2^2,1^2]$ and $[5^2,1^2]$. Before imposing the constraints, we have the vanishing orders

\begin{center}
\begin{tabular}{|c|c|c|c|c|c|c|c|}
\hline
&$\chi_2$&$\chi_4$&$\chi_6$&$\chi_8$&$\chi_{10}$&$\tilde{\chi}_6$\\
\hline
$p_1$&1&1&2&2&3&2\\
$p_2$&1&2&2&3&4&3\\
$p_3$&1&1&2&2&2&2\\
\hline
\end{tabular}
\end{center}

So
\[
\begin{split}
\phi_2(x,y)&=\tilde{\phi}_6(x,y)=0\\
\phi_4(x,y)&=x^2y(x-y) c\\
\phi_6(x,y)&=x^2y^2(x-y)^2 d\\
\phi_8(x,y)&=x^3 y^2(x-y)^2(e_1 x+e_2 y)\\
\phi_{10}(x,y)&=x^4y^3(x-y)^2(f_1x-f_2 (x-y))\\
\end{split}
\]

The \emph{even} constraint at $p_1$ sets $c=0$. The \emph{even} constraint at $p_2$ sets 
\[
\begin{split}
d&=\alpha_3^2\\
e_2&=2\alpha_3\alpha_5\\
f_2&=\alpha_5^2
\end{split}
\]
Finally, the \emph{odd} constraint at $p_3$ says that $f_1=\beta_5^2$ is a perfect square. Whether to impose the constraint or not is encoded in the dual Nahm nilpotent orbit at $p_3$. If it is the special nilpotent $O_N=[3^2,2^2,1^2]$, then we impose the 
constraint $f_2=\beta_5^2$. This yields the spectral curve
\begin{subequations}
\begin{equation}\label{ex1curvea}
\Sigma =\Bigl\{
0=w^2\Bigl[
w^{10} +x^2y^2(x-y)^2\bigl(
(\alpha_3 w^2+\alpha_5 xy)^2 + e x^2 w^2 +x^3y(\beta_5^2-\alpha_5^2)
\bigr)
\Bigr]
\Bigr\}
\end{equation}
If it is the non-special nilpotent  $O_N=[3,2^4,1]$ (in the same special piece as $[3^2,2^2,1^2]$) then we don't impose the constraint and the spectral curve is 
\begin{equation}\label{ex1curveb}
\Sigma =\Bigl\{
0=w^2\Bigl[
w^{10} +x^2y^2(x-y)^2\bigl(
(\alpha_3 w^2+\alpha_5 xy)^2 + e x^2 w^2 +x^3y(f_2-\alpha_5^2)
\bigr)
\Bigr]
\Bigr\}
\end{equation}
\end{subequations}
Away from $\beta_5=0$, the spectral curve \eqref{ex1curveb}  is simply a double cover of \eqref{ex1curvea}. But, as explained in \S\ref{oddtype}, the fiber of \eqref{ex1curvea} over $f_2=0$ (and the fiber of \eqref{ex1curveb} over $\beta_5=0$) need to be resolved. The upshot is that \eqref{ex1curveb} is merely birational to the double cover of \eqref{ex1curvea}. And, indeed, they admit different sets of mass deformations, as explained in Appendix $\ref{massappendix}$ and are physically distinct (they correspond to different families of SCFTs).

\subsection{\texorpdfstring{$n=4$}{n=4}}

Recall from \S\ref{C04} that $\overline{\mathcal{M}}_{0,4}\simeq \mathbb{CP}^1$ and the universal curve $\mathcal{C}_{0,4}$ is $\mathbb{CP}^1\times \mathbb{CP}^1$ blown up at 3 points or, equivalently $\mathbb{CP}^2$ blown up at 4 points. We let $(x,y,z)$ be homogeneous coordinates on $\mathbb{CP}^2$ (or, with a slight abuse of notation, their pullbacks to its blowup, $\mathcal{C}_{0,4}$) and let the exceptional divisors of the blowup be $E_1\to (1,0,0)$, $E_2\to (0,1,0)$, $E_3\to(0,0,1)$ and $E_4\to(1,1,1)$. The projection $\mathcal{C}_{0,4}\to \overline{\mathcal{M}}_{0,4}$ is given by
\begin{equation}
\lambda_1 x(y-z) + \lambda_2y(z-x) = 0
\label{universalcurve}\end{equation}
which determines $\lambda_{1,2}$ up to a common scaling.  

The Hitchin bases, as we vary $\lambda\coloneqq\lambda_1/\lambda_2$, fit together to form a nontrivial fiber bundle over $\overline{\mathcal{M}}_{0,4}$.  Hence so do the family of spectral curves. In order to write down explicit formul\ae, we trivialize the bundle over $\overline{\mathcal{M}}_{0,4}\backslash\{\lambda=\infty\}$ (``the northern hemisphere''). Then we can write the spectral curve as
\[
\Sigma=\Bigl\{
0=w^{2N}+x(y-z)\bigl(
\sum_{k=1}^{N-1} \mathcal{P}_{2(N-k-1)}(x,y,z) w^{2(N-k)}
\bigr)
+x^2(y-z)^2 \tilde{\mathcal{P}}_{N}(x,y,z)^2
\Bigr\}
\]
The form of the spectral curve over the southern hemisphere is then dictated by the transition functions for the bundle of Hitchin bases.

\refstepcounter{example}\paragraph{Example \theexample:\label{SO12fourpointexample}} As an example, consider the 4-punctured sphere, with Hitchin nilpotents
\goodbreak{\noindent}$[2^2,1^8][3^2,2^2,1^2]{\color{red}[4^2,2^2]}[3^4]$. Before imposing the constraints and before any twisting at the boundary, the line bundles on the universal curve are
\[
\begin{aligned}
\mathcal{L}_2&= \KcD^{\otimes 2}\otimes \mathcal{O}(-E_{1}-E_{2}-E_{3}-E_{4})\\
\mathcal{L}_4&= \KcD^{\otimes 4}\otimes \mathcal{O}(-2E_{1}-2E_{2}-E_{3}-2E_{4})\\
\mathcal{L}_6&= \KcD^{\otimes 6}\otimes \mathcal{O}(-4E_{1}-2E_{2}-2E_{3}-2E_{4})\\
\mathcal{L}_8&= \KcD^{\otimes 8}\otimes \mathcal{O}(-6E_{1}-3E_{2}-2E_{3}-3E_{4})\\
\mathcal{L}_{10}&= \KcD^{\otimes 10}\otimes \mathcal{O}(-8E_{1}-4E_{2}-3E_{3}-4E_{4})\\
\tilde{\mathcal{L}}&= \KcD^{\otimes 6}\otimes \mathcal{O}(-5E_{1}-3E_{2}-2E_{3}-2E_{4})
\end{aligned}
\]
The constraint at $p_1$ sets
\[
c'_4=\tfrac{1}{4}c_2^2
\]
The constraints at $p_2$ set 
\begin{equation}\label{complicatedexamplep2}
\begin{split}
c''_6&=\alpha_3^2\\
c'_8&=2\alpha_3\alpha_5\\
c_{10}&=\alpha_5^2
\end{split}
\end{equation}
The constraint at $p_3$ set
\[
\begin{split}
c''_8&=\tfrac{1}{4}c_4^2\\
c'_{10}&=+c_4\tilde{c}_6
\end{split}
\]
If we replace $\color{red}[4^2,2^2]$ with $\color{blue}[4^2,2^2]$, this would change the second constraint to $c'_{10}=-c_4\tilde{c}_6$.

The constraints \eqref{complicatedexamplep2} at $p_2$ introduce the torsion sheaves 
\[
\mathcal{S}_{p_2,k}^{\otimes 2}= \mathcal{L}_{2k}\vert_{E_{2}}
\]
for $k=3,5$. We seek corresponding line bundles $\mathcal{J}_{2,3}$ and $\mathcal{J}_{2,5}$ on the universal curve such that
\begin{itemize}
\item $(\mathcal{J}_{2,k})^{\otimes2}$ is a subsheaf of $\mathcal{L}_{2k}$.
\item $\mathcal{J}_{2,k}\vert_{E_{2}} = \mathcal{S}_{p_2,k}$
\item $\pi_*(\mathcal{J}_{2,k})=(I_2)^{\otimes\chi^{(2)}_{2k}/2} \otimes\mathcal{O}(-\sum_{a\neq 2}\ell_a D_{2,a}) $, where $\ell_a \coloneqq \max\bigl(0,(\chi^{(2)}_{2k}+\chi^{(a)}_{2k})/2-k\bigr)$ and $D_{2,a}$ denotes the divisor (point) on $\overline{\mathcal{M}}_{0,4}$ where $p_2$ and $p_a$ collide.
\end{itemize}
This yields
\begin{subequations}
\begin{align}
\mathcal{J}_{p_2,3}&=\KcD^{\otimes3}\otimes \mathcal{O}(-2E_{1}-E_{2}-2E_{3}-E_{4})\label{complicatedexampleJimageChoice}\\
\mathcal{J}_{p_2,5}&=\KcD^{\otimes5}\otimes \mathcal{O}(-4E_{1}-2E_{2}-2E_{3}-2E_{4})
\end{align}
\end{subequations}
which satisfy
\begin{equation}\label{complicatedexampleJimage}
\begin{split}
\pi_*(\mathcal{J}_{p_2,3})&=\pi_*( \mathcal{S}_{p_2,3})= \mathcal{O}(1)\\
\pi_*(\mathcal{J}_{p_2,5})&=\pi_*( \mathcal{S}_{p_2,5})\otimes \mathcal{O}(-1)= \mathcal{O}(1)\\
\end{split}
\end{equation}

In addition to $\mathcal{J}_{p_2,3},\mathcal{J}_{p_2,5}$, the constraints replace the $\mathcal{L}_{2k}$ by $\mathcal{L}'_{2k}$ where (after including the twist at the boundary)
\begin{equation}\label{complicatedexampleLp}
\begin{aligned}
\mathcal{L}'_2&= \mathcal{L}_2&\pi_*(\mathcal{L}_2)&=\mathcal{O}(1)\\
\mathcal{L}'_4&= \KcD^{\otimes 4}\otimes \mathcal{O}(-3E_{1}-2E_{2}-E_{3}-2E_{4})&\pi_*(\mathcal{L}'_4)&=\mathcal{O}(1)\\
\mathcal{L}'_6&= \KcD^{\otimes 6}\otimes \mathcal{O}(-4E_{1}-3E_{2}-2E_{3}-2E_{4})&\pi_*(\mathcal{L}'_6)&=2\mathcal{O}(2)\\
\mathcal{L}'_8&= \KcD^{\otimes 8}\otimes \mathcal{O}(-6E_{1}-4E_{2}-3E_{3}-3E_{4}-\mathcal{C}_{12})&\pi_*(\mathcal{L}'_8)&=\mathcal{O}(2)\\
\mathcal{L}'_{10}&= \KcD^{\otimes 10}\otimes \mathcal{O}(-8E_{1}-5E_{2}-4E_{3}-4E_{4}-2\mathcal{C}_{12}-\mathcal{C}_{13}-\mathcal{C}_{14})&\quad\pi_*(\mathcal{L}'_{10})&=0\\
\tilde{\mathcal{L}}'&= \tilde{\mathcal{L}}&\pi_*(\tilde{\mathcal{L}}')&=\mathcal{O}(1)
\end{aligned}
\end{equation}

Trivializing the bundle of Hitchin bases away from $\lambda=\infty$ (i.e. trivializing the bundle in the ``northern hemisphere'' of $\overline{\mathcal{M}}_{0,4}$), we get the spectral curve
\begin{equation}\label{complicatedexampleSCn}
\begin{split}
\Sigma&= \Bigl\{
0=w^{12}+w^{10}x(y-z)c_2 +w^8x(y-z)^2\bigl( zc_4+\tfrac{1}{4}x c_2^2\bigr)
+w^6x^2(y-z)^2z(yc_6+zc'_6)\\
&\qquad +w^4x^2(y-z)^3z^2\bigl(y c_8+\tfrac{1}{4}(y-z)c_4^2\bigr)
+w^2 x^2(y-z)^4y z^3 c_4 \tilde{c}_6\\
&\qquad +w^2x^2(y-z)^2y^2\bigl(w^2\alpha_3+z(y-z)\alpha_5\bigr)^2+x^2(y-z)^4y^2z^4 \tilde{c}_6^2
\Bigr\}
\end{split}
\end{equation}
Using the transition functions dictated by \eqref{complicatedexampleJimage},\eqref{complicatedexampleLp},
\[
c_2=\lambda\hat{c}_2,\quad
c_4=\lambda\hat{c}_4,\quad
c_6=\lambda^2\hat{c}_6,\quad
c'_6=\lambda^2\hat{c}'_6,\quad
c_8=\lambda^2\hat{c}_8,\quad
\tilde{c}_6=\lambda\hat{\tilde{c}}_6,\quad
\alpha_3=\lambda\hat{\alpha}_3,\quad
\alpha_5=\lambda\hat{\alpha}_5
\]
we can write the expression for the spectral curve in the southern hemisphere of $\overline{\mathcal{M}}_{0,4}$
\begin{equation}\label{complicatedexampleSCs}
\begin{split}
\Sigma&= \Bigl\{
0=w^{12}-w^{10}y(z-x)\hat{c}_2 
-w^8y(z-x)\bigl(-\tfrac{1}{4}y(z-x) \hat{c}_2^2 +z(y-z)\hat{c}_4\bigr)
+w^6y^2(z-x)^2z(y\hat{c}_6+z\hat{c}'_6)\\
&\qquad +w^4y^2(z-x)^2(y-z)z^2\bigl(y \hat{c}_8+\tfrac{1}{4}(y-z)\hat{c}_4^2\bigr)
+w^2 y^3(z-x)^2(y-z)^2 z^3 \hat{c}_4 \hat{\tilde{c}}_6\\
&\qquad +w^2y^4(z-x)^2\bigl(w^2\hat{\alpha}_3+z(y-z)\hat{\alpha}_5\bigr)^2+y^4(z-x)^2(y-z)^2z^4 \hat{\tilde{c}}_6^2
\Bigr\}
\end{split}
\end{equation}

We could just as well replace \eqref{complicatedexampleJimageChoice} with
\begin{equation}\label{complicatedexampleOtherJimageChoice}
\mathcal{J}_{p_2,3}=\KcD^{\otimes3}\otimes \mathcal{O}(-2E_{1}-E_{2}-E_{3}-2E_{4})
\end{equation}
This (obviously)  changes the third line of \eqref{complicatedexampleSCn}
\begin{equation}\label{complicatedexampleSCnAlt}
\begin{split}
\Sigma&= \Bigl\{
0=w^{12}+w^{10}x(y-z)c_2 +w^8x(y-z)^2\bigl( zc_4+\tfrac{1}{4}x c_2^2\bigr)
+w^6x^2(y-z)^2z(yc_6+zc'_6)\\
&\qquad +w^4x^2(y-z)^3z^2\bigl(y c_8+\tfrac{1}{4}(y-z)c_4^2\bigr)
+w^2 x^2(y-z)^4y z^3 c_4 \tilde{c}_6\\
&\qquad +{\color{red}w^2x^2(y-z)^4\bigl(w^2\alpha_3+yz\alpha_5\bigr)^2}+x^2(y-z)^4y^2z^4 \tilde{c}_6^2
\Bigr\}
\end{split}
\end{equation}
Passing from \eqref{complicatedexampleSCn} to \eqref{complicatedexampleSCnAlt} (i.e., replacing \eqref{complicatedexampleJimageChoice} with \eqref{complicatedexampleOtherJimageChoice}) simply amounts to the change of coordinates on the fibers of bundle of Hitchin bases
\begin{equation}
\begin{split}
c_6&\to c_6-2\alpha_3^2\\
c'_6&\to c'_6+\alpha_3^2\\
c_8&\to c_8-2\alpha_3\alpha_5
\end{split}
\end{equation}

Now let's examine the nodal degenerations of $C$.

\paragraph{\texorpdfstring{$\lambda=0$}{λ=0}:}
$C_L=\mathcal{C}_{13}=\{y=0\}$ and $C_R=\mathcal{C}_{24}=\{z=x\}$. The center parameters (the parameters common to $\Sigma_L$ and $\Sigma_R$) are  $c_2,c_4,c'_6$, which are the Casimirs of $\mathfrak{so}(7)$. Setting the center parameters to zero, we have the spectral curves:
\[
\begin{split}
\Sigma_L&=\{0=w^{12}\}\\
\Sigma_R&=\bigl\{
0=w^{12}+w^6x^3y(x-y)^2 c_6 - w^4x^4 y (x-y)^3 c_8 \\
&\qquad\qquad+w^2x^2y^2(x-y)^2(w^2\alpha_3-x(x-y)\alpha_5)^2 +x^6y^2(x-y)^4 \tilde{c}^2
\bigr\}
\end{split}
\]
The Hitchin moduli space for the theory on the left is a point; the theory consists of 8 free hypermultiplets transforming as the spinor of $\mathfrak{so}(7)$. The spectral curve $\Sigma_R$ is the one which arises for the 3-punctured sphere, with Hitchin nilpotents $[9,3] [3^2,2^2,1^2] [3^4] $. So we say that the nilpotent at the node is $\bigl([9,3],\mathfrak{so}(7)\bigr)$.

\paragraph{\texorpdfstring{$\lambda=1$}{λ=1}:}
$C_L=\mathcal{C}_{12}=\{z=0\}$ and $C_R=\mathcal{C}_{34}=\{y=x\}$. The center parameters are $c_2,\alpha_3$, which are the Casimirs of $\mathfrak{su}(3)$. Setting the center parameters to zero, we have the spectral curves
\[
\begin{split}
\Sigma_L&=\{0=w^{12}\}\\
\Sigma_R&=\bigl\{
0=w^{12}+w^8xz(x-z)^2c_4 + w^6 x^2z(x-z)^2(xc_6+z c'_6)
+w^4x^2z^2(x-z)^3(xc_8+\tfrac{1}{4}(x-z)c_4^2)\\
&\qquad+w^2x^3z^3(x-z)^4 c_4\tilde{c}_6 +w^2 x^4z^2(x-z)^4 \alpha_5^2 +x^4z^4(x-z)^4 \tilde{c}_6^2
\bigr\}
\end{split}
\]
The theory on the left is trivial. On the right, we have the spectral curve for the 3-punctured sphere with Hitchin nilpotents $[7,3,1^2] {\color{red}[4^2,2^2]}[3^4]$. More precisely, since $[7,3,1^2]$ has an odd-type constraint, we want the Hitchin system with $\alpha_5$ rather than $c_{10}$ as a base parameter. That is, we want the theory where the dual Nahm partition is the special one $[3^2,1^6]_N$ (rather than the non-special $[3,2^2,1^5]_N$). So the nilpotent at the node is $\bigl(([7,3,1^2],\Bid), \mathfrak{su}(3)\bigr)$.

Note that $\alpha_5$ appears as a parameter in the spectral curve $\Sigma_R$, even though $[3^2,2^2,1^2]$ appears on $C_L$. That's because the unique holomorphic section of\goodbreak\noindent $\mathcal{J}_{p_2,5}=\KcD^{\otimes5}\otimes \mathcal{O}(-4E_1-2E_2-2E_3-2E_4)$ is supported on $C_R$, not on $C_L$. Also note that, in this case, the Nahm nilpotent at the node is the special, rather than non-special nilpotent in the special piece.

\paragraph{\texorpdfstring{$\lambda=\infty$}{λ=∞}:}
To study the behaviour near $\lambda=\infty$, we need to work in the trivialization \eqref{complicatedexampleSCs} good in the southern hemisphere. $C_L=\mathcal{C}_{14}=\{z=y\}$, $C_R=\mathcal{C}_{23}=\{x=0\}$ and the center parameters are $\hat{c}_2,(\hat{c}'_6+\hat{c}_6+\hat{\alpha}^2_3)$ which are the Casimirs of $\mathfrak{g}_2$. Setting the center parameters to zero
\[
\begin{split}
\Sigma_L&=\{0=w^{12}\}\\
\Sigma_R&=\bigl\{
0=w^{12}-w^8yz^2(y-z)\hat{c}_4 +
w^6y^2z^3(y-z)\tfrac{1}{2}(\hat{c}_6+\hat{\alpha}_3^2)\\
&\qquad\qquad +w^4 y^2 z^4 (y-z) (y\hat{c}_8 +\tfrac{1}{4}(y-z)\hat{c}_4^2
+w^2y^3 z^5 (y-z)^2 \hat{c}_4\hat{\tilde{c}}_6\\
&\qquad\qquad +
 w^2 y^3 z^2 (y-z)\bigl(w^4 \hat{\alpha}_3^2+2w^2 yz\hat{\alpha}_3\hat{\alpha}_5+yz^2(y-z)\hat{\alpha}_5^2\bigr)
+y^4z^6(y-z)^2 \hat{\tilde{c}}_6^2
\bigr\}
\end{split}
\]
where the theory on the left is trivial and the Hitchin system on the right is the one that arises from the 3-punctured sphere with Hitchin nilpotents $[3^2,2^2,1^2] {\color{red}[4^2,2^2]} [9,3]$. The nilpotent at the node is $([9,3],\mathfrak{g}_2)$.

\refstepcounter{example}\paragraph{Example \theexample:\label{SO8veryevenexample}} Consider ${\color{red}[2^4]}{\color{red}[4^2]}[5,1^3][7,1]$ versus ${\color{red}[2^4]}{\color{blue}[4^2]}[5,1^3][7,1]$. Before imposing the constraints, we have
\[
\begin{split}
\mathcal{L}_2&= \KcD^{\otimes2}\otimes\mathcal{O}(-E_{1}-E_{2}-E_{3}-E_{4})\\
\mathcal{L}_4&= \KcD^{\otimes4}\otimes\mathcal{O}(-2E_{1}-E_{2}-E_{3}-E_{4})\\
\mathcal{L}_6&= \KcD^{\otimes6}\otimes\mathcal{O}(-3E_{1}-2E_{2}-2E_{3}-E_{4})\\
\tilde{\mathcal{L}}&= \KcD^{\otimes4}\otimes\mathcal{O}(-2E_{1}-E_{2}-2E_{3}-E_{4})
\end{split}
\]
Imposing the constraints replaces $\mathcal{L}_6$ with
\[
\mathcal{L}'_6= \KcD^{\otimes6}\otimes\mathcal{O}(-4E_{1}-2E_{2}-2E_{3}-E_{4})\\
\]
with $\pi_*(\mathcal{L}'_6)=3\mathcal{O}(2)+\mathcal{O}(1)$.

In the case of {\color{red}red}-{\color{red}red}, $\mathcal{L}_4\oplus \tilde{\mathcal{L}}$ is replaced by
\begin{subequations}\label{ff}
\begin{equation}\label{frr}
\mathcal{F} = \mathcal{L}_4(-E_{1}-E_{3})\oplus \tilde{\mathcal{L}}
\end{equation}
For  {\color{red}red}-{\color{blue}blue}, we seem to have to decide between 
\begin{equation}\label{frb1}
\mathcal{F} \stackrel{?}{=} \mathcal{L}_4(-E_{1})\oplus \tilde{\mathcal{L}}(-E_{2})
\end{equation}
and 
\begin{equation}\label{frb2}
\mathcal{F} \stackrel{?}{=} \mathcal{L}_4(-E_{2})\oplus \tilde{\mathcal{L}}(-E_{1})
\end{equation}
\end{subequations}
All three choices have the same direct image, so that the $k$-differentials end up being (in the conventions of Appendix \ref{so8complete})
\[
\begin{split}
\phi_2&=x(y-z)\; a\\
\phi_4&= x^2(y-z)^2(2\tilde{b}_1+\tfrac{1}{4}a^2) +x(y-z)[\pm 2y(y-z)\tilde{b}_2 +yzb_3+z(y-z)b_4]\\
\phi_6&= x^3(y-z)^3\; a\tilde{b}_1 +x^2(y-z)^2 [y^2 c_1+z^2c_2+(y-z)^2 c_3]+x(y-z) y^2z^2 c_4\\
\tilde{\phi}&=x^2(y-z)^2\tilde{b}_1 +x(y-z)[y(y-z)\tilde{b}_2+yz\tilde{b}_3]
\end{split}
\]
where the $\pm$ sign corresponds to $\color{red}[4^2]$ and $\color{blue}[4^2]$ respectively.

In all three cases of \eqref{ff}, $\pi_*(\mathcal{F}) = \mathcal{O}(2)\oplus 4\mathcal{O}(1)$. But the assignments are slightly different.
\begin{itemize}
\item For \eqref{frr}, $b_3,b_4$ are the global sections of $ \mathcal{L}_4(-E_{1}-E_{2})$, while $\tilde{b}_1,\tilde{b}_2,\tilde{b}_3$ are the global sections of $\tilde{\mathcal{L}}$.
\item For \eqref{frb1}, $\tilde{b}_2,b_3,b_4$ are the global sections of $\mathcal{L}_4(-E_{1})$, while $\tilde{b}_1,\tilde{b}_3$ are the global sections of $\tilde{\mathcal{L}}(-E_{2})$.
\item For \eqref{frb2}, $\tilde{b}_1,b_3,b_4$ are the global sections of $\mathcal{L}_4(-E_{2})$, while $\tilde{b}_2,\tilde{b}_3$ are the global sections of $\tilde{\mathcal{L}}(-E_{1})$.
\end{itemize}
Whichever way we do things, $\tilde{b}_1,c_1,c_2,c_3$ are fiber coordinates on $\mathcal{O}(2)$s and $a,\tilde{b}_2,\tilde{b}_3,b_3,b_4,c_4$ are fiber coordinates on $\mathcal{O}(1)$s.

\refstepcounter{example}\paragraph{Example \theexample:\label{SO8veryevennilpotentatnodeexample}} Consider ${\color{red}[2^4]} {\color{red}[2^4]} [3^2,1^2] [7,1]$. Before imposing any constraints, we have
\[
\begin{split}
\mathcal{L}_2&= \KcD^{\otimes 2}\otimes \mathcal{O}(-E_1-E_2-E_3-E_4)\\
\mathcal{L}_4&= \KcD^{\otimes 4}\otimes \mathcal{O}(-2E_1-2E_2-2E_3-E_4)\\
\mathcal{L}_6&= \KcD^{\otimes 6}\otimes \mathcal{O}(-3E_1-3E_2-2E_3-E_4)\\
\tilde{\mathcal{L}}&= \KcD^{\otimes 4}\otimes \mathcal{O}(-2E_1-2E_2-2E_3-E_4)\\
\end{split}
\]
When we impose the constraints, $\mathcal{L}_4\oplus \tilde{\mathcal{L}}$ is replaced by the rank-2 vector bundle $\mathcal{F}'$ which splits as
\begin{equation}
\mathcal{F}' = \mathcal{L}_4(-E_1-E_2-\mathcal{C}_{12})\oplus \tilde{\mathcal{L}}
\end{equation}
$[3^2,1^2]$ has two nilpotent orbits in the dual Nahm special piece. If we choose $[3,2^2,1]_N$, then we don't impose the constraint and
\begin{subequations}
\begin{equation}\label{ex4oddtypeL6}
\mathcal{L}'_6 = \KcD^{\otimes 6}\otimes\mathcal{O}(-4E_1-4E_2-2E_1 -E_4-\mathcal{C}_{12})
\end{equation}
If, instead, we choose $[3^2,1^2]_N$, then we impose the constraint and replace \eqref{ex4oddtypeL6} by the pair
\begin{equation}
\begin{split}
\mathcal{L}'_6 &= \KcD^{\otimes 6}\otimes\mathcal{O}(-4E_1-4E_2-3E_1 -E_4-\mathcal{C}_{12})\\
\mathcal{J}_3&= \KcD^{\otimes 3}\otimes\mathcal{O}(-2E_1-2E_2-E_3-E_4)
\end{split}
\end{equation}
\end{subequations}

Following the conventions of Appendix \ref{so8complete}, this yields the spectral curve
\begin{subequations}
\begin{equation}
\begin{split}
\Sigma=\Bigl\{
0=w^8+x(y-z)\bigr[w^6a&+w^4\bigl(x(y-z)(\tfrac{1}{4}a^2-2\tilde{b}_1)-2yz\tilde{b}_2\bigr)\\
&+w^2\bigl[
-x^2(y-z)^2 a\tilde{b}_1+x(y-z)z(z{\color{red}c_3}-ya\tilde{b}_2)+xyz^2c_6
\bigr]\\
&+\bigl(x(y-z)(x(y-z)\tilde{b}_1+yz\tilde{b}_2)\bigr)^2
\Bigr\}
\end{split}
\end{equation}
for the case of $[3,2^2,1]_N$ or 
\begin{equation}
\begin{split}
\Sigma=\Bigl\{
0=w^8+x(y-z)\bigr[w^6a&+w^4\bigl(x(y-z)(\tfrac{1}{4}a^2-2\tilde{b}_1)-2yz\tilde{b}_2\bigr)\\
&+w^2\bigl[
-x^2(y-z)^2 a\tilde{b}_1+x(y-z)z(z{\color{red}\alpha^2}-ya\tilde{b}_2)+xyz^2c_6
\bigr]\\
&+\bigl(x(y-z)(x(y-z)\tilde{b}_1+yz\tilde{b}_2)\bigr)^2
\Bigr\}
\end{split}
\end{equation}
for the case of $[3^2,1^2]_N$.
\end{subequations}
To be really precise (since we will need it later)
\[
\tilde{\phi}= x^2(y-z)^2 \tilde{b}_1+ x(y-z)yz\tilde{b}_2
\]
where $\tilde{b}_1,\tilde{b}_2$ are, respectively the fiber coordinates on $\mathcal{O}(2)\oplus\mathcal{O}(1)=\pi_*(\tilde{\mathcal{L}})$.

Since we chose the same nilpotents at $p_{1,2}$, there are really only two distinct degenerations to consider.

\paragraph{\texorpdfstring{$\lambda=0$}{λ=0}:}
$C_L=\{y=0\}$ and $C_R=\{z=x\}$. The center parameters are $a,\tilde{b}_1,c_3$ for $[3,2^2,1]_N$ and $a,\alpha, \tilde{b}_1$ for $[3^2,1^2]_N$. Setting the center parameters to zero,
\begin{equation}
\begin{split}
\Sigma_L&=\bigl\{0=w^8\bigr\}\\
\Sigma_R&=\Bigl\{0=w^8+2w^4x^2y(x-y)\tilde{b}_2-w^2 x^4y(x-y)c_6 +(x^2y(x-y)\tilde{b}_2)^2\Bigr\}\\
\end{split}
\end{equation}
$\phi_4,\phi_6$ and $\tilde{\phi}$ each have a simple zero at the node ($(x,y)=(1,0)$). So $\Sigma_R$ is the spectral curve for the Hitchin system associated to the 3-punctured sphere with $[7,1]{\color{red}[2^4]}[7,1]$ at the punctures. 
The Hitchin moduli space for the theory on the left is a point.
\begin{itemize}
\item For $[3,2^2,1]_N$, the center parameters are the independent Casimirs of $\mathfrak{so}(7)$, the nilpotent at the node is\footnote{Here, ${\color{red}\mathfrak{so}(7)}\subset \mathfrak{so}(8)$ is the embedding under which the three 8-dimensional irreps of $\mathfrak{so}(8)$ decompose as $({\color{midgreen}8_v},{\color{red}8_s},{\color{blue}8_c})=(8,7+1,8)$ under ${\color{red}\mathfrak{so}(7)}$.} $(O,\mathfrak{h})=([7,1],{\color{red}\mathfrak{so}(7)})$ and the theory on the left consists of 15 free hypermultiplets, transforming as $8+7$ under ${\color{red}\mathfrak{so}(7)}$.
\item For $[3^2,1^2]_N$, the center parameters are the independent Casimirs of $\mathfrak{so}(6)$, the nilpotent at the node is $(O,\mathfrak{h})=([7,1],\mathfrak{so}(6))$ and the theory on the left consists of 8 free hypermultiplets, transforming as $2(4)$ under $\mathfrak{so}(6)=\mathfrak{su}(4)$.
\end{itemize}

\paragraph{\texorpdfstring{$\lambda=1$}{λ=1}:}
$C_L=\{z=0\}$ and $C_R=\{y=x\}$. The center parameters are $a,\tilde{b}_1$, which are the Casimirs of $\mathfrak{sp}(2)$. Setting the center parameters to zero, we have the spectral curves
\begin{equation}
\begin{split}
\Sigma_L&=\bigl\{0=w^8\bigr\}\\
\Sigma_R&=\begin{cases}
\vphantom{\Biggl[}\Bigl\{0=w^8-2w^4x^2 z(x-z)\tilde{b}_2+w^2x^2z^2(x-z)[(x-z) {\color{red}c_3} + xc_6]+\bigl(x^2z(x-z)\tilde{b}_2\bigr)^2\Bigr\}\\
\vphantom{\Biggl[}\Bigl\{0=w^8-2w^4x^2 z(x-z)\tilde{b}_2+w^2x^2z^2(x-z)[(x-z) {\color{red}\alpha^2} + xc_6]+\bigl(x^2z(x-z)\tilde{b}_2\bigr)^2\Bigr\}
\end{cases}
\end{split}
\end{equation}
for $[3,2^2,1]_N$ and $[3^2,1^2]$ respectively.

The Hitchin moduli space for the theory on the left is a point. The theory on the left consists of 8 free hypermultiplets transforming as $2(4)$ of $\mathfrak{sp}(2)$. 

On the right, both $\phi_4$ and $\tilde{\phi}$ have simple zeros at the node ($(x,z)=(1,0))$, while $\phi_6$ has a double zero there. But $\phi_4+2\tilde{\phi}=0$, so the nilpotent at the node is $(O,\mathfrak{h})=({\color{red}[4^2]},\mathfrak{sp}(2))$ The spectral curve(s) $\Sigma_R$ is the one associated to the Hitchin system on the 3-punctured sphere with nilpotents ${\color{red}[4^2]}[3^2,1^2][7,1]$ (with the two choices of Nahm nilpotent corresponding to $[3^2,1^2]$).

\refstepcounter{example}\paragraph{Example \theexample:\label{fouralphas}} 

While it seems that there is no obstruction, at least at genus-0, to finding a collection of line bundles $\mathcal{J}_{p,k}$ on the universal curve satisfying the criteria set out in \S\ref{Jpk}, as we saw in Example \ref{SO12fourpointexample} there may be no preferred choice.


Consider $[3^2,1^2][3^2,1^2][3^2,1^2][3^2,1^2]$. Before imposing the constraints, we have
\begin{equation}\label{four3311Ls}
\begin{split}
\mathcal{L}_2&= \KcD^{\otimes 2}\otimes\mathcal{O}(-E_1-E_2-E_3-E_4)\\
\mathcal{L}_4&= \KcD^{\otimes 4}\otimes\mathcal{O}(-2E_1-2E_2-2E_3-2E_4)\\
\mathcal{L}_6&= \KcD^{\otimes 6}\otimes\mathcal{O}(-2E_1-2E_2-2E_3-2E_4)\\
\tilde{\mathcal{L}}&= \KcD^{\otimes 4}\otimes\mathcal{O}(-2E_1-2E_2-2E_3-2E_4)\\
\end{split}
\end{equation}
The odd-type constraints (assuming we impose the odd-type constraints at all four punctures) replace $\mathcal{L}_6$ with
\begin{equation}\label{four3311Lp6}
\mathcal{L}'_6= \KcD^{\otimes 6}\otimes\mathcal{O}(-3E_1-3E_2-3E_3-3E_4)
\end{equation}
and introduce four torsion sheaves
\[
\mathcal{S}_{a,3} =\mathcal{O}_{E_a}(-E_a),\qquad a=1,\dots,4
\]
 These push-forward as expected
 \[
 \begin{split}
\pi_*(\mathcal{L}_2)&=\mathcal{O}(1)\\
\pi_*(\mathcal{L}_4)&=\mathcal{O}(2)\\
\pi_*(\mathcal{L}'_6)&=\mathcal{O}(3)\\
\pi_*(\tilde{\mathcal{L}})&=\mathcal{O}(2)\\
\pi_*(\mathcal{S}_{a,3})&=\mathcal{O}(1)
\end{split}
 \]
 In complete accord with the table in Appendix \ref{so8complete}, this yields
 \label{four3311sol}
\begin{equation}
 \begin{split}
\phi_2&=x(y-z)a\\
\phi_4&=x^2(y-z)^2b\\
\tilde{\phi}&=x^(y-z)^2\tilde{b}\\
\end{split}
\end{equation}
To implement the odd-type constraints at the four punctures $p_a$ we would like to find line bundles $\mathcal{J}_{a,3}$ satisfying the criteria set out in \S\ref{Jpk}. There are six line bundles which satisfy those criteria
\begin{equation}
\begin{split}
\mathcal{J}_1&=\KcD^{\otimes 3}\otimes \mathcal{O}(-E_1-E_2-2E_3-2E_4)\\
\mathcal{J}_2&=\KcD^{\otimes 3}\otimes \mathcal{O}(-2E_1-E_2-E_3-2E_4)\\
\mathcal{J}_3&=\KcD^{\otimes 3}\otimes \mathcal{O}(-E_1-2E_2-E_3-2E_4)\\
\mathcal{J}_4&=\KcD^{\otimes 3}\otimes \mathcal{O}(-E_1-2E_2-2E_3-E_4)\\
\mathcal{J}_5&=\KcD^{\otimes 3}\otimes \mathcal{O}(-2E_1-E_2-2E_3-E_4)\\
\mathcal{J}_6&=\KcD^{\otimes 3}\otimes \mathcal{O}(-2E_1-2E_2-E_3-E_4)\\
\end{split}
\end{equation}
with no obvious subset of four to pick. Instead, we can use all six to write (in most symmetrical fashion)
\begin{equation}
\begin{split}
\phi_6= x^3(y-z)^3 c +\tfrac{1}{6}x^2(y-z)^2 \bigl[
(x-y)^2&(\;\;2\alpha_1^2+2\alpha_2^2-\alpha_3^2-\alpha_4^2)\\
+(y-z)^2&(-\alpha_1^2+2\alpha_2^2+2\alpha_3^2-\alpha_4^2)\\
+(z-x)^2&(\;\;2\alpha_1^2-\alpha_2^2+2\alpha_3^2-\alpha_4^2)\\
+x^2&(\;\;2\alpha_1^2-\alpha_2^2-\alpha_3^2+2\alpha_4^2)\\
+y^2&(-\alpha_1^2+2\alpha_2^2-\alpha_3^2+2\alpha_4^2)\\
+z^2&(-\alpha_1^2-\alpha_2^2+2\alpha_3^2+2\alpha_4^2)
\bigr]
\end{split}
\end{equation}
where the $\alpha_a$ are fiber coordinates on $\mathcal{O}(1)$ and $c$ is a fiber coordinate on $\mathcal{O}(3)$.

On the other hand, the table yields
\begin{equation}
 \phi_6= x^3(y-z)^3 c_1+x^2(y-z)^2\bigl[-(x-y)(z-x)\alpha_1^2-y(x-y)\alpha_2^2+z(z-x)\alpha_3^2+yz\alpha_4^2\bigr]
\end{equation}
 These differ by a change of fiber coordinates on the bundle of Hitchin bases
 \begin{equation}
c_1= c+\tfrac{1}{3}\bigl(\lambda-(1-\lambda)\bigr)(\alpha_1^2+\alpha_2^2+\alpha_3^2+\alpha_4^2)
\end{equation}
where, as in Appendix \ref{so8complete}, you should view ``$\lambda$'' and ``$(1-\lambda)$'' as very particular holomorphic sections of $\mathcal{O}(1)$ which we use to map the total space of $\mathcal{O}(2)$ to $\mathcal{O}(3)$.
 
Note that this peculiar situation resulted from having to solve \emph{four} odd-type constraints. If we replaced the nilpotent at $p_4$ with $[5,1^3]$, we wouldn't have this complaint. We would replace $\mathcal{L}_4$ in \eqref{four3311Ls} by
\[
\mathcal{L}_4=\KcD^{\otimes 4}\otimes\mathcal{O}(-2E_1-2E_2-2E_3-E_4)\\
\]
 and the lack of an odd-type constraint at $p_4$ would modify $\mathcal{L}'_6$ in \eqref{four3311Lp6} to
 \[
 \mathcal{L}'_6= \KcD^{\otimes 6}\otimes\mathcal{O}(-3E_1-3E_2-3E_3-2E_4)
 \]
 But now we \emph{can} find a nice choice of $\mathcal{J}_{a,3}$ for $a=1,2,3$:
 \[
 \begin{split}
\mathcal{J}_{1,3}&= \KcD^{\otimes3}\otimes\mathcal{O}(-E_1-2E_2-2E_3-E_4)\\
\mathcal{J}_{1,2}&= \KcD^{\otimes3}\otimes\mathcal{O}(-2E_1-E_2-2E_3-E_4)\\
\mathcal{J}_{1,3}&= \KcD^{\otimes3}\otimes\mathcal{O}(-2E_1-2E_2-E_3-E_4)\\
\end{split}
 \]
 and replace $\phi_4,\phi_6$ in \eqref{four3311sol} by
 \begin{equation}
\begin{split}
\phi_4&= x^2(y-z)^2 b_1+x(y-z)yz b_2\\
\phi_6&= x^3(y-z)^3 c_1 +x^2(y-z)^2\bigl[yz c_2 +x^2\alpha_1^2+y^2\alpha_2^2+z^2\alpha_3^2\bigr]
\end{split}
\end{equation}
where $\alpha_a$ is indeed the fiber coordinate on $\pi_*(\mathcal{J}_{a,3})$, as desired.

Using the Appendix, we would have written
\[
 \phi_6= x^3(y-z)^3 c'_1+x^2(y-z)^2\bigl[-(x-y)(z-x)c'_2-y(x-y)\alpha_2^2+z(z-x)\alpha_3^2+yz\alpha_4^2\bigr]
\]
where
\[
\begin{split}
c'_1&=c_1+\bigl(\lambda-(1-\lambda)\bigr)\alpha_1^2 +\lambda\alpha_2^2 -(1-\lambda)\alpha_3^2\\
c'_2&= c_2+\alpha_1^2+\alpha_2^2+\alpha_3^2
\end{split}
\]
  
\subsection{\texorpdfstring{$n=5$}{n=5}}\label{fivepuncturedexamples}
We will content ourselves with two examples of a 5-punctured sphere. 

\refstepcounter{example}\paragraph{Example \theexample:\label{SO12fivepointexample}} Consider $[3^2,2^2,1^2] [3^2,2^2,1^2][3^2,2^2,1^2][2^2,1^8][2^2,1^8]$. Before imposing any constraints, we have
\begin{equation}
\begin{split}
\mathcal{L}_2&=\KcD^{\otimes 2}\otimes \mathcal{O}(-E_1-E_2-E_3-E_4-E_5)\\
\mathcal{L}_4&=\KcD^{\otimes 4}\otimes \mathcal{O}(-2E_1-2E_2-2E_3-2E_4-2E_5)\\
\mathcal{L}_6&=\KcD^{\otimes 6}\otimes \mathcal{O}(-2E_1-2E_2-2E_3-4E_4-4E_5)\\
\mathcal{L}_8&=\KcD^{\otimes 8}\otimes \mathcal{O}(-3E_1-3E_2-3E_3-6E_4-6E_5)\\
\mathcal{L}_{10}&=\KcD^{\otimes 10}\otimes \mathcal{O}(-4E_1-4E_2-4E_3-8E_4-8E_5)\\
\tilde{\mathcal{L}}&=\KcD^{\otimes 6}\otimes \mathcal{O}(-3E_1-3E_2-3E_3-5E_4-5E_5)
\end{split}
\end{equation}
After imposing the constraints, these are replaced by\footnote{Here, we depart slightly from the notation of \S\ref{Jpk} in labeling the $\mathcal{J}_{a,5}$. \emph{Both} $\mathcal{J}_{2,5}$ and $\mathcal{J}_{3,5}$ satisfy the criteria set out there to replace $\mathcal{S}_{1,5}$. Similarly, we could use either $\mathcal{J}_{1,5}$ or $\mathcal{J}_{3,5}$ to replace $\mathcal{S}_{2,5}$ and either $\mathcal{J}_{1,5}$ or $\mathcal{J}_{2,5}$ to replace $\mathcal{S}_{3,5}$. Regardless of the slight awkwardness of labeling, we have three line bundles $\mathcal{J}_{a,5}$ with the requisite properties set out in \S\ref{Jpk} to replace the three torsion sheaves $\mathcal{S}_{a,5}$.}
\begin{equation}
\begin{split}
\mathcal{L}'_2&=\mathcal{L}_2\\
\mathcal{L}'_4&=\KcD^{\otimes 4}\otimes \mathcal{O}(-2E_1-2E_2-2E_3-3E_4-3E_5-\mathcal{C}_{45})\\
\mathcal{L}'_6&=\KcD^{\otimes 6}\otimes \mathcal{O}(-3E_1-3E_2-3E_3-4E_4-4E_5-\mathcal{C}_{45})\\
\mathcal{L}'_8&=\KcD^{\otimes 8}\otimes \mathcal{O}(-4E_1-4E_2-4E_3-6E_4-6E_5-\mathcal{C}_{14}-\mathcal{C}_{15}-\mathcal{C}_{24}-\mathcal{C}_{25}-\mathcal{C}_{34}-\mathcal{C}_{35}-3\mathcal{C}_{45})\\
\mathcal{L}'_{10}&=\KcD^{\otimes 10}\otimes \mathcal{O}(-5E_1-5E_2-5E_3-8E_4-8E_5-2\mathcal{C}_{14}-2\mathcal{C}_{15}-2\mathcal{C}_{24}-2\mathcal{C}_{25}-2\mathcal{C}_{34}-2\mathcal{C}_{35}-5\mathcal{C}_{45})\\
\tilde{\mathcal{L}}'&=\tilde{\mathcal{L}}\\
\mathcal{J}_{1,3}&=\KcD^{\otimes 3}\otimes \mathcal{O}(-E_1-2E_2-2E_3-2E_4-2E_5)\\
\mathcal{J}_{2,3}&=\KcD^{\otimes 3}\otimes \mathcal{O}(-2E_1-E_2-2E_3-2E_4-2E_5)\\
\mathcal{J}_{3,3}&=\KcD^{\otimes 3}\otimes \mathcal{O}(-2E_1-2E_2-E_3-2E_4-2E_5)\\
\mathcal{J}_{1,5}&=\KcD^{\otimes 5}\otimes \mathcal{O}(-3E_1-2E_2-2E_3-4E_4-4E_5-\mathcal{C}_{14}-\mathcal{C}_{15}-2\mathcal{C}_{45})\\
\mathcal{J}_{2,5}&=\KcD^{\otimes 5}\otimes \mathcal{O}(-2E_1-3E_2-2E_3-4E_4-4E_5-\mathcal{C}_{24}-\mathcal{C}_{25}-2\mathcal{C}_{45})\\
\mathcal{J}_{3,5}&=\KcD^{\otimes 5}\otimes \mathcal{O}(-2E_1-2E_2-3E_3-4E_4-4E_5-\mathcal{C}_{34}-\mathcal{C}_{35}-2\mathcal{C}_{45})\\
\end{split}
\end{equation}
As usual, let $I_a$ be the point line bundle corresponding to the $a^{\text{th}}$ point. Let
\[
\begin{split}
L&= I_1^{\otimes2}\otimes \mathcal{O}(-D_{14}-D_{15})
=I_2^{\otimes2}\otimes \mathcal{O}(-D_{24}-D_{25})
=I_3^{\otimes2}\otimes \mathcal{O}(-D_{34}-D_{35})\\
&=\mathcal{O}(D_{12}+D_{23}+D_{13}+D_{45})
\end{split}
\]
and let $\mathcal{T}^\vee$ be the log-cotangent bundle of $\overline{\mathcal{M}}_{0,5}$, as  defined in \eqref{logTangent}. Then
\begin{equation}
\begin{split}
\pi_*(\mathcal{L}_2)&=\mathcal{T}^\vee\\
\pi_*(\mathcal{L}'_4)&=L\\
 \pi_*(\mathcal{L}'_6)&=\mathcal{T}^\vee\otimes L\\
\pi_*(\mathcal{L}'_8)&= L^{\otimes 2}\\
\pi_*(\mathcal{L}'_{10})&=0\\
\pi_*(\tilde{\mathcal{L}}_6)&=0\\
\pi_*(\mathcal{J}_{a,3})&=I_a,\qquad a=1,2,3\\
\pi_*(\mathcal{J}_{a,5})&= L,\qquad a=1,2,3\\
\end{split}
\end{equation}
Altogether $\mathcal{B}\to \overline{\mathcal{M}}_{0,5}$ is a rank-12 graded vector bundle.

The spectral curve (over the patch with $x_1\neq 0$) is
\begin{equation}
\begin{split}
\Sigma =\Bigl\{0= w^{12} &+ w^{10}  \bigl[y_2(y_4-y_1) a_1+y_3(y_1-y_4)a_2 \bigr]\\
&+w^8  \bigl[\tfrac{1}{4}y_2^2(y_4-y_1)^2 a_1^2+\tfrac{1}{4}y_3^2(y_1-y_4)^2a_2^2+(y_1-y_2)(y_2-y_3)(y_3-y_1)y_4 b\bigr]\\
&+w^6\bigl[y_1^2(y_4-y_1)^2(y_2-y_3)^2\alpha_1^2+y_2^2(y_4-y_2)^2(y_3-y_1)^2\alpha_2^2+y_3^2(y_4-y_3)^2(y_1-y_2)^2\alpha_3^2\\
&\qquad
+(y_1-y_2)(y_2-y_3)(y_3-y_1)y_4\bigl(y_2(y_4-y_1)c_1+y_3(y_1-y_4)c_2\bigr)
\bigr]\\
& +w^4(y_1-y_2)(y_2-y_3)(y_3-y_1)y_4\bigl[ (y_1-y_2)(y_2-y_3)(y_3-y_1)y_4 d\\
&\qquad+2y_1(y_4-y_1)(y_2-y_3)\alpha_1\bigl((y_3-y_1)\beta_2+(y_1-y_2)\beta_3\bigr)\\
&\qquad+2y_2(y_4-y_2)(y_3-y_1)\alpha_2\bigl((y_1-y_2)\beta_3+(y_2-y_3)\beta_1\bigr)\\
&\qquad+2y_3(y_4-y_3)(y_1-y_2)\alpha_3\bigl((y_2-y_3)\beta_1+(y_3-y_1)\beta_2\bigr)
\bigr]\\
&+w^2 (y_1-y_2)^2(y_2-y_3)^2(y_3-y_1)^2 y_4^2\bigl[(y_2-y_3)\beta_1+(y_3-y_1)\beta_2+(y_1-y_2)\beta_3\bigr]^2
\Bigr\}
\end{split}
\end{equation}
Here $a_{1,2}$ are the fiber coordinates on $\pi_*(\mathcal{L}_2)=\mathcal{T}^\vee$, $b$ is the fiber coordinate on $\pi_*(\mathcal{L}'_4)$, $c_{1,2}$ are the fiber coordinates on $\pi_*(\mathcal{L}'_6)$, $d$ is the fiber coordinate on $\pi_*(\mathcal{L}'_8)$, $\alpha_a$ are the fiber coordinates on $\pi_*(\mathcal{J}_{a,3})$ and $\beta_a$ are the fiber coordinates on $\pi_*(\mathcal{J}_{a,5})$.

\refstepcounter{example}\paragraph{Example \theexample:\label{SO12fivepointsplittingexample}} Here we wish to explore the splitting of the rank-2 vector bundle $\mathcal{F}\to \mathcal{C}_{0,5}$ which arises, as in \S\ref{veryevensingle}, when we have very-even partition(s) consisting of a single distinct even part. To that end, consider ${\color{red}[6^2][2^6]}[2^2,1^8][2^2,1^8][5,1^7]$ or ${\color{red}[6^2]}{\color{blue}[2^6]}[2^2,1^8][2^2,1^8][5,1^7]$. 

Before imposing any constraints, we have
\begin{equation}
\begin{split}
\mathcal{L}_2&=\KcD^{\otimes 2}\otimes\mathcal{O}(-E_1-E_2-E_3-E_4-E_5)\\
\mathcal{L}_4&=\KcD^{\otimes 4}\otimes\mathcal{O}(-E_1-2E_2-2E_3-2E_4-E_5)\\
\mathcal{L}_6&=\KcD^{\otimes 6}\otimes\mathcal{O}(-E_1-3E_2-4E_3-4E_4-2E_5)\\
\mathcal{L}_8&=\KcD^{\otimes 8}\otimes\mathcal{O}(-2E_1-4E_2-6E_3-6E_4-4E_5)\\
\mathcal{L}_{10}&=\KcD^{\otimes 10}\otimes\mathcal{O}(-2E_1-5E_2-8E_3-8E_4-6E_5)\\
\tilde{\mathcal{L}}&=\KcD^{\otimes 6}\otimes\mathcal{O}(-E_1-3E_2-5E_3-5E_4-4E_5)\\
\end{split}
\end{equation}
There's no constraint at $p_5$. Imposing the constraints at $p_{3,4}$ and twisting at the boundary replaces $\mathcal{L}_4$ by
\begin{equation}
\mathcal{L}'_4=\KcD^{\otimes 4}\otimes\mathcal{O}(-E_1-2E_2-3E_3-3E_4-E_5-\mathcal{C}_{34})
\end{equation}
$\pi_*(\mathcal{L}'_4)$ is a rank-3 bundles with\footnote{Recall that $\psi_a\coloneqq c_1(I_a)$.}
\begin{equation}
\begin{split}
c_1(\pi_*(\mathcal{L}'_4))&= 2\psi_1+\psi_5+D_{25}\\
c_2(\pi_*(\mathcal{L}'_4))&= 6P
\end{split}
\end{equation}

We now have to impose the constraints at $p_{1,2}$. The effect on $\mathcal{L}_8$ and $\mathcal{L}_{10}$ from the constraints at $p_2$ are easy to incorporate. After twisting at the boundary, they are replaced by
\begin{equation}
\begin{split}
\mathcal{L}'_8&=\KcD^{\otimes 8}\otimes \mathcal{O}(-2E_1-5E_2-6E_3-6E_4-4E_5-2\mathcal{C}_{23}-2\mathcal{C}_{24}-3\mathcal{C}_{34}-\mathcal{C}_{35}-\mathcal{C}_{45})\\
\mathcal{L}'_{10}&=\KcD^{\otimes 10}\otimes \mathcal{O}(-2E_1-6E_2-8E_3-8E_4-6E_5-3\mathcal{C}_{23}-3\mathcal{C}_{24}-\mathcal{C}_{25}\\
&\qquad\qquad\qquad\qquad-5\mathcal{C}_{34}-3\mathcal{C}_{35}-3\mathcal{C}_{45}-\mathcal{C}_{234}-\mathcal{C}_{345})
\end{split}
\end{equation}
The constraint at $p_1$ and the remaining constraint at $p_2$ are trickier. They involve a rank-2 vector bundle $\mathcal{F}$, which fits into a short exact sequence
\begin{equation}
0\to\mathcal{F}\to \mathcal{L}_6\oplus\tilde{\mathcal{L}}\to \mathcal{O}_{E_1}(-E_1)\oplus \mathcal{O}_{E_2}(-3E_2)\to 0
\end{equation}
For ${\color{red}[6^2][2^6]}\dots$, we expect a splitting of the form
\begin{equation}
\mathcal{F}= \mathcal{L}_6(-E_1-E_2)\oplus \tilde{\mathcal{L}}
\end{equation}
This requires twisting at the boundary
\begin{equation}
\begin{split}
\mathcal{F}'&=\KcD^{\oplus 6}\otimes\Bigl[\mathcal{O}(-2E_1-4E_2-4E_3-4E_4-2E_5-\mathcal{C}_{23}-\mathcal{C}_{24}-\mathcal{C}_{34})\\
&\qquad\qquad\oplus
\mathcal{O}(-E_1-3E_2-5E_3-5E_4-4E_5-\mathcal{C}_{23}-\mathcal{C}_{24}-3\mathcal{C}_{34}-2\mathcal{C}_{35}-2\mathcal{C}_{45}-\mathcal{C}_{345})\Bigr]\\
&\coloneqq \mathcal{F}'_1\oplus\tilde{\mathcal{L}}'
\end{split}
\end{equation}
The push-forward $\pi_*(\mathcal{F}'_1)$ is a rank-3 vector bundle $V_3$ with
\begin{equation}\label{F1pforward}
\begin{split}
c_1(V_3)&=3(\psi_1+\psi_5)\\
c_2(V_3)&=20P
\end{split}
\end{equation}
The push-forward\footnote{Here we use that
\begin{equation*}
\begin{split}
\pi_*\bigl(
\KcD^{\oplus 6}&\otimes
\mathcal{O}(-E_1-3E_2-5E_3-5E_4-4E_5-\mathcal{C}_{23}-\mathcal{C}_{24}-3\mathcal{C}_{34}-2\mathcal{C}_{35}-2\mathcal{C}_{45}-\mathcal{C}_{345})\bigr)\\
&=\pi_*\bigl(
\KcD^{\oplus 5}\otimes
\mathcal{O}(-E_1-3E_2-4E_3-4E_4-3E_5-\mathcal{C}_{23}-\mathcal{C}_{24}-2\mathcal{C}_{34}-\mathcal{C}_{35}-\mathcal{C}_{45})\bigr)\\
&= I_1
\end{split}
\end{equation*}
} $\pi_*(\tilde{\mathcal{L}}')=I_1$.

For ${\color{red}[6^2]}{\color{blue}[2^6]}\dots$, we couldn't decide between the splittings
\begin{subequations}
\begin{equation}\label{rbtwistposa}
\mathcal{F}= \mathcal{L}_6(-E_1)\oplus \tilde{\mathcal{L}}(-E_2)\coloneqq \mathcal{F}_1\oplus \mathcal{F}_2
\end{equation}
and
\begin{equation}\label{rbtwistposb}
\mathcal{F}= \mathcal{L}_6(-E_2)\oplus \tilde{\mathcal{L}}(-E_1)\coloneqq \mathcal{F}_1\oplus \mathcal{F}_2
\end{equation}
\end{subequations}
In either case, we should have $\pi_*(\mathcal{F}'_2)=0$ and $\pi_*(\mathcal{F}'_1)$ a rank-4 bundle. 

The choice \eqref{rbtwistposa} yields a rank-4 bundle with Chern classes
\begin{subequations}
\begin{equation}
\begin{split}
c_1(\pi_*(\mathcal{F}'_1)) &= 5D_{12}+7D_{15}+5D_{25}+8D_{34}=3\psi_1+\psi_2+4\psi_5+D_{12}\\
c_2(\pi_*(\mathcal{F}'_1))&=44P
\end{split}
\end{equation}

For \eqref{rbtwistposb}, $\pi_*(\mathcal{F}'_1)\coloneqq V_4$ with Chern classes
\begin{equation}
\begin{split}
c_1(V_4)&=4\psi_1+3\psi_5\\
c_2(V_4)&=29P
\end{split}
\end{equation}
\end{subequations}

When we actually explicitly solve the constraints, we see that \eqref{rbtwistposb} is the correct choice. The difference, in the bundle of Hitchin bases, between ${\color{red}[6^2]}{\color{red}[2^6]}$ and ${\color{red}[6^2]}{\color{blue}[2^6]}$ will come down to whether the sequence
\begin{equation}\label{doesitsplit}
0\to V_3\to V_4\xrightarrow{\;\beta\;}  I_1\to 0
\end{equation}
(where $V_3$ is given by \eqref{F1pforward}) splits.

Independent of these choices, $V_2\coloneqq \pi_*(\mathcal{L}'_8)$ is a rank-2 bundle whose Chern classes are
\begin{equation}
\begin{split}
c_1(V_2)&= 4\psi_1+D_{25}\\
c_2(V_2)&=6P
\end{split}
\end{equation}
and $\pi_*(\mathcal{L}'_{10})= I_1^{\otimes 2}$.

We can explicitly write the corresponding $k$-differentials that result from solving the constraints. The constraints at $p_{3,4}$ set the leading coefficients $\left.(\phi_4-\tfrac{1}{4}\phi_2^2)\right\vert_{E_3}=\left.(\phi_4-\tfrac{1}{4}\phi_2^2)\right\vert_{E_4}=0$, which yields
\begin{equation}\label{phi24choice}
\begin{split}
\phi_2&= y_1\bigl(a_1(y_2-y_3) +a_2(y_2-y_4)\bigr)\\
\phi_4&= \phi_4'+ \tfrac{1}{4}\phi_2^2\quad\text{where} \\
\phi'_4&=  y_2 (y_1-y_3)(y_1-y_4)(c_1y_1+c_2y_2) +c_3 y_1 y_3 y_4(y_1-y_2)\end{split}
\end{equation}
where $a_1,a_2$ are fiber coordinates on the log-cotangent bundle $\mathcal{T}^\vee$ and $c_1,c_2,c_3$ are fiber coordinates on $\pi_*(\mathcal{L}'_4)$.

Before imposing the constraints at $p_{1,2}$, we can write
\begin{equation}\label{phi6phitildechoice}
\begin{split}
\tilde{\phi}&= \tilde{d} y_1 y_2 (y_1-y_2)^2(y_1-y_3)(y_1-y_4)\\
\phi_6&=y_1 y_2 (y_1-y_3)(y_1-y_4)\bigl[(d_4y_1+d_5y_2+d_1y_3+d_2y_4)(y_1-y_2) +d_3 y_1y_2\bigr]
\end{split}
\end{equation}
The constraint at $p_1$ simply says that the leading coefficient $(\phi_6-2\tilde{\phi})\vert_{E_1}=0$ or
\begin{equation}\label{p1constraint6}
d_4-2\tilde{d}=0
\end{equation}
At $p_2$, one of the constraints says
\[
\left.\left(\phi_6\mp 2\tilde{\phi}- \tfrac{1}{2}\phi_2\phi'_4\right)\right\vert_{E_2}=0
\]
which yields
\begin{equation}\label{p2constraint6}
d_5\pm 2\tilde{d} +\tfrac{1}{2}(a_1+a_2)c_2=0
\end{equation}
Putting these together,
\begin{equation}\label{phi6phitildeimposed}
\begin{split}
\tilde{\phi}&= \tilde{d} y_1 y_2 (y_1-y_2)^2(y_1-y_3)(y_1-y_4)\\
\phi_6&=y_1 y_2 (y_1-y_3)(y_1-y_4)\Bigl[\bigl(2\tilde{d}y_1
-(\tfrac{1}{2}(a_1+a_2)c_2\mp2\tilde{d})y_2+d_1y_3+d_2y_4\bigr)(y_1-y_2)\\
&\hspace{3in} +d_3 y_1y_2\Bigr]
\end{split}
\end{equation}
For ${\color{red}[6^2]}{\color{red}[2^6]}$, we choose the $-$ sign above. The sequence \eqref{doesitsplit} splits and  $d_1,d_2,d_3$ are the fiber coordinates on the rank-3 bundle $V_3$ in \eqref{F1pforward} while  $\tilde{d}$ is the fiber coordinate on $I_1$.  For the ${\color{red}[6^2]}{\color{blue}[2^6]}$, we choose the $+$ sign,  \eqref{doesitsplit} doesn't split and  $d_1,d_2,d_3,\tilde{d}$ are fiber coordinates on $V_4$.

Solving the remaining constraints is straightforward. The next constraint at $p_2$ is
\[
\left.\left(\phi_8 \mp \phi_2\tilde{\phi} -\tfrac{1}{4}(\phi'_4)^2\right)\right\vert_{E_2}=0
\]
which can be solved to give
\begin{equation}\label{phi8sol}
\begin{split}
\phi_8= y_2(y_1-y_2)(y_1-y_3)(y_1-y_4)\Bigl[&y_1^2(y_1-y_2)(e_1y_3+e_2 y_4)\\
&+y_2\bigl(\tfrac{1}{4}y_2(y_1-y_3)(y_1-y_4)c_2^2 \mp(a_1+a_2)\tilde{d} y_1^2(y_1-y_2)\bigr)\Bigr]
\end{split}
\end{equation}
where $e_1,e_2$ are the fiber coordinates on the rank-2 bundle $V_2\coloneqq \pi_*(\mathcal{L}'_8)$ whose Chern classes are
\begin{equation}
\begin{split}
c_1(V_2)&= 4\psi_1+D_{25}\\
c_2(V_2)&=6P
\end{split}
\end{equation}
For ${\color{red}[6^2]}{\color{blue}[2^6]}$, ``$\tilde{d}$ '' in \eqref{phi8sol} and in \eqref{phi10sol} below should be interpreted as the image of $\beta$ in \eqref{doesitsplit}.

The final constraint at $p_2$ imposes
\begin{equation}\label{phi10sol}
\phi_{10}= y_1 y_2^2(y_1-y_2)^2(y_1-y_3)^2(y_1-y_4)^2[f y_1\pm \tilde{d}c_2 y_2]
\end{equation}
where $f$ is the fiber coordinate on $\pi_*(\mathcal{L}'_{10})= I_1^{\otimes 2}$.

\section*{Acknowledgements}
We would like to thank Sean Keel, Sam Grushevsky and especially Andres Fernandez Herrero and Victor Alekseev for many helpful discussions. The work of JD and CP was supported in part by the National Science Foundation under Grant No.~PHY--2210562. CP was also partially supported by the Robert N.~Little Fellowship and by the Heising-Simons Foundation under the ``Observational Signatures
of Quantum Gravity” collaboration grants 2021-2818 and 2024-5305. The research of RD was partially supported by NSF grant DMS--2244978, FRG: Collaborative Research: New birational invariants; by NSF grant DMS–2401422; and by Simons Foundation Collaboration grant \#390287 ``Homological Mirror Symmetry.'' Some of this research was conducted by JD while at the Aspen Center for Physics, which is supported by National Science Foundation grant PHY-2210452. JD is grateful for the hospitality of  the Simons Center for Geometry \& Physics, where some of this work was performed.

\section*{Appendices}
\appendix

\section{Splitting the Sequence for Line Bundles on \texorpdfstring{$\mathcal{C}_{0,4}$}{C₀,₄}}\label{foursplitting}

Here we address the question of whether the sequence \eqref{hopeitsplits} splits in the particular case of $\mathcal{C}_{0,4}$. Given the short exact sequence of line bundles on $\mathcal{C}_{0,4}$,
\[
0\to \mathcal{L}\to  \mathcal{L}(E_{a})\to \mathcal{L}(E_{a})\otimes\mathcal{O}_{E_{p_a}}\to 0
\]
and assuming $R^1\pi_*(\mathcal{L})=0$ (i.e.~that the OK condition holds for $\mathcal{L}$), then we induce the short exact sequence
\begin{equation}\label{splits}
0\to \pi_*(\mathcal{L})\to  \pi_*(\mathcal{L}(E_{a}))\to \pi_*(\mathcal{L}(E_{a})\otimes\mathcal{O}_{E_{a}})\to 0
\end{equation}
We want to claim that \eqref{splits} splits. Without loss of generality, we can always use
\[
\mathcal{O}(-\,\mathcal{C}_{ab})= \KcD^{-1}\otimes\mathcal{O}(E_{a}+E_{b})
\]
to write $\mathcal{L}$ in a form without any twisting by the boundary, at the cost of a shift in the effective values of $k$ and $\tau^{(a)}$.

Assume we have already used the algorithm of \S\ref{directimage} to compute $\pi_*(\mathcal{L})$. What is $\pi_*(\mathcal{L}(E_{a}))$? That is, what is the effect of substituting $\tau^{(a)}\to \tau^{(a)}-1$ in the algorithm? It's clear that, for $i>\tau^{(a)}-1$, the $f_i$ of \eqref{fidef} are unchanged, whereas $f_i \to f_i+1$ for $i\leq \tau^{(a)}-1$. Plugging into \eqref{midef}, we see $m_{\tau^{(a)}-1}\to m_{\tau^{(a)}-1}+1$, with all the other $m_i$ unaffected. Thus
\begin{equation}
\pi_*(\mathcal{L}(E_{a}))= \pi_*(\mathcal{L})\oplus \mathcal{O}(\tau^{(a)}-1)
\end{equation}
On the other hand, we have from \eqref{pointrestrictions},\eqref{pointlineimage} that 
\[
\pi_*(\mathcal{L}(E_{a})\otimes \mathcal{O}_{E_{a}}) = \mathcal{O}(\tau^{(a)}-1)
\]
Thus the sequence \eqref{splits} splits. The proof relied on very special features of $\mathcal{C}_{0,4}$. Nevertheless, we will see in \S\ref{fivepuncturedexamples} that we also get a splitting for $\mathcal{C}_{0,5}$.

\section{Dimensions of Adjoint Orbits in Type-D}
\label{dimensionappendix}

\subsection{Nilpotent orbits}

Here, we recall some formul\ae for calculating the dimensions of nilpotent orbits in type-D (from Th 6.1.3 in \cite{CollingwoodMcGovern}) . 

Let $[P]$ denote the Hitchin partition. Let $[S]$ denote its transpose $[S]=[P]^T$ and let $m_k$ be the multiplicity of the part $k$ in the partition $[P]$ (ie $m_k$ is the number of parts $P_i$ that obey $P_i=k$). 

The complex dimension of the nilpotent orbit with partition label $[P]$ is now given by 

\begin{equation}
\dim(\mathcal{O}_{[P]}) = \dim(\mathfrak{so}_{2N}) - \frac{1}{2} \bigl( \sum_i S_i^2 - \sum_{i \in \text{odd}} m_i \bigr) 
\end{equation}
As an example, let us compute the dimension of $[3^2,1^2]$ in $D_4$. We have $[S] = [4,2^2]$. So, the dimension is 
\begin{equation}\label{example1}
\begin{split}
    \dim(\mathcal{O}_{[3^2,1^2]}) &= 28 - \frac{1}{2} ( 4^2 + 2^2 + 2^2 - 2 - 2 ) \\
    &= 28 - (10) \\ 
    &= 18  
\end{split}
\end{equation}

\subsection{Semi-simple orbits}

For a semi-simple orbit, one can write a formula for the dimension in terms of the sheet Levi. Recall that any semi-simple orbit $\mathcal{O}_{ss}$ will occur as the ``dense orbit'' in a unique sheet. Let this sheet have sheet Levi $\mathfrak{l}$. By definition, this means that the centralizer (in $\mathfrak{g}$) of the semi-simple orbit $\mathcal{O}_{ss}$ is the sub-algebra $\mathfrak{l}$. The semi-simple orbit can be thought of as the quotient $\mathfrak{g} /  \mathfrak{l} $.  The dimension of the semi-simple orbit is given by 

\begin{equation}
    \dim (\mathcal{O}_{ss}) = \dim(\mathfrak{so}_{2N}) - \dim(\mathfrak{l}),
\end{equation}

where $\text{dim}(\mathfrak{l})$ is the dimension of the Levi subalgebra. The convention is to label Levi sub-algebras by the Cartan type of their semi-simple parts. The full dimension of the Levi is given by the sum of dimensions of its center $Z_{\mathfrak{l}}$ and that of its semi-simple part $\mathfrak{l}_{ss}$: 

\begin{equation}
    \text{dim}(\mathfrak{l}) = \text{dim} (Z_{\mathfrak{l}} ) + \text{dim} (\mathfrak{l}_{ss} ) 
\end{equation}

and the dimension of the center of the Levi is given by 

\begin{equation}
    \text{dim} (Z_{\mathfrak{l}})  = \text{rank} (\mathfrak{g}) - \text{rank} (\mathfrak{l}_{ss} )
\end{equation}

As an example, the semi-simple orbit in the $A_2$ sheet in $D_4$ has the following dimension:

\begin{equation}\label{example2}
\begin{split}
    \dim (\mathcal{O}_{ss}^{A_2}) &=28 - ( 2 + 8  )\\ 
     &= 18 
\end{split}
\end{equation}

As a second example, consider the semi-simple orbit in the $D_2 + A_1$ sheet in $D_4$. 

\begin{equation}\label{example3}
\begin{split}
    \text{dim} (\mathcal{O}_{ss}^{D_2 + A_1}) &= 28 - ( 1 + 3 \times 3  ) \\
     &= 18 
\end{split}
\end{equation}

Note : The orbit $[3^2,1^2]$ occurs at the boundary of the $A_2$ sheet \textit{and} the $D_2+A_1$ sheet. The matching of the dimension counts in Eqs \eqref{example1}, \eqref{example2} and \eqref{example3} is a necessary condition for this fact to be true. 

\section{Mass Deformations}
\label{massappendix}
Recall that in the Hitchin system corresponding to the conformal theory, the residues at every puncture are nilpotent. We are interested in describing what happens to the Hitchin system when we turn on mass deformations. Physically, these are $\mathcal{N}=2$ preserving mass deformations in the 4d Class S superconformal theory or $\mathcal{N}=4$ preserving mass deformations in the 3d superconformal theory. The purpose of this appendix is not comprehensive but diagnostic; it illustrates how mass deformations retain information about the choice of Nahm orbit and associated sheet that may be hard to see in the conformal limit spectral curve.

Let us focus on a single puncture for the present discussion. Let this nilpotent residue live in a nilpotent orbit $O_H$. Let the dual Nahm orbit be $O_N$. We have $d_S(O_N)  = O_H$, where $d_S$ is the Spaltenstein duality map.  ``Turning on a mass deformation''  corresponds to deforming this residue to a non-nilpotent orbit that lives in a sheet $\mathcal{S}_{O}$ in the Lie algebra for which the original nilpotent orbit  $O_H$ occurs as the boundary orbit. The procedure for identifying this sheet is explained in detail (for any $\mathfrak{g}$) in \cite{Balasubramanian:2018pbp}. As explained in \cite{Balasubramanian:2018pbp}, sheets are labeled by a pair $(\mathfrak{l},O_{\mathfrak{l}})$ where $\mathfrak{l}$ is the Levi subalgebra that centralizes the generic orbit  in the sheet and $O_{\mathfrak{l}}$ is a rigid nilpotent orbit in $\mathfrak{l}$.  It can happen that a given nilpotent orbit occurs at the boundary of more than one sheet. In such cases, we first restrict to a class of \textit{special} sheets (Def.~A.3.12 in \cite{Balasubramanian:2018pbp}). We then identify the particular special sheet corresponding to the mass deformation using the condition that
\begin{equation}
\begin{split}
    \mathfrak{l}_{\textit{sheet}}&= \mathfrak{l}^\vee_{BC} (O_N)\\
   (O_{\mathfrak{l}})_{\textit{sheet}}&=d_S ({O^d_{\mathfrak{l}^\vee_{BC}} })
\end{split}
\label{levi-matching-condition}
\end{equation}
where  $\mathfrak{l}^\vee_{BC} (O_N)  $ is the Bala-Carter Levi of $O_N$ and $O^d_{\mathfrak{l}^\vee_{BC}}$ is the distinguished orbit in the Levi $\mathfrak{l}^\vee_{BC}$ which is the restriction of the Nahm orbit $O_N$ to the Levi. 

When $O_{\mathfrak{l}}$ is the zero orbit, the sheet is known as a Dixmier sheet and we abbreviate $(\mathfrak{l},0)$ by $(\mathfrak{l})$. In type-A, all sheets are Dixmier sheets and, as it turns out, in the type-D examples we discuss below the sheets onto which we can mass-deform are also Dixmier.

We will describe the mass deformed global Hitchin system in a follow-up paper.  But, we include here a brief discussion with the goal of introducing the rich set of phenomena that one encounters. 

\subsection{Mass deformed residues in type-A}

First, we describe how the residues of a tame Hitchin system of type $A$ are deformed when we turn on mass deformations. 

Here, we provide canonical representative(s) for such semi-simple orbits in the specified sheet.  Let the partition label for the Hitchin nilpotent $\mathcal{O}$ be $[P]$ with $\sum P_i = N$.  Let the dual Nahm partition be $[R] = [P]^T$. Note that $P_1$ counts the number of parts in $[R]$ . 

We choose to work with the canonical representatives for nilpotent orbits used in \cite{CollingwoodMcGovern}. To this end, let us define $e_{P_i},s_{P_i}$ to be the following matrices : 

\begin{equation}
\begin{split}
   e_{P_i}& \coloneqq \begin{pmatrix}
       0  & 1  & 0 & \ldots &  \ldots \\
       0  & 0 &  1 & 0 & \ldots  \\
      \vdots & \vdots   &  \vdots  & \vdots & \vdots  \\ 
      0 & 0  & \ldots &  \ldots &  1 \\
       0 & 0  & \ldots & \ldots &  0 \\
      \end{pmatrix} _{P_i \times P_i}\\
    s_{P_i}&\coloneqq \begin{pmatrix}
       m_1  & 1  & 0 & \ldots &  \ldots \\
       0  & m_2 &  1 & 0 & \ldots  \\
      \vdots & \vdots   &  \vdots  & \vdots & \vdots  \\ 
      0 & 0  & \ldots &  \ldots &  1 \\
       0 & 0  & \ldots & \ldots &  m_{P_i} \\
      \end{pmatrix} _{P_i \times P_i}  = e_{P_i} + \operatorname{diag}(m_1,m_2,\ldots,m_{P_i})
\end{split}
\end{equation}

The canonical representative $e$ for the nilpotent orbit $\mathcal{O}$ has following simple form :
\begin{equation}
     \mathcal{O} \ni  e  = \begin{pmatrix}
       e_{P_1}  & 0  & 0 & \ldots &  \ldots \\
       0  & e_{P_2} &  0 & 0 & \ldots  \\
      \vdots & \vdots   &  \vdots  & \vdots & \vdots  \\ 
      0 & 0  & \ldots &  \ldots &  e_{P_{R_1}}\\
       \end{pmatrix}_{N \times N}
\end{equation}
where $R_1$ counts the total number of parts in $[P]$. 

We choose the following family of representatives for the semi-simple orbits in the sheet $\mathcal{S}_{\mathcal{O}}$  :
\begin{equation}
     \mathcal{S}_{\mathcal{O}} \ni  s  = \begin{pmatrix}
       s_{P_1}  & 0  & 0 & \ldots &  \ldots \\
       0  & s_{P_2} &  0 & 0 & \ldots  \\
      \vdots & \vdots   &  \vdots  & \vdots & \vdots  \\ 
      0 & 0  & \ldots &  \ldots &  s_{P_{R_1}}\\
        \end{pmatrix}_{N \times N}
\end{equation}
subject to the condition that $Tr(s) = 0$ which translates to the following linear condition on the diagonal entries,
\begin{equation}
  \sum_{i=1}^{P_1} R_i m_i =0.
\end{equation}
We also define the mass matrix $M$ to be the deformation 
\begin{equation}
    M = s - e 
\end{equation}
Since we have $P_1 \geq P_2 \geq P_3 , \ldots \geq P_{R_1} $ and there is one linear relation among the diagonal entries, the total number of independent complex numbers in $M$ is $P_1 - 1$. 

Note that any re-ordering of the entries along the diagonal \textit{within} a Jordan block leads to an equally good parameterization of the relevant class of semi-simple orbits in the sheet $\mathcal{S}_{\mathcal{O}}$. The finite group generated by such re-orderings generates a particular subgroup $H \subset W_{G}$ where $W_G$ is the Weyl group associated to the Lie algebra $\mathfrak{g}$. The Weyl group of the Flavor symmetry $W_F$, when non-trivial, is a subgroup of the discrete group $H$. To summarize, we have the following inclusions :
 \begin{equation}
     W_F \subset H \subset W_G
 \end{equation}

\subsection{Mass deformed residues in type-D}
We choose a parametrization of the elements of the Lie algebra $\mathfrak{so}(2N)$ by $2N \times 2N$ matrices of the form 
\begin{equation}
   \begin{pmatrix}
      A   &  B \\
      C   &  -A^T
    \end{pmatrix}
\end{equation}
where $A$ is an arbitrary $N\times N$ matrix and $B,C$ are skew-symmetric $N\times N$ matrices. 

For nilpotent orbits in $\mathfrak{so}(2N)$, we again work with the canonical representatives from \cite{CollingwoodMcGovern}. For the algorithm to write down these representatives for an arbitrary nilpotent orbits, we refer the reader to \cite{CollingwoodMcGovern,Chacaltana:2011ze}. Here, we will only be concerned with a particular class of nilpotent orbits whose partition label is such that every part occurs with even multiplicity. Let us denote such orbits as \textit{square} nilpotent orbits. 

Square nilpotent orbits have a property that allows us to write their mass deformed versions in way that is quite similar to the type-A situation recalled above. 

By definition, the partition label for a square nilpotent orbit in type-D is of the form
\begin{equation}
    [P] = [a_1 ^{2n_1}, a_2^{2n_2},\ldots, a_k^{2n_k}]
    \label{square-nilpotent-partition}
\end{equation}
where $a_i,n_i$ are positive integers and $a_1 > a_2 > a_3> \ldots$. Obviously, all square nilpotents are special.  Moreover, if $a_i$ and $a_{i+1}$ are both odd, then there is a marked pair in $[P]$ and a nontrivial dual special piece on the Nahm side.

For convenience, let us also define the following partition of $N$ that is constructed out of the partition $[P]$ by halving the multiplicities,
\begin{equation}
    [Q] \coloneqq [a_1 ^{n_1}, a_2^{n_2},\ldots, a_k^{n_k}].
\end{equation}
where $a_i,n_i$ are exactly the integers occurring in \eqref{square-nilpotent-partition}. Clearly we have that $\sum_i Q_i = N$. 

From the algorithm for canonical representatives outlined in \cite{CollingwoodMcGovern,Chacaltana:2011ze}, we learn that for any nilpotent orbit $\mathcal{O}$ in type-D with partition label as in \eqref{square-nilpotent-partition}, it is possible to pick a nilpotent element $e \in \mathcal{O}$ which is of the following form :
\begin{equation}
    e = \begin{pmatrix}
       A_e  & 0 \\
       0  & -A_e^T
    \end{pmatrix}
\end{equation}
with $A$ being the following $N\times N$ matrix
  
\begin{equation}
    A_e  = \begin{pmatrix}
       e_{Q_1}  & 0  & 0 & \ldots &  \ldots \\
       0  & e_{Q_2} &  0 & 0 & \ldots  \\
      \vdots & \vdots   &  \vdots  & \vdots & \vdots  \\ 
      0 & 0  & \ldots &  \ldots &  e_{Q_k}\\
       \end{pmatrix} _{N \times N} 
\end{equation}
Now, as in the type-A case, the mass deformed residue has the following form 
\begin{equation}
    A_s = \begin{pmatrix}
       s_{Q_1}  & 0  & 0 & \ldots &  \ldots \\
       0  & s_{Q_2} &  0 & 0 & \ldots  \\
      \vdots & \vdots   &  \vdots  & \vdots & \vdots  \\ 
      0 & 0  & \ldots &  \ldots &  s_{Q_k}\\
       \end{pmatrix}_{N \times N} 
\end{equation}
but with some important differences.


As before, we define the mass matrix $M$ to be deformation of the full $so(2N)$ element
\begin{equation}
    M  = \begin{pmatrix}
       A_s  & 0 \\
       0  & -A_s^T
   \end{pmatrix}  - \begin{pmatrix}
       A_e  & 0 \\
       0  & -A_e^T
    \end{pmatrix}
\end{equation}
The entries in this diagonal matrix are determined as follows. Let $[R]$ be a partition in the (``Nahm'') dual special piece dual to $[P]$. Since $[P]$ is square, it is easy to see\footnote{The transpose of $[P]$ has only even parts. D-collapse produces pairs of adjacent odd parts. The odd parts are equal if and only if the erstwhile even parts differed by 2. A Kraft-Procesi small degeneration \cite{kraft1989special} removes odd parts in pairs. The upshot is that we never obtain an odd part with multiplicity $>2$. Example: consider $[P]=[5^4,3^2]$. $[P]^t= [6^3,4^2]$. D-collapse yields the special orbit $[R]_s=[6^2,5^2,3,1]$. A small degeneration then leads to the non-special orbit $[R]_{ns}=[6^2,5,4^2,1]$. In either case, the odd parts of $[R]$ occur with multiplicity at most 2.} that the odd parts of $R$ occur with multiplicity 1 or 2. For each part of $[R]$ which occurs with even multiplicity, $[R]= [\dots, (b_i)^{2l_i},\dots]$, the diagonal matrix $M$ will contain entries $m_{i,1},\dots,m_{i,l_i},-m_{i,1},\dots,-m_{i,l_i}$, each with multiplicity $b_i$. For each odd entry which appears with multiplicity 1, $[R]=[\dots,(2b+1),\dots]$, the diagonal matrix $M$ will contain $(2b+1)$ zeroes. We now arrange those $2N$ entries in $M$  as
\[
M= \operatorname{diag}(M_{Q_1},M_{Q_2},\dots, M_{Q_k}, -M_{Q_1},-M_{Q_2},\dots, -M_{Q_k})
\]
where the entries in each $M_{Q_i}$ are all distinct. For instance, if $[P]=[5^2,1^2]$, then $[R]$ is either $[3^2,2^2,1^2]$ (the special Nahm orbit), in which case
\[
M= \operatorname{diag}(m_1,m_2,m_3,-m_1,-m_2,m_1,-m_1,-m_2,-m_3,m_1,m_2,-m_1)
\]
or $[R]=[3,2^4,1]$ (the Kraft-Procesi small degeneration \cite{kraft1989special} of $[3^2,2^2,1^2]$ in the same special piece), in which case 
\[
M= \operatorname{diag}(m_1,m_2,0,-m_1,-m_2,0,-m_1,-m_2,0,m_1,m_2,0)
\]

When the special piece on the Nahm side is nontrivial, this choice of which Nahm orbit $[R]$ goes into defining the mass matrix $M$ corresponds to inequivalent mass-deformations along \textit{different} special sheets. \footnote{What is relevant for the story of mass deformations is a certain refinement of the usual notion of sheets, see \cite{Balasubramanian:2018pbp} for a more detailed description. } The special sheet that is dual to a particular Nahm orbit can be identified using the conditions in \eqref{levi-matching-condition}. 

Let the sheet along which we are deforming have sheet label $(\mathfrak{l},\mathcal{O}_{\mathfrak{l}})$ and let the rank of the sheet Levi $\mathfrak{l}$ be $t$ (for a quick introduction to the theory of sheets, see  \cite{Balasubramanian:2018pbp}). Then, the number of linearly independent complex numbers in $M$ will equal $N - t$ which is also the dimension of the center $Z(\mathfrak{l})$ of the Levi $\mathfrak{l}$.  Furthermore, the multiplicities of the eigenvalues are determined by the multiplicities in the dual Nahm partition. We leave a complete discussion of these aspects to a future paper and instead include below a specific example from $\mathfrak{so}(8)$ to illustrate how the \textit{local} phenomenon of multiple sheets meeting at the same nilpotent orbit manifests itself in the geometry of the spectral curves describing the \textit{global} Hitchin system. 
 
\subsection{Mass deformed spectral curve in type-D}\label{MassDeformSpectral}

As noted in  example \ref{so12oddtypeexample} in \S\ref{3pointexamples}, the spectral curve in the conformal limit generally will not distinguish whether we do or do not impose the odd-type constraint. The difference may only be revealed when we mass-deform. Deforming the residue onto distinct sheets whose boundaries contain the given Hitchin nilpotent yields distinct mass-deformed spectral curves. Which sheet we can deform onto is dictated by whether or not we impose the odd-type constraint. As an example, consider the three punctured sphere with punctures $[7,1]$ $[3^2,1^2]$ $[5,1^3]$ at $(x,y)=(1,0),(0,1),(1,1)$ respectively. The $[3^2,1^2]$ has an odd type constraint. Depending on whether we impose the constraint, we get  --- in the conformal limit --- one of the following two spectral curves:

\begin{subequations}
\begin{align}
    \Sigma &=\Bigl\{
0=w^8+w^4x^2y(x-y)b+w^2 y x^2(x-y)^2(x d_1-y d_2)
\Bigr\}\label{3221curve} \\
\Sigma &=\Bigl\{
0=w^8+w^4x^2y(x-y)b+w^2 y x^2(x-y)^2(x d_1-y \alpha^2)
\Bigr\}\label{3311curve}
\end{align}
\end{subequations}
The first corresponds to the dual Nahm orbit $[3,2^2,1]_N$, the second to the dual Nahm orbit $[3^2,1^2]_N$. The corresponding sheets onto which we can deform $[3^2,1^2]_H$ are labeled by the sheet Levis  $(D_2+A_1)$ and $(A_2)$. \footnote{These match the Bala-Carter Levis of the two Nahm orbits $[3,2^2,1]_N$ and $[3^2,1^2]_N$ respectively. } Naively, this is a boring double cover $d_2=\alpha^2$. In fact, after properly resolving the fibers over $d_2=0$ ($\alpha=0$), it is merely birational to a double cover.  The difference becomes manifest when we mass-deform.

For the $(D_2+A_1)$ sheet, the nilpotent residue is deformed by adding the mass matrix
\begin{equation}
    M = \operatorname{diag}(m,0,-m,0,-m,0,m,0)
\end{equation}
where we interpret $m$ as the Cartan of the $SU(2)$ flavour symmetry of the Nahm orbit $[3,2^2,1]_N$.

The mass deformation changes the local behaviour near the puncture. Introducing a local coordinate $t$ and writing
\[
\det(X+M+t(R+O(t))-\lambda\Bid)=\lambda^8+\lambda^6\phi_2(t)+\lambda^4\phi_4(t)+\lambda^2\phi_6(t)+\tilde{\phi}(t)^2
\]
we have
\begin{equation}
\begin{split}
\phi_2(t)&= -2m^2+ c_2t+\dots\\
\phi_4(t)&= m^4 -  m^2 c_2t + c_4t^2 +\dots\\
\phi_6(t)&=  c_6t^2+\dots\\
\tilde{\phi}(t)&= \tilde{c}t^2+\dots
\end{split}
\end{equation}
This deforms \eqref{3221curve} to
\begin{equation}
    \Sigma= \bigl\{
0=w^8+2w^6y(x-y)m^2+w^4y(x-y)[x^2b+y(x-y)m^4]
+w^2  x^2y(x-y)^2(x d_1-y d_2)
\bigr\}
\end{equation}
This mass-deformed spectral curve is invariant under the Weyl-group of the $SU(2)$ flavour symmetry (which takes $m\to-m$).

On the other hand for $[3^2,1^2]_N$ (\emph{i.e.}~deforming onto the $A_2$ sheet) 
\begin{equation} \label{eq:massdefA2}
    M = \operatorname{diag}(m_1,m_2,-m_1,m_1,-m_1,-m_2,m_1,-m_1)
\end{equation}
where we interpret $m_1,m_2$ as the generators of the $U(1)\times U(1)$ flavour symmetry of the Nahm orbit $[3^2,1^2]_N$. The local behaviour
\begin{equation}
\begin{split}
\phi_2(t)&= -(3m_1^2+m_2^2) +c_2 t+\dots\\
\phi_4(t)&= 3m_1^2(m_1^2+m_2^2) -2(m_1^2 c_2+ m_2\tilde\alpha)t +c_4 t^2+\dots\\
\phi_6(t)&= -m_1^4(m_1^2+3m_2^2) +m_1^2(m_1^2c_2+4 m_2 \tilde\alpha) t-\bigl[(\tilde\alpha+2m_1^2m_2)^2+m_1^4(c_2-8 m_1^2)\bigr]t^2+\dots\\
\tilde{\phi}(t)&= m_1^3m_2-m_1\tilde\alpha t +\tilde{c}t^2+\dots
\end{split}
\end{equation}
deforms \eqref{3311curve}  to
\begin{equation}
\begin{split}
\Sigma&=\Bigl\{
0=w^8
 + w^6 y(x-y)(3m_1^2+m_2^2) 
 +w^4y(x-y)[x^2b+2x y m_2\alpha  + y(x-y)3m_1^2(m_1^2+m_2^2)]\\
&\qquad\qquad+ w^2 y(x-y)^2\bigl[x^3 d_1 -x^2 y \alpha^2 + 4x y^2 m_1^2 m_2\alpha +y^2 (x-y)m_1^4(m_1^2+3m_2^2)\bigr]\\
&\qquad\qquad+ \bigl(y(x-y)^2m_1(y m_1^2 m_2  - x \alpha )\bigr)^2
\Bigr\}
\end{split}
\end{equation}
Note that this deformation only made sense for \eqref{3311curve}, because it has explicit dependence on $\alpha$, rather than $\alpha^2$. Also note that it is invariant under $\mathbb{Z}_2\times \mathbb{Z}_2$, where the first $\mathbb{Z}_2: (m_1,m_2,\alpha)\mapsto (-m_1,m_2,\alpha)$ (while also flipping the sign of the Pfaffian) and the second $\mathbb{Z}_2: (m_1,m_2,\alpha)\mapsto (-m_1,-m_2,-\alpha)$ (preserving the sign of the Pfaffian).

\section{Notes on \texorpdfstring{$\overline{\mathcal{M}}_{0,n}$}{ℳ̅₀,ₙ}}\label{notes_on_}

As noted in \S\ref{sec:C0n}, $\overline{\mathcal{M}}_{0,n}$ is a smooth projective variety of dimension $n-3$, birational to $(\mathbb{CP}^1)^{n-3}$, via a succession of smooth blowups. As a consequence,
\begin{itemize}%
\item $H^{\text{odd}}(\overline{\mathcal{M}}_{0,n})=0$
\item $H^{2i}(\overline{\mathcal{M}}_{0,n})=A^i(\overline{\mathcal{M}}_{0,n})$
\end{itemize}
In particular, $Pic(\overline{\mathcal{M}}_{0,n})=H^2(\overline{\mathcal{M}}_{0,n})=A^1(\overline{\mathcal{M}}_{0,n})$ is freely-generated of dimension

\begin{displaymath}
\dim(A^1(\overline{\mathcal{M}}_{0,n}))= 2^{n-1} -n(n-1)/2 -1
\end{displaymath}
It is generated by the boundary divisors $D_S$, subject to the linear relations

\begin{equation}
\begin{gathered}
D_{S} = D_{S^\vee}\\
\sum_{\mathclap{\begin{smallmatrix}a,b\in S\\c,d\in S^\vee\end{smallmatrix}}} D_S =
\sum_{\mathclap{\begin{smallmatrix}a,c\in S\\b,d\in S^\vee\end{smallmatrix}}} D_S =
\sum_{\mathclap{\begin{smallmatrix}a,d\in S\\b,c\in S^\vee\end{smallmatrix}}} D_S,\qquad \forall\; a,b,c,d\; \text{distinct}
\end{gathered}
\label{DivisorRelations}\end{equation}
where $S\subset\{1,2,\dots,n\}$ and $|S|,|S^\vee| \geq 2$.

The trick to solving the relations \eqref{DivisorRelations} is to use the isomorphism\goodbreak\noindent $\overline{\mathcal{M}}_{0,n+1}\simeq \mathcal{C}_{0,n}\xrightarrow{\pi} \overline{\mathcal{M}}_{0,n}$.

We'll denote the point divisors (the images of the $n$ sections $\sigma_a$) by

\begin{displaymath}
E_a=D_{a (n+1)},
\end{displaymath}
the class of the fiber by $F$, and the class of the vertical canonical bundle by

\begin{displaymath}
K=c_1(\Kc)
\end{displaymath}
Let's work out the first few examples.

\subsubsection*{\texorpdfstring{$\mathcal{C}_{0,4} =\overline{\mathcal{M}}_{0,5}$}{C₀,₄=ℳ̅₀,₅}}

We have

\begin{displaymath}
\begin{split}
D_{a5}&= E_a,\qquad a=1,\dots,4\\
D_{ab}&= K+D_{a5}+D_{b5}
\end{split}
\end{displaymath}
$H^2(\overline{\mathcal{M}}_{0,5})$ is freely-generated by the $E_a$ and $K$, with the ring relations

\begin{displaymath}
\begin{split}
K \cap D_{a5}&=P\\
K\cap K &= -3P\\
D_{a5}\cap D_{b5}&= -\delta_{ab}P
\end{split}
\end{displaymath}
where $P$ is the class of a point. The class of the fiber

\begin{displaymath}
\begin{split}
F&= D_{12}+D_{34}\\
&= 2K +\sum_{a=1}^4 E_a
\end{split}
\end{displaymath}
satisfies

\begin{displaymath}
\begin{split}
F\cap E_a&= P,\quad a=1,\dots,4\\
F\cap K&=-2P
\end{split}
\end{displaymath}
The point line bundles $I_a=\sigma_a^*(\Kc),\; a=1,\dots,5$  have first Chern classes
\[
\psi_a= c_1(I_a)=\sum_{\mathclap{\begin{aligned}a&\in S\\b,c&\in S^\vee\end{aligned}}} D_S
\]
which, in this basis, are
\begin{displaymath}
\begin{split}
\psi_a&= E_a +2K +\sum_{b=1}^4 E_b,\qquad a=1,\dots,4\\
\psi_5&= K + \sum_{b=1}^4 E_b
\end{split}
\end{displaymath}
and satisfy

\begin{displaymath}
\psi_a\cap\psi_b =(2 -\delta_{ab})P
\end{displaymath}

\subsubsection*{\texorpdfstring{$\mathcal{C}_{0,5}=\overline{\mathcal{M}}_{0,6}$}{C₀,₅=ℳ̅₀,₆}}

Now we have that $H^2(\mathcal{C}_{0,5})$ is freely-generated by

\begin{displaymath}
\begin{split}
E_a&= D_{a6}\quad (\text{a dP}_4)\\
\mathcal{C}_{ab}&= D_{ab 6}\quad (\text{a}\; \mathbb{CP}^1\times\mathbb{CP}^1)\\
K&
\end{split}
\end{displaymath}
where the remaining boundary divisors are (for $a,b,c,d,e$ distinct)

\begin{equation}\label{remaining}
\begin{split}
D_{cde}&= D_{ab6} = \mathcal{C}_{ab}\\
D_{ab}&= K - (\mathcal{C}_{cd}+\mathcal{C}_{ce}+\mathcal{C}_{de})+E_a+E_b
\end{split}
\end{equation}
From the obvious intersections of boundary divisors, we have the relations

\begin{equation}
\begin{split}
E_a\cap E_b&=0\quad\text{unless}\; a=b\\
E_a\cap \mathcal{C}_{b c}&=0\quad\text{unless}\; a=b\; \text{or}\; a=c\\
\mathcal{C}_{a b}\cap \mathcal{C}_{c d}&=0\quad\text{unless}\; \{a,b\}=\{c,d\}
\end{split}
\label{rels1}\end{equation}
Similarly, we get

\begin{equation}
\begin{split}
0&=E_a\cap(K+E_a)\\
0&=\mathcal{C}_{a b}\cap(K+E_a)\\
0&= K\cap K + K\cap(E_a+E_b+E_c -2 \mathcal{C}_{d e}) +\mathcal{C}_{d e}\cap \mathcal{C}_{d e}
\end{split}
\label{rels2}\end{equation}
(for $a,b,c,d,e$ distinct) from demanding $D_{a6}\cap D_{a b}= D_{a b 6}\cap D_{a c}= D_{a b}\cap D_{a c}=0$. Using the relations \eqref{rels1},\eqref{rels2} and

\begin{displaymath}
\begin{split}
P&= D_{a6}\cap D_{a b 6}\cap D_{c d}\\
P&= D_{a b}\cap D_{a b 6}\cap D_{c d}\\
P&= D_{a6}\cap D_{b c}\cap D_{d e}
\end{split}
\end{displaymath}
we get that the only nonvanishing triple intersections are

\begin{equation}
\begin{split}
K\cap K\cap K&=-4P\\
K\cap K\cap E_a&=P\\
K\cap K\cap\mathcal{C}_{a b}&=0\\
K\cap E_a\cap E_a&=-P\\
K\cap E_a\cap \mathcal{C}_{a b}&=0\\
K\cap\mathcal{C}_{a b}\cap\mathcal{C}_{a b}&=P\\
E_a\cap E_a\cap E_a&=P\\
E_a\cap E_a\cap\mathcal{C}_{a b}&=0\\
E_a\cap\mathcal{C}_{a b}\cap \mathcal{C}_{a b}&=-P\\
\mathcal{C}_{a b}\cap\mathcal{C}_{a b}\cap\mathcal{C}_{a b}&=2P
\end{split}
\label{tripleintersection}\end{equation}
From \eqref{tripleintersection} and \eqref{rels2}, we construct the dual basis for $H_2(\mathcal{C}_{0,5})$:

\begin{equation}
\begin{split}
s &= K\cap K -\sum_{b=1}^5 E_b\cap E_b\\
s_a &= s+ E_a\cap E_a\\
s_{a b} &= K\cap\mathcal{C}_{a b}
\end{split}
\label{dualbasis}\end{equation}
The fiber

\begin{displaymath}
\begin{split}
F&=D_{1 2 6}\cap D_{3 4}+D_{1 2}\cap D_{3 4} + D_{1 2}\cap D_{3 4 6}\\&= K\cap K -\mathcal{C}_{1 2}\cap(E_1+E_2 +\mathcal{C}_{1 2})
-\mathcal{C}_{3 4}\cap(E_3+E_4 +\mathcal{C}_{3 4})-\sum_{a=1}^4 E_a\cap E_a\\
&=s_1+s_2+s_3+s_4+s_5-2s
\end{split}
\end{displaymath}
satisfies

\begin{displaymath}
\begin{split}
F\cap E_a&=P,\quad a=1,\dots 5\\
F\cap K&=-2P
\end{split}
\end{displaymath}
\hypertarget{grothendieckriemannroch}{}\subsubsection*{{Grothendieck-Riemann-Roch}}\label{grothedieckriemannroch}

We are interested in the direct image under the projection

\begin{displaymath}
\pi: \mathcal{C}_{0,n}\to \overline{\mathcal{M}}_{0,n}
\end{displaymath}
The GRR Theorem says

\begin{equation}
ch(\pi_* V) = \pi_*[ch(V)Td(\pi)]
\label{GRR}\end{equation}
We can compute following Zvonkine's notes \cite{MR2952773} as follows.

Here $Td(\pi)$ is obtained from $Td(\pi^\vee)$ by reversing the signs of all of the odd entries and

\begin{equation}
Td(\pi^\vee)=\frac{Td(\Kc)}{Td(\mathcal{O}_\Delta)}
\label{ToddPiVee}\end{equation}
where $\Delta$ is the singularity set of the fiber (a set of codimension-2 in $\mathcal{C}_{0,n}$).

\paragraph*{\texorpdfstring{$\overline{\mathcal{M}}_{0,4}$}{ℳ̅₀,₄}\;:}

As a warmup, let's apply this to $\mathcal{C}_{0,4}\xrightarrow{\pi} \overline{\mathcal{M}}_{0,4}$. $\Delta$ consists of 3 points

\begin{displaymath}
\Delta_1=(0,1,1),\quad\Delta_2=(1,0,1),\quad\Delta_3=(1,1,0)
\end{displaymath}
Each has a Koszul resolution

\begin{displaymath}
0\to\mathcal{O}(-2H)\to\mathcal{O}(-H)\oplus\mathcal{O}(-H)\to \mathcal{O}\to \mathcal{O}_{\Delta_i}\to 0
\end{displaymath}
So

\begin{displaymath}
\begin{split}
Td(\mathcal{O}_{\Delta_i})&=Td(\mathcal{O}(-H))^{-2}Td(\mathcal{O}(-2H))\\
&=\left(\frac{1-e^H}{-H}\right)^{2}\left(\frac{-2H}{1-e^{2H}}\right)\\
&= 1-\tfrac{1}{12}P
\end{split}
\end{displaymath}
and

\begin{displaymath}
Td(\mathcal{O}_{\Delta})=\prod_{i=1}^3 Td(\mathcal{O}_{\Delta_i})= 1-\tfrac{1}{4}P
\end{displaymath}
Plugging this into \eqref{ToddPiVee}, we obtain

\begin{displaymath}
\begin{split}
Td(\pi^\vee)&=\frac{Td(\Kc)}{Td(\mathcal{O}_\Delta)}\\
&=\left(\frac{H-\sum E_a}{1-e^{-H+\sum E_a}}\right)\left(1+\frac{1}{4}P\right)\\
&=1+\tfrac{1}{2}(H-\sum_a E_a)
\end{split}
\end{displaymath}
so that, finally, we obtain

\begin{displaymath}
Td(\pi)= 1-\tfrac{1}{2}(H-\sum_a E_a)
\end{displaymath}
Now let $D=\sum_a E_a$ and 

\begin{displaymath}
\mathcal{L}_k= \KcD^{\otimes k}\otimes\mathcal{O}(-\sum E_a)
\end{displaymath}
We then have

\begin{displaymath}
\begin{split}
ch(\mathcal{L}_k)&=1+k H -\sum E_a +\tfrac{1}{2}(k^2-4) P\\
ch(\mathcal{L}_k)Td(\pi)&= 1+\tfrac{2k-1}{2}H-\tfrac{1}{2}\sum E_a +\tfrac{1}{2}k(k-1)P
\end{split}
\end{displaymath}
Then using

\begin{displaymath}
\pi_* 1=0,\qquad\pi_*H=2\; (\text{equivalently}\; \pi_*K=-2),\qquad\pi_* E_a=1,\qquad \pi_* P = P
\end{displaymath}
we plug into \eqref{GRR} to obtain

\begin{displaymath}
\begin{split}
ch(\pi_*(\mathcal{L}_k))&=\pi_*[ch(\mathcal{L}_k)Td(\pi)]\\
&= (2k-3) +\tfrac{1}{2}k(k-1)P
\end{split}
\end{displaymath}
More generally, for

\begin{displaymath}
\mathcal{L}_k= \KcD^{\otimes k} \otimes\mathcal{O}(-\sum \chi^{(a)}_k E_a)
\end{displaymath}
we have

\begin{displaymath}
ch(\mathcal{L}_k)= 1+k H -\sum\chi^{(a)}_k E_a +\tfrac{1}{2}\left(k^2-\sum\bigl(\chi^{(a)}_k\bigr)^2\right)P
\end{displaymath}
and hence

\begin{displaymath}
ch(\pi_*(\mathcal{L}_k))= (2k+1-\sum_a \chi^{(a)}_k)+\tfrac{1}{2}\bigl[k(k-1)-\sum_a\chi^{(a)}_k(\chi^{(a)}_k-1)\bigr]P
\end{displaymath}
Of course, since $\overline{\mathcal{M}}_{0,4}=\mathbb{CP}^1$, $\pi_*(\mathcal{L}_k)$ splits as a direct sum of line bundles, so there's more refined (and harder-to-compute) discrete information available. The holomorphic structures of vector bundles on $\overline{\mathcal{M}}_{0,5}=\text{dP}_4$ vary continuously, so the Chern classes we compute using Grothendieck-Riemann-Roch lose even more information.

\paragraph*{\texorpdfstring{$\overline{\mathcal{M}}_{0,5}$}{ℳ̅₀,₅}\;:}

Let's turn to the case of interest, $\mathcal{C}_{0,5}\to\overline{\mathcal{M}}_{0,5}$.

The singularity sets are labeled by the pair of points that collide

\begin{displaymath}
\begin{split}
\Delta_{a b}&= D_{a b} \cap D_{a b6}\\
&= \bigl(K-(\mathcal{C}_{c d}+\mathcal{C}_{c e}+\mathcal{C}_{d e})+E_a+E_b\bigr)\cap\mathcal{C}_{a b}\\
&=(K+E_a+E_b)\cap\mathcal{C}_{a b}
\end{split}
\end{displaymath}
Using the Koszul resolution

\begin{displaymath}
0\to \mathcal{O}(-D_{a b}-D_{a b 6}) \to
\mathcal{O}(-D_{a b})\oplus \mathcal{O}(-D_{a b 6}) \to 
\mathcal{O}\to \mathcal{O}_{\Delta_{a b}}\to 0
\end{displaymath}
we find

\begin{displaymath}
Td(\Delta_{a b}) = 1 - \tfrac{1}{12} c_{a b}
\end{displaymath}
where, using \eqref{remaining},\eqref{rels1},\eqref{rels2},\eqref{dualbasis},

\begin{displaymath}
\begin{split}
c_{a b}&= D_{a b}\cap D_{a b 6}\\
&= (K+E_a+E_b)\cap\mathcal{C}_{a b}\\
&= -s_{a b}
\end{split}
\end{displaymath}
So the total Todd class of the singularity set is

\begin{displaymath}
Td(\Delta) = 1+\tfrac{1}{12} \sum_{a<b} s_{a b}
\end{displaymath}
and

\begin{displaymath}
\begin{split}
Td(\pi^\vee)&=\frac{Td(\Kc)}{Td(\Delta)}\\
&= 1 +\tfrac{1}{2} K +\tfrac{1}{12}(K\cap K -\sum_{a<b} s_{a b}) -\tfrac{1}{24} K\cap\sum_{a<b} s_{a b}\\
&= 1 +\tfrac{1}{2} K + \tfrac{1}{12} (K\cap K -\sum_{a<b} s_{a b})
\end{split}
\end{displaymath}
Hence

\begin{displaymath}
\begin{split}
Td(\pi)&=1 -\tfrac{1}{2} K + \tfrac{1}{12} (K\cap K -\sum_{a<b} s_{a b})\\
&=1 -\tfrac{1}{2} K + \tfrac{1}{12} (-4s+\sum_a s_a -\sum_{a<b} s_{a b})
\end{split}
\end{displaymath}
As before, denote $D=\sum_a E_a$. Then, for the line bundle

\begin{equation}\label{Ltopush}
\mathcal{L}= \KcD^{\otimes k}\otimes\mathcal{O}(-\sum_a\chi^{(a)}_k E_a-\sum_{a\lt b}\chi^{(a b)}_k\mathcal{C}_{a b})
\end{equation}
we have

\begin{displaymath}
\begin{split}
ch(\mathcal{L})Td(\pi)&=exp\Bigl(k K +\sum_a(k-\chi^{(a)}_k)E_a-\sum_{a\lt b} \chi^{(a b)}_k\mathcal{C}_{a b}\Bigr) Td(\pi)\\
&=1+\left(\tfrac{2k-1}{2} K + \sum_a(k-\chi^{(a)}_k) E_a -\sum_{a\lt b}\chi^{(a b)}_k\mathcal{C}_{a b}\right)\\
&\quad +\tfrac{1}{12}\Bigl[
\bigl(-4+6k(k-1)-6\sum_a \chi^{(a)}_k(\chi^{(a)}_k-1)+6\sum_{a<b} (\chi^{(a b)}_k)^2\bigr)s\\
&\qquad\qquad+
\sum_a \bigl(1+6\chi^{(a)}_k(\chi^{(a)}_k-1)-6\sum_{b\neq a} (\chi^{(a b)}_k)^2\bigr)s_a\\
&\qquad\qquad+\sum_{a\lt b} s_{a b}\left(-1+6\chi^{(a b)}_k(2\chi^{(a b)}_k-2\chi^{(a)}_k-2\chi^{(b)}_k+2k-1)\right)
\Bigr]\\
&\quad+\tfrac{1}{12}\Bigl[k(k-1)(2k-1) -\sum_a \chi^{(a)}_k(\chi^{(a)}_k-1)(2\chi^{(a)}_k-1)\\
&\qquad\qquad-\sum_{a\lt b} \bigl(\chi^{(a b)}_k(\chi^{(a b)}_k+1)(4\chi^{(a b)}_k-1) -6(\chi^{(a b)}_k)^2(\chi^{(a)}_k+\chi^{(b)}_k-k)\bigr) \Bigr]P
\end{split}
\end{displaymath}
The direct image of our homology basis is

\begin{equation}
\begin{split}
\pi_*(1)&=0\\
\pi_*(K)&= -2\\
\pi_*(E_a)&=1\\
\pi_*(\mathcal{C}_{a b})&=0\\
\pi_*(s)&= \tfrac{1}{2}\sum_{a<b} D_{a b}=\tfrac{1}{2}\delta\\
\pi_*(s_a)&= D_{b c} +D_{d e}\qquad a,b,c,d,e\;\text{distinct}\\
\pi_*(s_{a b})&=-D_{a b}\\
\pi_*(P)&= P
\end{split}
\label{dirim}\end{equation}
Note that $\delta$ is \emph{twice} an integer class and $(\tfrac{1}{2}\delta)\cap(\tfrac{1}{2}\delta)=5P$.

The nontrivial bit to calculate here are $\pi_*(s,s_a,s_{a b})$. Write

\begin{displaymath}
\pi^*(D_{a b})=\alpha K +\beta \sum_{c\neq a,b} E_c +\gamma(E_a+E_b) +\epsilon\Bigl(\mathcal{C}_{a b} -\sum_{\begin{smallmatrix}\scriptstyle c\lt d\\ \scriptstyle c,d\neq a,b\end{smallmatrix}}\mathcal{C}_{c,d}\Bigr)
\end{displaymath}
as the most general ansatz compatible with the symmetries and the triviality of \hbox{$D_{a b}+D_{c d} - D_{a c} -D_{b d}$}. From the projection formula $\pi_*(a)\cap b=a \cap \pi^*(b)$, we get

\begin{displaymath}
\begin{aligned}
\pi_*(s)&= \alpha(\tfrac{1}{2}\delta)\\
\pi_*(s_a)&= \tfrac{1}{2}\beta \sum_{b\neq a} D_{a b} +\tfrac{1}{6}(\beta +2\gamma) \sum_{\begin{smallmatrix}b\lt c\\ b,c\neq a\end{smallmatrix}}D_{b c}\\
\pi_*(s_{a b})&= -\epsilon D_{a b}
\end{aligned}
\end{displaymath}
Checking various cases of $(k;\vec{\chi}_k)$ lets us fix parameters. For $(k;\vec{\chi}_k)=(0;0,0,0,0,0)$, we have $ch(\pi_*\mathcal{L})=1$, which yields

\begin{displaymath}
\epsilon =\tfrac{1}{2}(4\alpha-3\beta-2\gamma)
\end{displaymath}
For $(k;\vec{\chi}_k)=(2;1,1,1,1,1)$, $\pi_*(\mathcal{L})=\mathcal{T}^\vee\coloneqq T^*(\overline{\mathcal{M}}_{0,5})(\log\delta)$ which fits into the exact sequence (see \eqref{logTangent})
\[
0\to T^*(\overline{\mathcal{M}}_{0,5}) \to T^*(\overline{\mathcal{M}}_{0,5})(\log\delta) \to \bigoplus \mathcal{O}_{D_{a b}}\to 0
\]
Taking Chern characters yields

\begin{displaymath}
\begin{split}
ch(\pi_*\mathcal{L}) &= ch(T^*(\overline{\mathcal{M}}_{0,5})) + \delta +5P\\
&= (2-\tfrac{1}{2}\delta -\tfrac{9}{2}P) +\delta + 5P\\
&= 2+\tfrac{1}{2}\delta + \tfrac{1}{2}P
\end{split}
\end{displaymath}
which says $\alpha=1$.

Finally, for $(k;\vec{\chi}_k)=(3;1,2,2,2,2)$, we have $\pi_*\mathcal{L}= I_1$, the point line bundle, so

\begin{displaymath}
ch(\pi_*\mathcal{L})=1+(D_{1 2}+D_{1 3} +D_{4 5}) +\tfrac{1}{2} P
\end{displaymath}
which fixes $\beta=0$, $\gamma=1$. Putting all these together yields \eqref{dirim}.

Using \eqref{dirim}, for general $\mathcal{L}$ of the form \eqref{Ltopush}, we get

\begin{equation}
\begin{split}
ch(\pi_*\mathcal{L})
&= \bigl(3k+1-\sum_a\chi^{(a)}_k\bigr)\\
&\qquad +
\tfrac{1}{6}\Bigl[
3\bigl(k(k-1)-\sum_a \chi^{(a)}_k(\chi^{(a)}_k-1)+\sum_{a<b} (\chi^{(a b)}_k)^2\bigr)(\tfrac{1}{2}\delta)\\
&\qquad\qquad+
\sum_{a\lt b}\Bigl(\sum_{c\neq a,b}\bigl(\chi^{(c)}_k(\chi^{(c)}_k-1)-(\chi^{(a c)}_k)^2-(\chi^{(b c)}_k)^2\bigr)
-2\sum_{\mathclap{\begin{smallmatrix}c,d\neq a,b\\c<d\end{smallmatrix}}}(\chi^{(c d)}_k)^2\\
&\qquad\qquad 
-\chi^{(a b)}_k(2\chi^{(a b)}_k-2\chi^{(a)}_k-2\chi^{(b)}_k+2k-1)\Bigr)D_{a b}
\Bigr]\\
&\qquad+\tfrac{1}{12}\Bigl[k(k-1)(2k-1) -\sum_a \chi^{(a)}_k(\chi^{(a)}_k-1)(2\chi^{(a)}_k-1) \\
&\qquad\qquad
-\sum_{a\lt b} \bigl(\chi^{(a b)}_k(\chi^{(a b)}_k+1)(4\chi^{(a b)}_k-1) -6(\chi^{(a b)}_k)^2(\chi^{(a)}_k+\chi^{(b)}_k-k)\bigr)\Bigr]P
\end{split}
\label{M05directimage}\end{equation}
In general, the direct image $\pi_*(\mathcal{L})$ is not a vector bundle, but a torsion-free sheaf.  As in \S\ref{AFamilies},\S\ref{twisting}, we need to twist $\mathcal{L}$ by a combination of components of the nodal curves at the boundary.
\begin{itemize}
\item If $\chi^{(a)}_k+\chi^{(b)}_k-k-1\coloneqq n^{(ab)}>0$, then we replace $\mathcal{L}$ by $\mathcal{L}\otimes \mathcal{O}(-n^{(ab)}\mathcal{C}_{ab})$.
\item If $\chi^{(a)}_k+\chi^{(b)}_k+\chi^{(c)}_k-2k-1\coloneqq n^{(abc)}>0$ then we further twist by tensoring with $\mathcal{O}(-n^{(abc)}\mathcal{C}_{abc})$ where
\[
\mathcal{C}_{abc} \simeq (K+D)-E_a -E_b -E_c -\mathcal{C}_{ab}-\mathcal{C}_{ac}-\mathcal{C}_{bc}
\]
\end{itemize}
For example, consider
\[
\mathcal{L}= \KcD^{\otimes 9}\otimes \mathcal{O}(-E_1-6E_2-6E_3-7E_4-7E_5)
\]
The direct image $\pi_*(\mathcal{L})$ has rank-1, but it's not a line bundle on $\overline{\mathcal{M}}_{0,5}$.  We have to twist
\[
\begin{split}
\mathcal{L}' &= \mathcal{L}\otimes \mathcal{O}(-2\mathcal{C}_{2 3}-3\mathcal{C}_{2 4}-3\mathcal{C}_{2 5}-3\mathcal{C}_{3 4}-3\mathcal{C}_{3 5}-4\mathcal{C}_{4 5}-\mathcal{C}_{245} -\mathcal{C}_{345})\\
&\simeq \KcD^{\otimes 7}\otimes \mathcal{O}(-E_1-5E_2-5E_3-5E_4-5E_5-2\mathcal{C}_{23}-2\mathcal{C}_{24}-2\mathcal{C}_{25}-2\mathcal{C}_{3 4}
-2\mathcal{C}_{3 5}-2\mathcal{C}_{4 5})
\end{split}
\]
and $\pi_*(\mathcal{L}')=I_1$, the point line bundle on $\overline{\mathcal{M}}_{0,5}$.

Of particular relevance are the line bundles $\mathcal{J}$ on the universal curve, whose existence was conjectured in \S\ref{imposing}. Without loss of generality, assume the constraint to be imposed is at $p_1$. Then we have a line bundle of the form
\begin{equation}\label{trialJ}
\mathcal{J}_0 = \KcD^{\otimes k}\otimes\mathcal{O}(-\sum_{a=1}^5 \chi_aE_a)
\end{equation}
with
\[
\sum_{a=1}^5 \chi_a = 3k
\]
(so that the direct image has rank-1) and
\[
\chi_1<\chi_a<k,\qquad a=2,3,4,5
\]
Without loss of generality, we might as well assume that
\[
\chi_a+\chi_b+\chi_c \leq 2k+1 \qquad\forall 1<a<b<c<5
\]
Otherwise, we would have to twist by a multiple of $\mathcal{C}_{abc}$, which lowers the effective value of $k$ (without changing $\chi_1$). Given such a $\mathcal{J}_0$, then the twisting at the boundary yields
\begin{equation}
\mathcal{J} = \mathcal{J}_0 \otimes\mathcal{O}(-\sum_{1<a<b}\chi_{ab} \mathcal{C}_{ab}), \qquad \chi_{ab}\coloneqq\max(0,\chi_a+\chi_b-k-1)
\end{equation}
which has a direct image which is a line bundle on $\overline{\mathcal{M}}_{0,5}$. That line bundle is a subsheaf of a power of the point line bundle
\begin{equation}\label{Jdirectimage}
\pi_*(\mathcal{J})= (I_1)^{\otimes \chi_1}\otimes \mathcal{O}(-\sum_{a=2}^5 \ell_a D_{1a}),\qquad \ell_a\coloneqq\max(0,\chi_1+\chi_a -k)
\end{equation}
as desired.

\section{\texorpdfstring{$\mathfrak{so}(8)$}{so(8)} on the 4-Punctured Sphere}\label{so8complete}

Since the prescription for implementing the constraints and constructing the resulting family of spectral curves is a bit complicated, we present here the complete solution for any collection of 4 nilpotent orbits in $D_4$.

We first construct a basis of $\phi_k$, global sections of $\KcD^{\otimes k}\otimes \mathcal{O}(-E_1-E_2-E_3-E_4)$ good over the northern hemisphere $\overline{\mathcal{M}}_{0,4}\backslash\{\lambda=\infty\}$.
\begin{equation}\label{so8basisNorth}
    \begin{split}
    \phi_2&=x(y-z)a\\
    \phi_4&=x^2(y-z)^2 b_1\\
    &\quad + x(y-z)[y z b_2+ z(z-x) b_3 - y(x-y) b_4 -(x-y)(z-x) b_5]\\
    \phi_6&=x^3(y-z)^3 c_1\\
    &\quad+ x^2(y-z)^2[y z c_2 +z(z-x) c_3 - y(x-y) c_4 - (x-y)(z-x) c_5]\\
    &\quad+ x(y-z)[x^2 y z c_6+ z(z-x)(y-z)^2 c_7 - y(x-y)(y-z)^2 c_8 - x^2(x-y)(z-x)c_9]\\
    \tilde{\phi}&= x^2(y-z)^2 \tilde{b}_1\\
    &\quad + x(y-z)[y z \tilde{b}_2+ z(z-x) \tilde{b}_3 - y(x-y) \tilde{b}_4 -(x-y)(z-x) \tilde{b}_5]\\
    \end{split}
\end{equation}
Over the southern hemisphere, we write
\begin{equation}\label{so8basisSouth}
    \begin{split}
    \phi_2&=-y(z-x)\hat{a}\\
    \phi_4&=y^2(z-x)^2 \hat{b}_1\\
    &\quad -y(z-x)[y z \hat{b}_2+ z(z-x) \hat{b}_3 - y(x-y) \hat{b}_4 -(x-y)(z-x) \hat{b}_5]\\
    \phi_6&=-y^3(z-x)^3 \hat{c}_1\\
    &\quad+ y^2(z-x)^2[y z \hat{c}_2 +z(z-x) \hat{c}_3 - y(x-y) \hat{c}_4 - (x-y)(z-x) \hat{c}_5]\\
    &\quad-y(z-x)[x^2 y z \hat{c}_6+ z(z-x)(y-z)^2 \hat{c}_7 - y(x-y)(y-z)^2 \hat{c}_8 - x^2(x-y)(z-x)\hat{c}_9]\\
    \tilde{\phi}&= y^2(z-x)^2 \hat{\tilde{b}}_1\\
    &\quad -y(z-x)[y z \hat{\tilde{b}}_2+ z(z-x) \hat{\tilde{b}}_3 - y(x-y) \hat{\tilde{b}}_4 -(x-y)(z-x) \hat{\tilde{b}}_5]\\
    \end{split}
\end{equation}
On the overlap between the two hemispheres, we have
\begin{equation}
\begin{gathered}
c_1=\lambda^3 \hat{c}_1\\
b_1=\lambda^2\hat{b}_1,\;\; \tilde{b}_1=\lambda^2\hat{\tilde{b}}_1,\\
c_2=\lambda^2\hat{c}_2,\;\;c_3=\lambda^2\hat{c}_3,\;\;c_4=\lambda^2\hat{c}_4,\;\;c_5=\lambda^2\hat{c}_5,\\
a=\lambda\hat{a},\\
b_2=\lambda\hat{b}_2,\;\;b_3=\lambda\hat{b}_3,\;\;b_4=\lambda\hat{b}_4,\;\;b_5=\lambda\hat{b}_5,\\
\tilde{b}_2=\lambda\hat{\tilde{b}}_2,\;\;b_3=\lambda\hat{\tilde{b}}_3,\;\;b_4=\lambda\hat{\tilde{b}}_4,\;\;b_5=\lambda\hat{\tilde{b}}_5,\\
c_6=\lambda\hat{c}_6,\;\;c_7=\lambda\hat{c}_7,\;\;c_8=\lambda\hat{c}_8,\;\;c_9=\lambda\hat{c}_9
\end{gathered}
\end{equation}
where we used \eqref{C04fiber}
\[
C_\lambda=\{\lambda x(y-z)+y(z-x)=0\}
\]
Here we see that $c_1$ is the fiber coordinate on $\mathcal{O}(3)$ (in its trivialization over the northern hemisphere), $b_1,\tilde{b}_1,c_2,c_3,c_4,c_5$ are fiber coordinates on $\mathcal{O}(2)$ and the remaining 13 coefficients are fiber coordinates on $\mathcal{O}(1)$. The corresponding hatted coefficients are the fiber coordinates in the southern hemisphere.

For 4 regular nilpotents, the family of spectral curves is 
\begin{equation}\label{so84full}
    \Sigma= \left\{0=w^8+w^6\phi_2+w^4\phi_4+w^2\phi_6+\tilde{\phi}^2\right\}
\end{equation}
where we use \eqref{so8basisNorth} and \eqref{so8basisSouth}, respectively, in the northern and southern hemispheres of $\overline{\mathcal{M}}_{0,4}$.

For every other choice of nilpotents, the spectral curve is some specialization of \eqref{so84full}, given by the table below.

{
\setlength\LTleft{-.125in}
\scriptsize
\renewcommand{\arraystretch}{4}
\renewcommand{\tabcolsep}{2pt}

\begin{longtable}{|c|c|c|c|c|c|}
\hline
$O_N$&$O_H$&$E_1$&$E_2$&$E_3$&$E_4$\\
\hline 
\endhead
$[1^8]$&$[7,1]$ & -- & -- & -- & -- \\
\hline
$[2^2,1^4] $&$[5,3]$ &
$c_9=0$ &
$c_8=0$ &
$c_7=0$ &
$c_6=0$ \\
\hline
$\textcolor{midgreen}{[3,1^5]}$&$\textcolor{midgreen}{[5,1^3]}$ &
$\begin{gathered}c_9=0\\ \tilde{b}_5=0\end{gathered}$ &
$\begin{gathered}c_8=0\\ \tilde{b}_4=0\end{gathered}$ &
$\begin{gathered}c_7=0\\ \tilde{b}_3=0\end{gathered}$ &
$\begin{gathered}c_6=0\\ \tilde{b}_2=0\end{gathered}$ \\
\hline

$\textcolor{red}{[2^4]}$&$\textcolor{red}{[4^2]}$ &
$\begin{gathered}c_9=0\\ b_5+2\tilde{b}_5=0\end{gathered}$ &
$\begin{gathered}c_8=0\\ b_4+2\tilde{b}_4=0\end{gathered}$ &
$\begin{gathered}c_7=0\\ b_3+2\tilde{b}_3=0\end{gathered}$ &
$\begin{gathered}c_6=0\\ b_2+2\tilde{b}_2=0\end{gathered}$ \\
\hline
$\textcolor{blue}{[2^4]}$&$\textcolor{blue}{[4^2]}$ &
$\begin{gathered}c_9=0\\ b_5-2\tilde{b}_5=0\end{gathered}$ &
$\begin{gathered}c_8=0\\ b_4-2\tilde{b}_4=0\end{gathered}$ &
$\begin{gathered}c_7=0\\ b_3-2\tilde{b}_3=0\end{gathered}$ &
$\begin{gathered}c_6=0\\ b_2-2\tilde{b}_2=0\end{gathered}$ \\
\hline
$[3,2^2,1]$&$[3^2,1^2]$ &
$\begin{gathered}c_9=0\\b_5=\tilde{b}_5=0\end{gathered}$ &
$\begin{gathered}c_8=0\\b_4=\tilde{b}_4=0\end{gathered}$ &
$\begin{gathered}c_7=0\\b_3=\tilde{b}_3=0\end{gathered}$ &
$\begin{gathered}c_6=0\\b_2=\tilde{b}_2=0\end{gathered}$ \\
\hline
$[3^2,1^2]$&$[3^2,1^2]$ &
$\begin{gathered}c_9=0\\ c_5=\alpha_1^2\\b_5=\tilde{b}_5=0\end{gathered}$ &
$\begin{gathered}c_8=0\\ c_4=\alpha_2^2\\b_4=\tilde{b}_4=0\end{gathered}$ &
$\begin{gathered}c_7=0\\ c_3=\alpha_3^2\\b_3=\tilde{b}_3=0\end{gathered}$ &
$\begin{gathered}c_6=0\\ c_2=\alpha_4^2\\b_2=\tilde{b}_2=0\end{gathered}$ \\
\hline
$\color{midgreen}[5,1^3]$&$\color{midgreen}[3,1^5]$&$
\begin{gathered}
c_9=c_5=0\\ 
\begin{aligned}
c_1+(1-\lambda)&c_3- \lambda c_4 \\ 
 &- \lambda(1-\lambda)c_6=0
 \end{aligned}\\ 
b_5=\tilde{b}_5=0\\ 
\tilde{b}_1+(1-\lambda)\tilde{b}_3  -\lambda\tilde{b}_4=0\\
\end{gathered}$&$
\begin{gathered}c_8=c_4=0\\ 
\begin{aligned}
c_1+(1-\lambda)&c_2-\lambda c_5 \\
& -\lambda(1-\lambda)c_7=0
\end{aligned}\\ 
b_4=\tilde{b}_4=0\\ 
\tilde{b}_1+(1-\lambda)\tilde{b}_2  -\lambda\tilde{b}_5=0
\end{gathered}$&$
\begin{gathered}c_7=c_3=0\\ 
\begin{aligned}
c_1+(1-\lambda)&c_5-\lambda c_2 \\
&  -\lambda(1-\lambda)c_8=0
\end{aligned}\\ 
b_3=\tilde{b}_3=0\\ 
\tilde{b}_1+(1-\lambda)\tilde{b}_5 -\lambda\tilde{b}_2=0
\end{gathered}$&$
\begin{gathered}c_6=c_2=0\\ 
\begin{aligned}
c_1+(1-\lambda)&c_4-\lambda c_3 \\
& -\lambda(1-\lambda)c_9=0
\end{aligned}\\ 
b_2=\tilde{b}_2=0\\ 
\tilde{b}_1+(1-\lambda)\tilde{b}_4  -\lambda\tilde{b}_3=0
\end{gathered}$\\ 
\hline
$\color{red}[4^2]$&$\color{red}[2^4]$&
$\begingroup\setlength{\arraycolsep}{2pt}
\begin{gathered}
c_9=c_5=0\\[-2pt]
\begin{aligned}
c_1 &{}+ (1-\lambda)c_3\\
&{}- \lambda c_4 - \lambda(1-\lambda)c_6\\
&= -a\left(
\tilde b_1 {}+ (1-\lambda)\tilde b_3
{}- \lambda \tilde b_4
\right)
\end{aligned}\\[-2pt]
b_5=\tilde b_5=0\\[-2pt]
\begin{aligned}
(b_1 &{}+ (1-\lambda)b_3 - \lambda b_4)\\
&{}\;+\; 2\left(
\tilde b_1 {}+ (1-\lambda)\tilde b_3
{}- \lambda \tilde b_4
\right)
\end{aligned}\\[-2pt]
= \tfrac14 a^2
\end{gathered}
\endgroup$&$
\begingroup\setlength{\arraycolsep}{2pt}
\begin{gathered}
c_8=c_4=0\\[-2pt]
\begin{aligned}
c_1 &{}+ (1-\lambda)c_2\\
&{}- \lambda c_5 - \lambda(1-\lambda)c_7\\
&= -a\left(
\tilde b_1 {}+ (1-\lambda)\tilde b_2
{}- \lambda \tilde b_5
\right)
\end{aligned}\\[-2pt]
b_4=\tilde b_4=0\\[-2pt]
\begin{aligned}
(b_1 &{}+ (1-\lambda)b_2 - \lambda b_5)\\
&{}\;+\; 2\left(
\tilde b_1 {}+ (1-\lambda)\tilde b_2
{}- \lambda \tilde b_5
\right)
\end{aligned}\\[-2pt]
= \tfrac14 a^2
\end{gathered}
\endgroup$&$
\begingroup\setlength{\arraycolsep}{2pt}
\begin{gathered}
c_7=c_3=0\\[-2pt]
\begin{aligned}
c_1 &{}+ (1-\lambda)c_5\\
&{}- \lambda c_2 - \lambda(1-\lambda)c_8\\
&= -a\left(
\tilde b_1 {}+ (1-\lambda)\tilde b_5
{}- \lambda \tilde b_2
\right)
\end{aligned}\\[-2pt]
b_3=\tilde b_3=0\\[-2pt]
\begin{aligned}
(b_1 &{}+ (1-\lambda)b_5 - \lambda b_2)\\
&{}\;+\; 2\left(
\tilde b_1 {}+ (1-\lambda)\tilde b_5
{}- \lambda \tilde b_2
\right)
\end{aligned}\\[-2pt]
= \tfrac14 a^2
\end{gathered}
\endgroup$&$
\begingroup\setlength{\arraycolsep}{2pt}
\begin{gathered}
c_6=c_2=0\\[-2pt]
\begin{aligned}
c_1 &{}+ (1-\lambda)c_4\\
&{}- \lambda c_3 - \lambda(1-\lambda)c_9\\
&= -a\left(
\tilde b_1 {}+ (1-\lambda)\tilde b_4
{}- \lambda \tilde b_3
\right)
\end{aligned}\\[-2pt]
b_2=\tilde b_2=0\\[-2pt]
\begin{aligned}
(b_1 &{}+ (1-\lambda)b_4 - \lambda b_3)\\
&{}\;+\; 2\left(
\tilde b_1 {}+ (1-\lambda)\tilde b_4
{}- \lambda \tilde b_3
\right)
\end{aligned}\\[-2pt]
= \tfrac14 a^2
\end{gathered}
\endgroup$ \\
\hline
$\color{blue}[4^2]$&$\color{blue}[2^4]$
& $\begingroup\setlength{\arraycolsep}{2pt}
\begin{gathered}
c_9=c_5=0\\[-2pt]
\begin{aligned}
c_1 &{}+ (1-\lambda)c_3\\
&{}- \lambda c_4 - \lambda(1-\lambda)c_6\\
&= +a\left(
\tilde b_1 {}+ (1-\lambda)\tilde b_3
{}- \lambda \tilde b_4
\right)
\end{aligned}\\[-2pt]
b_5=\tilde b_5=0\\[-2pt]
\begin{aligned}
(b_1 &{}+ (1-\lambda)b_3 - \lambda b_4)\\
&{}\;-\; 2\left(
\tilde b_1 {}+ (1-\lambda)\tilde b_3
{}- \lambda \tilde b_4
\right)
\end{aligned}\\[-2pt]
= \tfrac14 a^2
\end{gathered}
\endgroup$
& $\begingroup\setlength{\arraycolsep}{2pt}
\begin{gathered}
c_8=c_4=0\\[-2pt]
\begin{aligned}
c_1 &{}+ (1-\lambda)c_2\\
&{}- \lambda c_5 - \lambda(1-\lambda)c_7\\
&= +a\left(
\tilde b_1 {}+ (1-\lambda)\tilde b_2
{}- \lambda \tilde b_5
\right)
\end{aligned}\\[-2pt]
b_4=\tilde b_4=0\\[-2pt]
\begin{aligned}
(b_1 &{}+ (1-\lambda)b_2 - \lambda b_5)\\
&{}\;-\; 2\left(
\tilde b_1 {}+ (1-\lambda)\tilde b_2
{}- \lambda \tilde b_5
\right)
\end{aligned}\\[-2pt]
= \tfrac14 a^2
\end{gathered}
\endgroup$
& $\begingroup\setlength{\arraycolsep}{2pt}
\begin{gathered}
c_7=c_3=0\\[-2pt]
\begin{aligned}
c_1 &{}+ (1-\lambda)c_5\\
&{}- \lambda c_2 - \lambda(1-\lambda)c_8\\
&= +a\left(
\tilde b_1 {}+ (1-\lambda)\tilde b_5
{}- \lambda \tilde b_2
\right)
\end{aligned}\\[-2pt]
b_3=\tilde b_3=0\\[-2pt]
\begin{aligned}
(b_1 &{}+ (1-\lambda)b_5 - \lambda b_2)\\
&{}\;-\; 2\left(
\tilde b_1 {}+ (1-\lambda)\tilde b_5
{}- \lambda \tilde b_2
\right)
\end{aligned}\\[-2pt]
= \tfrac14 a^2
\end{gathered}
\endgroup$
& $\begingroup\setlength{\arraycolsep}{2pt}
\begin{gathered}
c_6=c_2=0\\[-2pt]
\begin{aligned}
c_1 &{}+ (1-\lambda)c_4\\
&{}- \lambda c_3 - \lambda(1-\lambda)c_9\\
&= +a\left(
\tilde b_1 {}+ (1-\lambda)\tilde b_4
{}- \lambda \tilde b_3
\right)
\end{aligned}\\[-2pt]
b_2=\tilde b_2=0\\[-2pt]
\begin{aligned}
(b_1 &{}+ (1-\lambda)b_4 - \lambda b_3)\\
&{}\;-\; 2\left(
\tilde b_1 {}+ (1-\lambda)\tilde b_4
{}- \lambda \tilde b_3
\right)
\end{aligned}\\[-2pt]
= \tfrac14 a^2
\end{gathered}
\endgroup$
\\ \hline
$[5,3]$&$[2^2,1^4]$
& $\begingroup\setlength{\arraycolsep}{2pt}
\begin{gathered}
c_9=c_5=0\\[-2pt]
\begin{aligned}
c_1 + (1-\lambda)&{}c_3- \lambda c_4\\
&{} - \lambda(1-\lambda)c_6 = 0
\end{aligned}\\[-2pt]
b_5=\tilde b_5=0\\[-2pt]
b_1 + (1-\lambda)b_3 - \lambda b_4= \tfrac14 a^2\\
\tilde b_1+(1-\lambda)\tilde b_3-\lambda \tilde b_4=0
\end{gathered}
\endgroup$
& $\begingroup\setlength{\arraycolsep}{2pt}
\begin{gathered}
c_8=c_4=0\\[-2pt]
\begin{aligned}
c_1 + (1-\lambda)&{}c_2 - \lambda c_5\\
&{}- \lambda(1-\lambda)c_7= 0
\end{aligned}\\[-2pt]
b_4=\tilde b_4=0\\[-2pt]
b_1 + (1-\lambda)b_2 - \lambda b_5= \tfrac14 a^2\\
\tilde b_1+(1-\lambda)\tilde b_2-\lambda \tilde b_5=0
\end{gathered}
\endgroup$
& $\begingroup\setlength{\arraycolsep}{2pt}
\begin{gathered}
c_7=c_3=0\\[-2pt]
\begin{aligned}
c_1 + (1-\lambda)&{}c_5 - \lambda c_2\\
&{}- \lambda(1-\lambda)c_8 = 0
\end{aligned}\\[-2pt]
b_3=\tilde b_3=0\\[-2pt]
b_1 + (1-\lambda)b_5 - \lambda b_2= \tfrac14 a^2\\
\tilde b_1+(1-\lambda)\tilde b_5-\lambda \tilde b_2=0
\end{gathered}
\endgroup$
& $\begingroup\setlength{\arraycolsep}{2pt}
\begin{gathered}
c_6=c_2=0\\[-2pt]
\begin{aligned}
c_1 + (1-\lambda)&{}c_4 - \lambda c_3\\
&{}- \lambda(1-\lambda)c_9 = 0
\end{aligned}\\[-2pt]
b_2=\tilde b_2=0\\[-2pt]
b_1 + (1-\lambda)b_4 - \lambda b_3= \tfrac14 a^2\\
\tilde b_1+(1-\lambda)\tilde b_4-\lambda \tilde b_3=0
\end{gathered}
\endgroup$
\\ \hline
\end{longtable}

}
The $\lambda$-dependence of some of the entries in the table requires some explanation.  The expression ``$(1-\lambda)$'' should be interpreted as the unique holomorphic section of $\mathcal{O}(1)$ which vanishes at the point $\lambda=1$ and takes the value 1 at the point $\lambda=0$. Similarly a factor of ``$\lambda$'' should be interpreted as the unique holomorphic section of $\mathcal{O}(1)$ which vanishes at the point $\lambda=0$ and takes the value 1 at the point $\lambda=1$. Tensoring with one of these sections maps a point in the total space of $\mathcal{O}(j)$ to a point in the total space of $\mathcal{O}(j+1)$, which projects to the same point on $\overline{\mathcal{M}}_{0,4}$. In the same vein, $\lambda(1-\lambda)$ should be interpreted as the unique section of $\mathcal{O}(2)$ with zeroes at $0,1$ and which takes the value $-2$ at the point $\lambda=2$. More prosaically, we can use \eqref{C04fiber} to substitute
\begin{equation}
\begin{gathered}
\lambda x(y-z)\to -y(z-x)\\
(1-\lambda)x(y-z)\to - z(x-y)\\
\lambda(1-\lambda)x^2(y-z)^2 \to yz(x-y)(z-x)
\end{gathered}
\end{equation}
in our expressions \eqref{so8basisNorth} for the $\phi_k$.

As an example, consider $[2^2,1^4] [2^2,1^4] [7,1] [7,1]$. From the table, we have
\[
\begin{gathered}
c_4=c_5=c_8=c_9=0\\
b_4=b_5=\tilde{b}_4=\tilde{b}_5=0
\end{gathered}
\]
and the remaining constraints read
\begin{equation}
\begin{split}
b_1+(1-\lambda)b_2&=\tfrac{1}{4}a^2\\
b_1+(1-\lambda)b_3&=\tfrac{1}{4}a^2\\
c_1+(1-\lambda)c_3&=\lambda(1-\lambda)c_6\\
c_1+(1-\lambda)c_2&=\lambda(1-\lambda)c_7
\end{split}
\end{equation}
Obviously, these set $b_2=b_3$ and
\[
\begin{split}
x^2(y-z)^2b_1&= x^2(y-z)^2\bigl(\tfrac{1}{4}a^2 -(1-\lambda)b_3\bigr)\\
&= x^2(y-z)^2\tfrac{1}{4}a^2 + x(y-z)z(x-y)b_3
\end{split}
\]
or
\[
\phi_4= x^2(y-z)^2 \tfrac{1}{4}a^4 + x(y-z)z^2 b_3
\]
Similarly, they set
\[
\begin{split}
c_2&= c_3+\lambda (c_6-c_7)\\
c_1&= -(1-\lambda)[c_3-\lambda c_6]
\end{split}
\]
which lets us simplify
\[
\phi_6= x^2(y-z)^2 z^2 c +x(y-z)z^2\bigl[xyc_6+z(z-x)c_7\bigr]
\]
where we have absorbed
\[
c \coloneqq c_3+2\lambda c_7
\]
(since both terms transform in $\mathcal{O}(2)$). Putting all of this together, we have the spectral curve
\begin{equation}
\begin{split}
\Sigma=\Bigl\{
0=w^8 + x(y-z)\bigl[
w^6 a&+w^4 \bigl(x(y-z) \tfrac{1}{4}a^4 + z^2 b_3\bigr)\\
&+w^2z^2\bigl(
x(y-z) c +(xyc_6+z(z-x)c_7)
\bigr)\bigr]
+ \bigl(x(y-z)z^2\tilde{b}_3\bigr)^2
\Bigr\}
\end{split}
\end{equation}
which fibers over the bundle of Hitchin bases $\mathcal{B}$ whose graded components are
\begin{equation}
\begin{split}
\mathcal{B}_2&=\pi_*(\mathcal{L}_2)=\mathcal{O}(1)\\
\mathcal{B}_4&=\pi_*(\mathcal{L}'_4)=\mathcal{O}(1)\\
\mathcal{B}_6&=\pi_*(\mathcal{L}'_6)=\mathcal{O}(2)\oplus 2\mathcal{O}(1)\\
\tilde{\mathcal{B}}&=\pi_*(\tilde{\mathcal{L}}')=\mathcal{O}(1)
\end{split}
\end{equation}
where
\begin{equation}
\begin{split}
\mathcal{L}_2&= \KcD^{\otimes 2}\otimes\mathcal{O}(-E_1-E_2-E_3-E_4)\\
\mathcal{L}'_4&= \KcD^{\otimes 4}\otimes\mathcal{O}(-3E_1-3E_2-E_3-E_4-\mathcal{C}_{12})\\
\mathcal{L}'_6&= \KcD^{\otimes 6}\otimes\mathcal{O}(-4E_1-4E_2-E_3-E_4-\mathcal{C}_{12})\\
\tilde{\mathcal{L}}'&= \KcD^{\otimes 4}\otimes\mathcal{O}(-3E_1-3E_2-E_3-E_4-\mathcal{C}_{12})\\
\end{split}
\end{equation}

\section{Rank-1 Hitchin Systems}\label{Karayayla}

Here we illustrate the point emphasized in the main text: when one takes a finite quotient of a Hitchin system, we generally need to do a resolution of singularities in order to obtain the ``quotient'' Hitchin system.  This is easiest to illustrate for systems of rank 1, 
which are rational elliptic surfaces with a section,
$\pi: X \to \mathbb{P}^1$,  
from which the fiber over $\infty$ has been removed. 
All automorphism groups of such surfaces have been classified by T. Karayayla in \cite{MR2900539}, 
when the j-invariant of the elliptic fibers varies over the base, 
and in \cite{MR3232639}, when the j-invariant is constant.
The massless case is invariant under an action of $\mathbb{C}^*$, 
so the j-invariant is constant, 
and there are only two singular fibers, over $0$ and $\infty$. 
There are 4 such surfaces, with singular fibers of types:
\[
I_0^* + I_0^*, \qquad II^* + II, \qquad III^* + III, \qquad IV^* + IV.
\]
This produces 7 systems $X_A$ where $A$ is one of the 7 types of singular fibers that can occur. 
We place $A$ over $0$ and the other fiber is placed over $\infty$ (and removed). These 7 choices of singular fiber are the ones listed in the classification of \cite{Argyres:2016xua}.

Each of these surfaces $X=X_A$ can be obtained from a product $E \times \mathbb{P}^1$
by dividing by a finite group and resolving the singularities of the quotient (cf. section 6 of \cite{MR3232639}). 
We focus on $X_{II^*}$ and $X_{IV^*}$. For both of these, the elliptic curve $E$ is 
$\mathbb{C}/\Lambda$, where $\Lambda$ is the hexagonal lattice, 
so that $j(E)=0$ and the equation can be taken to be:
\[
E:\qquad v^2=u^3+1.
\]
Choose a primitive sixth root of unity $\zeta_6$,
put $\zeta_3=\zeta_6^2$, and let the order-six automorphism of $E$ be
\[
\alpha_6(u,v)=(\zeta_3^{-1}u,-v).
\]
Then $\alpha_6^2(u,v)=(\zeta_3u,v)$, 
the standard order-three automorphism of $E$.

Consider the diagonal actions on $E\times\mathbb{P}^1_b$
\[
\alpha_6:(u,v,b)\longmapsto
(\zeta_3^{-1}u,-v,\zeta_6 b)
\]
and
\[
\alpha_3=\alpha_6^2:(u,v,b)\longmapsto
(\zeta_3u,v,\zeta_3 b).
\]

\leftline{\underline{The quotient by $\langle\alpha_3\rangle$:}}

For the quotient by $\langle\alpha_3\rangle$, near $b=0$ introduce
the invariant functions
\[
x=ub^2,\qquad y=vb^3,\qquad t=b^3.
\]
They satisfy
\[
y^2=x^3+t^2.
\]
Thus the relatively minimal resolution of
\[
(E\times\mathbb{P}^1)/\langle\alpha_3\rangle
\]
is the isotrivial rational elliptic surface
\[
X_3:\qquad y^2=x^3+t^2.
\]
We claim that $X_3=X_{IV^*}$, in other words that the singular fibers are
\[
\mathrm{IV}+\mathrm{IV}^*.
\]

To identify the singular fibers, one can use Tate's algorithm in its
particularly simple form for these $j=0$ Weierstrass equations.  For
an equation
\[
y^2=x^3+a_6(t)
\]
we have
\[
\Delta=-432a_6^2.
\]
At $t=0$ the equation of $X_3$ has
\[
\operatorname{ord}_0(a_6)=2,\qquad
\operatorname{ord}_0(\Delta)=4.
\]
Tate's algorithm therefore gives a fiber of type $\mathrm{IV}$.

At $t=\infty$, put $r=t^{-1}$.  The equation becomes
\[
y^2=x^3+r^{-2}.
\]
After the change of variables
\[
x=r^{-2}X,\qquad y=r^{-3}Y,
\]
the equation becomes
\[
Y^2=X^3+r^4.
\]
Thus
\[
\operatorname{ord}_0(a_6)=4,\qquad
\operatorname{ord}_0(\Delta)=8,
\]
and Tate's algorithm gives type $\mathrm{IV}^*$.

The group $Aut_0(X)$ of automorphisms of $X$ that preserve the 0-section is 
$\mathbb{C}^* \times \mathbb{Z}/2$.
Indeed, any $\alpha \in Aut_0(X)$ is induced by an automorphism of 
$E \times \mathbb{P}^1$ 
which preserves the zero section, maps fibers to fibers, and respects the action of $\omega$,
and conversely any such automorphism induces an $\alpha$.
In other words, $Aut_0(X)$ consists of 
rescalings 
$b \mapsto cb$
and rotations of $E$ by multiples of $2\pi/6$, 
modulo the action of $\omega$.
In other words:
\[
Aut_0(X) = (\mathbb{C}^* \times (\mathbb{Z}/6)) \ / \ (\mathbb{Z}/3) = \mathbb{C}^* \times (\mathbb{Z}/2).
\]
Explicitly, these automorphisms are
\[
c\in\mathbb{C}^*:\qquad
(x,y,t)\longmapsto(c^2x,c^3y,c^3t),
\]
together with the fiberwise elliptic involution
\[
\iota:(x,y,t)\longmapsto(x,-y,t).
\]
These two actions commute.

\leftline{\underline{The quotient by $\langle\alpha_6\rangle$:}}

For the full order-six quotient, the corresponding relatively minimal
elliptic surface is
\[
X_6 = X_{II}:\qquad y^2=x^3+s.
\]
It has Kodaira fibers
\[
\mathrm{II}+\mathrm{II}^*,
\]
at $s=0$ and $s=\infty$, respectively.

Again Tate's algorithm identifies the two fibers immediately.  
At $s=0$,
\[
\operatorname{ord}_0(a_6)=1,\qquad
\operatorname{ord}_0(\Delta)=2,
\]
so the fiber is of type $\mathrm{II}$.

At $s=\infty$, put $r=s^{-1}$.  After the change
\[
x=r^{-2}X,\qquad y=r^{-3}Y,
\]
the equation becomes
\[
Y^2=X^3+r^5.
\]
Hence
\[
\operatorname{ord}_0(a_6)=5,\qquad
\operatorname{ord}_0(\Delta)=10,
\]
so the fiber is of type $\mathrm{II}^*$.

Here
\[
\operatorname{Aut}_0(X_6)\simeq\mathbb{C}^*,
\]
realized by
\[
c\in\mathbb{C}^*:\qquad
(x,y,s)\longmapsto(c^2x,c^3y,c^6s).
\]
Notice that the fiberwise involution $y\mapsto-y$ is already the
element $c=-1$ of this $\mathbb{C}^*$; this is the reason there is no
additional $\mathbf Z/2$ factor.

 \leftline{\underline{The double quotient:}}

Since
\[
\langle\alpha_3\rangle \subset\langle\alpha_6\rangle
,\qquad
\langle\alpha_6\rangle
/\langle\alpha_3\rangle
\simeq\mathbf Z/2,
\]
there is a residual involution on the $\langle\alpha_3\rangle$-quotient.  
In the coordinates above, $\alpha_6$ acts by
\[
(x,y,t)\longmapsto(x,y,-t).
\]
Indeed, $x=ub^2$ and $y=vb^3$ are fixed, while $t=b^3$ changes
sign.  The invariant coordinate is therefore
\[
s=t^2,
\]
and the equation
\[
y^2=x^3+t^2
\]
descends precisely to
\[
y^2=x^3+s.
\]
Consequently
\[
X_{IV^*}/(\mathbf Z/2)\dashrightarrow X_{II^*}
\]
is birational, and after passing to the corresponding relatively
minimal resolutions one obtains the $\mathrm{II} + \mathrm{II}^*$
surface.

We can identify this involution inside
\[
\operatorname{Aut}_0(X_{IV^*})\simeq\mathbb{C}^*\times\mathbf Z/2.
\]
The element $c=-1$ of the $\mathbb{C}^*$-factor acts as
\[
(x,y,t)\longmapsto(x,-y,-t),
\]
while the generator $\iota$ of the second factor acts as
\[
(x,y,t)\longmapsto(x,-y,t).
\]
Their product is therefore
\[
(-1,\iota):\qquad
(x,y,t)\longmapsto(x,y,-t).
\]
Thus the subgroup by which one must divide is the diagonal order-two
subgroup
\[
\;
\left\langle(-1,\iota)\right\rangle
\subset \mathbb{C}^*\times\mathbf Z/2.
\;
\]
We conclude that $X_{II^*}$ is birational to the quotient of $X_{IV^*}$ by the involution
$(x,y,t)\mapsto(x,y,-t)$.

To see how this is related to our choice of whether to impose an odd-type constraint, consider the 3-punctured sphere in the $D_4$ theory, with Hitchin poles given by ${\color{red}[4^2]}{\color{blue}[4^2]}[3^2,1^2]$. The spectral curve when we impose the odd-type constraint (i.e.~when the dual Nahm orbit at the $3^{\text{rd}}$ puncture is $[3^2,1^2]_N$) is
\begin{equation}\label{anE6}
\begin{split}
\Sigma&=\{0=w^2(w^6+a^2x^2y^2(x-y)^2)\}\\
&=\{0=w^2(w^3+i a xy(x-y))(w^3-i a xy(x-y))\}
\end{split}
\end{equation}
whereas, for the dual Nahm orbit $[3,2^2,1]_N$,
\begin{equation}\label{Sp5}
\Sigma=\{0=w^2(w^6+bx^2y^2(x-y)^2)\}
\end{equation}
In both cases, the Hitchin fiber is the Prym variety for the involution $w\to -w$ of $\Sigma$. In \eqref{anE6}, the Prym of the sextic can be identified with the Jacobian of one of the cubic factors. So we immediately recognize that the Hitchin system is $X_{IV^*}$. Indeed, the corresponding SCFT is the $E_6$ Minahan-Nemeschansky theory with 8 free hypermultiplets \cite{Chacaltana:2011ze}. In \eqref{Sp5}, we have the $Sp(5)_7$ SCFT  with 4 free hypermultiplets. The latter SCFT indeed has $X_{II^*}$ as its associated symplectic Hitchin system \cite{Argyres:2016xua} and a 5-dimensional space of mass-deformations (corresponding to the Cartan of $\mathfrak{sp}(5)$, of which an $\mathfrak{sp}(2)\times\mathfrak{sp}(2)\times\mathfrak{sp}(1)$ is the manifest flavour symmetry of the punctures.).

\bibliographystyle{utphys}
\bibliography{refs}
\end{document}